\documentclass[a4paper,onecolumn,11pt,accepted=2025-08-19]{quantumarticle}
\pdfoutput=1
\usepackage[utf8]{inputenc}
\usepackage[english]{babel}
\usepackage[T1]{fontenc}
\usepackage{hyperref}

\usepackage{tikz}
\usepackage{lipsum}

\usepackage[nointlimits]{amsmath}
\usepackage{amssymb, cancel}
\usepackage{exscale}
\usepackage{graphicx,nicefrac} 
\usepackage{color}
\usepackage{stackengine}
\usepackage{txfonts}
\usepackage{algorithm}
\usepackage{mathtools}
\usepackage{hyperref}
\hypersetup{colorlinks=true, citecolor=magenta, linkcolor=magenta, urlcolor=magenta}
\usepackage{footnote}
\usepackage{subfigure}
\usepackage{color}
\usepackage{blkarray}
\usepackage{dsfont}
\usepackage{mathrsfs}
\usepackage{bbold}
\usepackage{bbm}

\makeatletter
\DeclareFontFamily{OMX}{MnSymbolE}{}
\DeclareSymbolFont{MnLargeSymbols}{OMX}{MnSymbolE}{m}{n}
\SetSymbolFont{MnLargeSymbols}{bold}{OMX}{MnSymbolE}{b}{n}
\DeclareFontShape{OMX}{MnSymbolE}{m}{n}{
    <-6>  MnSymbolE5
   <6-7>  MnSymbolE6
   <7-8>  MnSymbolE7
   <8-9>  MnSymbolE8
   <9-10> MnSymbolE9
  <10-12> MnSymbolE10
  <12->   MnSymbolE12
}{}
\DeclareFontShape{OMX}{MnSymbolE}{b}{n}{
    <-6>  MnSymbolE-Bold5
   <6-7>  MnSymbolE-Bold6
   <7-8>  MnSymbolE-Bold7
   <8-9>  MnSymbolE-Bold8
   <9-10> MnSymbolE-Bold9
  <10-12> MnSymbolE-Bold10
  <12->   MnSymbolE-Bold12
}{}

\let\llangle\@undefined
\let\rrangle\@undefined
\DeclareMathDelimiter{\llangle}{\mathopen}%
                     {MnLargeSymbols}{'164}{MnLargeSymbols}{'164}
\DeclareMathDelimiter{\rrangle}{\mathclose}%
                     {MnLargeSymbols}{'171}{MnLargeSymbols}{'171}
\makeatother

\graphicspath{{./figures/}}
\def\bra#1{{\left\langle #1 \right|}}
\def\ket#1{{\left| #1 \right\rangle}}
\def\sbra#1{{\left\llangle #1 \right|}}
\def\sket#1{{\left| #1 \right\rrangle}}
\def\sdbra#1{{\left( #1 \right|}}
\def\sdket#1{{\left| #1 \right)}}

\usepackage{amsthm}
\newtheorem{definition}{Definition}
\newtheorem{proposition}{Proposition}
\newtheorem{lemma}{Lemma}
\newtheorem{theorem}{Theorem}

\begin{document}
\title{A Theory of Direct Randomized Benchmarking}
\author{Anthony M. Polloreno}
\email{ampolloreno@gmail.com}
\affiliation{JILA, NIST and Department of Physics, University of Colorado, 440 UCB, Boulder, Colorado 80309, USA}
\affiliation{Quantum Performance Laboratory, Sandia National Laboratories, Livermore, CA 94550 and Albuquerque, NM 87185 }
\author{Arnaud Carignan-Dugas}
\affiliation{Institute for Quantum Computing and the Department of Applied Mathematics, University of Waterloo, Waterloo, Ontario N2L 3G1, Canada}
\affiliation{Keysight Technologies Canada, Kanata, ON K2K 2W5, Canada}
\author{Jordan Hines}
\affiliation{Department of Physics, University of California, Berkeley, CA 94720}
\affiliation{Quantum Performance Laboratory, Sandia National Laboratories, Livermore, CA 94550 and Albuquerque, NM 87185 }
\author{Robin Blume-Kohout}
\author{Kevin Young}
\author{Timothy Proctor}
\email{tjproct@sandia.gov}
\affiliation{Quantum Performance Laboratory, Sandia National Laboratories, Livermore, CA 94550 and Albuquerque, NM 87185 }
\maketitle

\begin{abstract}
Randomized benchmarking (RB) protocols are widely used to measure an average error rate for a set of quantum logic gates. However, the standard version of RB is limited because it only benchmarks a processor's native gates indirectly, by using them in composite $n$-qubit Clifford gates. Standard RB's reliance on $n$-qubit Clifford gates restricts it to the few-qubit regime, because the fidelity of a typical composite $n$-qubit Clifford gate decreases rapidly with increasing $n$. Furthermore, although standard RB is often used to infer the error rate of native gates, by rescaling standard RB's error per Clifford to an error per native gate, this is an unreliable extrapolation. Direct RB is a method that addresses these limitations of standard RB, by directly benchmarking a customizable gate set, such as a processor's native gates. Here we provide a detailed introduction to direct RB, we discuss how to design direct RB experiments, and we present two complementary theories for direct RB. The first of these theories uses the concept of error propagation or scrambling in random circuits to show that direct RB is reliable for gates that experience stochastic Pauli errors. We prove that the direct RB decay is a single exponential, and that the decay rate is equal to the average infidelity of the benchmarked gates, under broad circumstances. This theory shows that group twirling is not required for reliable RB. Our second theory proves that direct RB is reliable for gates that experience general gate-dependent Markovian errors, using similar techniques to contemporary theories for standard RB. Our two theories for direct RB have complementary regimes of applicability, and they provide complementary perspectives on why direct RB works. Together these theories provide comprehensive guarantees on the reliability of direct RB.
\end{abstract}

\section{Introduction}\label{sec:introduction}
Reliable, efficient, and flexible methods for benchmarking quantum computers are becoming increasingly important as 5-50+ qubit processors become commonplace. Isolated qubits or coupled pairs can be studied in detail with tomographic methods \cite{merkel2013self, blume2017demonstration, nielsen2020gate}, but tomography of general $n$-qubit processes requires resources that scale exponentially with $n$. Randomized benchmarking (RB) methods \cite{emerson2005scalable, emerson2007symmetrized,  knill2008randomized, magesan2011scalable, magesan2012characterizing, carignan2015characterizing, cross2016scalable, brown2018randomized, hashagen2018real, magesan2011scalable, magesan2012characterizing, magesan2012efficient, carignan2015characterizing, cross2016scalable, brown2018randomized, hashagen2018real, helsen2018new, Helsen2020-it, helsen2020general, Claes2020-cy, McKay2020-no} were introduced partly to avoid the scaling problems that afflict tomography. RB avoids these scaling problems because, in the most commonly used RB protocols \cite{magesan2011scalable, magesan2012characterizing, magesan2012efficient}, both the necessary number of experiments \cite{helsen2017multiqubit} and the complexity of the data analysis \cite{magesan2012characterizing} are independent of $n$. RB methods efficiently estimate a single figure of merit---the average infidelity of a gate set---by (i) running random circuits of varied depths ($d$) that should implement an identity operation, (ii) observing the probability that the input is successfully returned ($S_d$) versus $d$, and then (iii) fitting this data to an exponential. The decay rate of this exponential ($r$) is an estimate of the gates' average infidelity.

However, many RB methods have a different scaling problem, introduced by a \emph{gate compilation} step. Most RB protocols benchmark an $n$-qubit gate set that forms a group \cite{magesan2011scalable,magesan2012characterizing,carignan2015characterizing,cross2016scalable,brown2018randomized,hashagen2018real, magesan2012efficient, helsen2018new, Helsen2020-it, helsen2020general, Claes2020-cy}, so that they can use group twirls as their theoretical foundations. But a quantum processor's native operations do not normally form a group. Instead, subsets of a processor's native gates can be used to \emph{generate} a variety of different groups (e.g., the unitary group, the Clifford group, the Pauli group, etc). This is a problem for implementing group-based RB methods because, for many $n$-qubit groups, the number of basic gates required to implement a typical group element is very large, meaning that many group elements can only be implemented with low fidelity.
 
\begin{figure}[t!]
\begin{center}
\includegraphics[width=12cm]{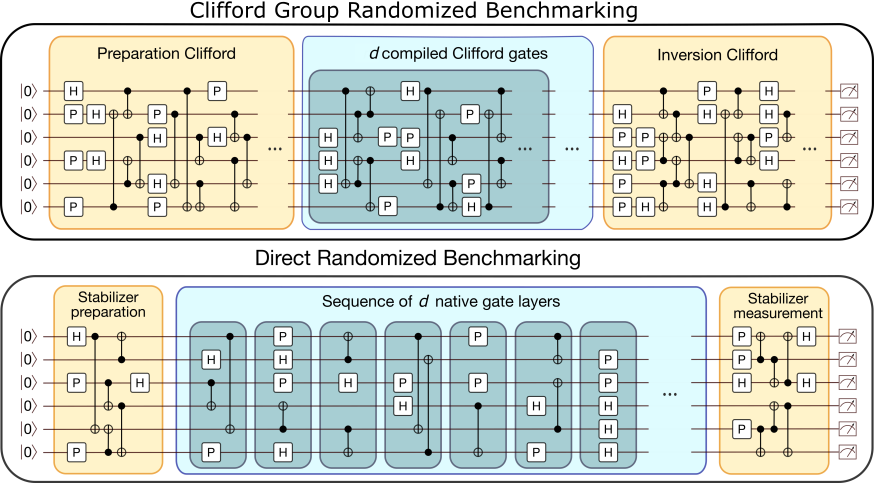}
\end{center}
\caption{An illustration (adapted from \cite{proctor2018direct}) comparing the varied-length random circuits used in the standard RB protocol of Magesan \emph{et al.}~\cite{magesan2011scalable,magesan2012characterizing} (``Clifford group RB'') and the streamlined direct RB protocol that was introduced in \cite{proctor2018direct} and that we investigate further here. Direct RB can be implemented on almost any $n$-qubit gate set (a set of circuit ``layers'' or ``cycles'') that generates a unitary 2-design. The case shown here is when that gate set is a set of Clifford gates (\textsc{cnot}, Hadamard, and the phase gate) that generate the $n$-qubit Clifford group. The circuits of direct RB can be shallower than those of Clifford group RB (e.g., consider the case of $d=0$ in each circuit), allowing for RB on more qubits.}
\label{Fig:schematic}
\end{figure}

The standard RB protocol---which we will call ``Clifford group RB''---benchmarks the $n$-qubit Clifford group \cite{magesan2011scalable,magesan2012characterizing}, estimating an error rate ($r$) that corresponds closely to the average infidelity of a gate from the $n$-qubit Clifford group \cite{proctor2017randomized, wallman2017randomized}. This protocol runs random sequences of $d + 2$ composite gates from that group (see the upper circuit in Fig.~\ref{Fig:schematic}), with a variable benchmark depth ($d$, with $d\geq 0$). The first $d + 1$ gates are sampled uniformly from the Clifford group, and the last gate is selected to invert the proceeding $d + 1$ gates. For a single qubit there are only 24 Clifford gates, so each Clifford gate can typically be compiled into a short sequence of native gates. But the Clifford group grows quickly with the number of qubits $n$: there are $2^{O(n^2)}$ $n$-qubit Clifford gates \cite{ozols2008clifford, koenig2014efficiently}, and a single typical $n$-qubit Clifford gate can be compiled into no fewer than $O(\nicefrac{n^2}{\log n})$ one- and two-qubit gates \cite{aaronson2004improved,patel2003efficient,bravyi2020hadamard}. This means that the fidelity of a random Clifford gate degrades rapidly with $n$ for a fixed quality of one- and two-qubit gates, i.e., $r \to 1$ quickly with increasing $n$. This makes it infeasible to reliably estimate Clifford group RB's error rate $r$ on many qubits---even with very high fidelity one- and two-qubit gates---because as $n$ increases even the shortest Clifford group RB circuits ($d = 0$, corresponding to a random $n$-qubit Clifford gate and its inverse) typically have such small success probabilities that they cannot be distinguished from $\nicefrac{1}{2^n}$ with a reasonable number of circuit repetitions (see Fig.~\ref{fig:simple-simulation}). Indeed, one- and two-qubit Clifford group RB has been widely implemented \cite{yoneda2018quantum,zajac2018resonantly,watson2018programmable,nichol2017high,veldhorst2014addressable,corcoles2013process,xia2015randomized,corcoles2015demonstration,chen2016measuring,muhonen2015quantifying,barends2014superconducting,raftery2017direct,rol2017restless,kelly2014optimal} but we are aware of only two publications presenting Clifford group RB experiments on more than two qubits \cite{mckay2017three, proctor2021scalable}, and none presenting Clifford group RB experiments on six or more qubits. 

An alternative to Clifford group RB is to benchmark a gate set that is a smaller group \cite{carignan2015characterizing, hashagen2018real, cross2016scalable, brown2018randomized, helsen2018new, helsen2020general}, enabling group elements to be implemented with fewer native gates. The gate compilation overhead can then be reduced from Clifford group RB, but often at a cost---the data analysis and the experiments become more complex. The simple behavior of Clifford group RB, i.e., the exponential decay of its success probability data, is guaranteed by a particular property of the Clifford group: conjugation of an error channel by a uniformly random Clifford group element ``twirls'' that channel into an $n$-qubit depolarizing channel. This follows because the Clifford group is a unitary 2-design \cite{gross2007evenly,dankert2009exact,wallman2014randomized} (or, equivalently, because the superoperator representation of Clifford gates consists of the direct sum of two irreducible representations of the Clifford group). The less like a unitary 2-design a group is (i.e., the more irreducible representations of the group its superoperator representation decomposes into) the weaker the effect of twirling over it is, and so the more complex the RB decay curve can be \cite{carignan2015characterizing, hashagen2018real, cross2016scalable, brown2018randomized, helsen2018new, helsen2020general} (see \cite{helsen2020general} for a theory of RB over general groups). For general groups, the RB decay curve is a sum of many exponentials, rather than the single exponential of Clifford group RB.

Another consequence of the gate compilation required in group-based RB protocols is that they measure an error \emph{per compiled group element}. There are infinitely many ways to compile each group element into native gates, and the observed error rate depends strongly on which group compilations is used. This is desirable if we aim to estimate the fidelity with which gates in that group can be implemented. But normally group-based RB error rates are used as a proxy for native gate performance \cite{muhonen2015quantifying,barends2014superconducting,raftery2017direct,mckay2017three}. In particular, it is common to estimate a native gate error rate from a Clifford group RB error rate, e.g., by dividing Clifford group RB's $r$ by the average circuit size of a compiled gate \cite{muhonen2015quantifying,barends2014superconducting,raftery2017direct} or by the average number of two-qubit gates in a compiled gate \cite{mckay2017three} (possibly after correcting for the estimated contribution of one-qubit gates). However, these rescaling methods have little theoretical justification \cite{epstein2014investigating,proctor2018direct}: in general, the resultant error rate is not a reliable estimate of the average infidelity of the native gates.

One solution to the limitations of Clifford group RB is \emph{direct randomized benchmarking} \cite{proctor2018direct}. Direct RB is a technique that is designed to retain the conceptual and experimental simplicity and efficiency of Clifford group RB while enabling direct benchmarking of a processor's native gates. Like all RB protocols, direct RB utilizes random circuits of variable length. Furthermore, like Clifford group RB, direct RB's data analysis simply consists of (1) fitting success probability data to a single exponential decay, and (2) rescaling the fit decay rate to obtain an estimate of the benchmarked gates' average infidelity. But, unlike Clifford group RB, the varied-length portion of direct RB circuits can consist of randomly sampled layers of a processor's native gates, rather than compiled group operations (see the lower circuit in Fig.~\ref{Fig:schematic}). Direct RB therefore enables estimating the average infidelity of native gate layers. The only restriction on the gate set that can be benchmarked with direct RB is that it must \emph{generate} a group that is a unitary 2-design (e.g., the $n$-qubit Clifford group). 

Direct RB addresses the main limitations of Clifford group RB, but it has been lacking a theory that proves that it is reliable. Those existing theories for RB that are applicable to direct RB \cite{helsen2020general, proctor2018direct, francca2018approximate} provide few guarantees on direct RB's behavior. In this paper we provide a detailed introduction to direct RB and two complementary theories of direct RB. These theories show that direct RB is reliable---the direct RB success probability follows an exponential decay and the direct RB error rate is the infidelity of the benchmarked gates---and explain why direct RB works. Shortly before the completion of this work, Heinrich \emph{et al.}~\cite{Heinrich2022-cs} presented a theory for \emph{filtered RB}, which is a technique that enables RB of general gate sets using random circuits containing i.d.d.~gates from that set, by using a classical post-processing of the data which they call \emph{filtering}. That work presents ideas and methods that are distinct from our work here (e.g., they analyze a different kind of RB) using related techniques.

This paper's main results and structure are as follows. In Section~\ref{sec:definitions} we introduce our notation and review the necessary background material. In Section~\ref{sec:drb} we provide a detailed definition and introduction to the direct RB protocol, and we compare it to other RB protocols for benchmarking native gate sets \cite{knill2008randomized, ryan2009randomized, francca2018approximate}. The definition for direct RB presented here generalizes that in \cite{proctor2018direct}. This section includes examples that illustrate how direct RB can be used to benchmark a variety of physically relevant gate sets---including universal gate sets on few qubits. We also present a simulation of direct RB that shows (approximately) how many qubits direct RB can \textit{feasibly} be used to benchmark, given the magnitude of current gate error rates. This limit depends on a variety of factors, but for realistic gate and readout error rates (between $0.1\%$ and $1\%$) our numerical results suggest that it is feasible to benchmark around 10-15 qubits with direct RB (although recent extensions of direct RB \cite{proctor2021scalable, hines2022demonstrating, Hines2024-qe}, not discussed in detail here, remain feasible well beyond this limit).

In Section~\ref{sec:stochastic-theory} we present a theory for direct RB on gates that experience stochastic Pauli errors, which is the relevant error model for gates that have undergone Pauli frame randomization \cite{Knill2005-xm, Ware2018-tc} or randomized compilation \cite{Wallman2016-rd, Hashim2020-qg}. This theory provides a physically intuitive underpinning for direct RB, by using the intuitive concepts of error propagation and scrambling in random circuits. Interestingly, this theory shows that group twirling is not necessary for reliable RB---direct RB can be reliable even when the benchmarked gate set is very far from approximating a unitary 2-design, and these insights are broadly applicable to benchmarks that use random circuits. This enables us to show that direct RB is reliable even in the $n\gg 1$ regime, where short sequences of layers of native gates cannot approximate a unitary 2-design.

One of the contributions of this paper is a theory for direct RB with general gate-dependent Markovian errors that is similar in nature to the group twirling (and Fourier analysis) theories for group-based RB methods. This theory uses the concept of a \emph{sequence-asymptotic unitary 2-design}, which we introduce in Section~\ref{sec:sequads}. A sequence-asymptotic unitary 2-design is any gate set for which length $d$ random sequences of gates from that set create a distribution over unitaries that converges to a unitary 2-design as $d\to\infty$. Sequence-asymptotic unitary 2-design are closely related to various existing notions of approximate or asymptotic unitary designs and the general theory of scrambling in random circuits \cite{dankert2009exact, gross2007evenly, brown2015decoupling, sekino2008fast, brandao2016local, Oszmaniec2022-gs, Hunter-Jones2018-zh,Von_Keyserlingk2018-ts,Nahum2018-zs, Heinrich2022-cs}. However, to our knowledge, the particular concept and technical results that we require for our theory of direct RB do not appear in the literature.

In Section~\ref{sec:general-theory} we present our theory for direct RB under general gate-dependent Markovian errors, which is based on twirling superchannels (or Fourier transforms on groups). First, we show that the depth-$d$ direct RB success probability can be computed exactly by taking the $d^{\textrm{th}}$ power of a superchannel ($\mathscr{R}$) constructed from the permutation matrix representation of the group generated by the benchmarked gate set. Then we show how the direct RB decay can be (approximately) rewritten in terms of a smaller superchannel ($\mathscr{L}$) that is constructed from the superoperator representation of the generated group. These results correspond to equivalent theories for Clifford group RB presented in \cite{proctor2017randomized, wallman2017randomized, merkel2018randomized, carignan2018randomized}. We then use the spectral decomposition of these superchannels, together with the theory of sequence-asymptotic unitary 2-designs, to show that the direct RB decay is approximately a singe exponential. Finally, we use the $\mathscr{L}$ superchannel to show that the direct RB error rate is equal to a gauge-invariant version of the mean fidelity of the benchmarked gates, building on Wallman's derivation of an equivalent result for Clifford group RB \cite{wallman2017randomized}.

Our two theories of direct RB are complementary, providing different perspectives as well as different regimes of applicability. We summarize these differences in Table~\ref{table:theories}. Our superchannel-based theory (Section~\ref{sec:general-theory}) is more mathematically rigorous than our scrambling-based theory (Section~\ref{sec:stochastic-theory}), and it applies to a more general class of errors. However, our superchannel-based theory relies on an approximation whose size increases with $n$, whereas our scrambling-based theory shows that direct RB is reliable even when $n\gg 1$. Therefore, together these theories provide comprehensive evidence for the reliability of direct RB.

\begin{table}
\setlength{\tabcolsep}{9pt} 
\renewcommand{\arraystretch}{1.2} 
\begin{center}
\begin{small}
\begin{tabular}{ l | c | c }
     & Theory in Section~\ref{sec:stochastic-theory} & Theory in Section~\ref{sec:general-theory} \\
  \hline
  \hline	
  Techniques & Error scrambling & Group Fourier transform \\
    & Stabilizer state theory & Unitary 2-designs \\
    \hline
  Assumptions & Clifford gates  & Any gates \\
    & Stochastic Pauli channels & General Markovian errors \\
    & $n \gg 1$ & Any $n$ \\
    & $\epsilon \ll 1$ & $\epsilon \ll 1$ \\
    & $\epsilon$ constant as $n$ increases & $\epsilon$ decreasing as $n$ increases \\
  \hline  
  Result & $r_{\Omega} \approx \epsilon_{\Omega}$ &  $r_{\Omega} \approx \epsilon_{\Omega}$ with caveats
\end{tabular}
\end{small}
\end{center}
\caption{A summary of the two complementary theories for direct RB that we present. The ``techniques'' row concisely describes the mathematical techniques used within each theory. The ``assumptions'' row describes the key assumptions made in each theory. Here $n$ denotes the number of qubits, and $\epsilon$ the error per $n$-qubit gate (a.k.a.~circuit layer) in the gate set being benchmarked. Both theories require that $\epsilon$ is small ($\epsilon \ll 1$), but our theory that is applicable to general Markovian errors has a much more stringent condition on how small $\epsilon$ is as a function of $n$. The ``result'' row summarizes the key result of each theory. Both theories show that direct RB's error rate ($r_{\Omega}$) is approximately equal to the weighted average of the benchmarked gates' infidelities ($\epsilon_{\Omega}$). However, in the case of general Markovian errors (the theory of Section~\ref{sec:general-theory}), this result has subtle caveats that are related to the difficulty in defining a gate's fidelity in the presence of general Markovian errors.}\label{table:theories}
\end{table}

\section{Definitions}\label{sec:definitions}
In this section we introduce our notation and review the necessary background material.

\subsection{Gates, gate sets, and circuits}
An $n$-qubit gate $G$ is a physical operation associated with an instruction to ideally implement a particular unitary evolution on $n$ qubits (we use the term ``gate'' for consistency with the RB literature---an $n$-qubit gate is also often referred to as a ``layer'' or a ``cycle''). We denote the unitary corresponding to $G$ by $U(G) \in \text{SU}(2^n)$. It will also often be convenient to use the superoperator representation of a unitary, so we define $\mathcal{G}(G)$ to be the linear map
\begin{equation}
\mathcal{G}(G)[\rho] := U(G) \rho U(G)^{\dagger},
\end{equation}
where $\rho$ is an $n$-qubit density operator. We consider a gate $G$ to be entirely defined by the unitary $U(G)$. RB methods are agnostic about how an $n$-qubit gate is implemented, except that an attempt must be made to faithfully implement the unitary it defines. We will use $\mathbb{G}$ to denote a set of $n$-qubit gates, which need not be a finite set, and, slightly abusing notation, we will also use $\mathbb{G}$ to denote the corresponding sets of unitaries (i.e., $\{U(G) \mid G \in \mathbb{G}\}$ and superoperators (i.e., $\mathcal\{\mathcal{G}(G) \mid G \in \mathbb{G}\}$) with the meaning implied by the context.

A quantum circuit $C$ over a $n$-qubit gate set $\mathbb{G}$ is a sequence of $d \geq 0$ gates from $\mathbb{G}$. We will write this as
\begin{equation}
C = G_d G_{d-1} \cdots G_2 G_1,
\end{equation}
 where each $G_i \in \mathbb{G}$, and we use a convention where the circuit is read from right to left. The circuit $C$ is an instruction to apply its constituent gates, $G_1$, $G_2$, $\dots$, in sequence, and it implements the unitary
 \begin{equation}
U(C) = U(G_d)U(G_{d-1}) \cdots U(G_2)U(G_1).
\end{equation}
 In direct RB, and most other RB methods, multiple gates in a circuit must \emph{not} be combined or compiled together by implementing a physical operation that enacts their composite unitary (there are ``barriers'' between gates).

\subsection{Fidelity}
In our theory for direct RB we will show how the direct RB error rate is related to the fidelity of the benchmarked gates. There are two commonly used version of a gate's fidelity: average gate fidelity and entanglement fidelity, defined below. Throughout this work, we will use the Markovian model for errors, whereby the imperfect implementation of a gate $G$ is represented by a complete positive and trace preserving (CPTP) superoperator $\tilde{\mathcal{G}}(G)$ [so, for low-error gates $\tilde{\mathcal{G}}(G) \approx \mathcal{G}(G)$].
The \emph{average gate fidelity} of $\tilde{\mathcal{G}}(G)$ to $\mathcal{G}(G)$ is defined by
\begin{align}
F_{\textsc{a}}(\tilde{\mathcal{G}},\mathcal{G})  : = \int d\psi \,\, \text{Tr} \left\{ \tilde{\mathcal{G}}[\ket{\psi}\bra{\psi}]\, \mathcal{G} [\ket{\psi}\bra{\psi}]  \right\},
\label{eq:AGI-def}
\end{align}
where $ d\psi$ is the normalized Haar measure \cite{nielsen2002simple}, and where here we have dropped the dependence of $\mathcal{G}$ and $\tilde{\mathcal{G}}$ on $G$ for brevity (we also do this elsewhere when convenient). 
The \emph{entanglement fidelity} is defined by
\begin{align}
F_{\textsc{e}}(\tilde{\mathcal{G}},\mathcal{G})  : = \langle \varphi |( \mathds{1} \otimes \Lambda)[\ket{\varphi}\bra{\varphi}]|\varphi \rangle,
\label{eq:EI-def}
\end{align}
where $\varphi$ is any maximally entangled state \cite{nielsen2002simple} and
\begin{equation}
\Lambda(G) = \tilde{\mathcal{G}}(G)\mathcal{G}(G)^{-1}. 
\end{equation}
The corresponding infidelities, \emph{average gate infidelity} ($\epsilon_{\textsc{a}}$) and \emph{entanglement infidelity} ($\epsilon_{\textsc{e}}$), are defined by
\begin{align}
\epsilon_{\textsc{a/e}}(\tilde{\mathcal{G}},\mathcal{G})  = 1 - F_{\textsc{a/e}}(\tilde{\mathcal{G}},\mathcal{G}).
\end{align}
These infidelities are related by \cite{nielsen2002simple}
\begin{equation}
\epsilon_{\textsc{e}} =  \frac{2^n+1}{2^n}\epsilon_{\textsc{a}}.
\end{equation}
The average gate [in]fidelity is more commonly used in the RB literature, but in this work we use the entanglement [in]fidelity. Typically we drop the subscript, letting $\epsilon \equiv \epsilon_{\textsc{e}}$. Note that a gate's [in]fidelity is not a ``gauge-invariant'' property of a gate set \cite{proctor2017randomized}, meaning that it is not technically measurable: we discuss this subtle point when presenting our theory for direct RB with general gate-dependent Markovian errors (Section~\ref{sec:general-theory}).

Randomized benchmarking methods are typically designed to measure the mean infidelity of a set of gates. Direct RB is designed to measure
\begin{equation}
\epsilon_{\Omega} = \sum_{G \in \mathbb{G}} \Omega(G) \epsilon\left[\tilde{\mathcal{G}}(G), \mathcal{G}(G)\right], \label{eq:epsilon-omega-def}
\end{equation}
where $\Omega$ is a probability distribution over the gate set $\mathbb{G}$. Note that, except where otherwise stated, $\mathbb{G}$ does not need to be a finite set, i.e., it can include gates with continuous parameters. The $\sum_{G\in \mathbb{G}}\Omega(G)$ notation is a shorthand for integrating and/or summing over $\mathbb{G}$ according to the measure $\Omega(G)$. When gates experience only stochastic Pauli errors, then $\epsilon_{\Omega}$ is equal to the probability that a Pauli error occurs on a gate sampled from $\Omega$.

\section{Direct randomized benchmarking}\label{sec:drb}
In this section we first define the direct RB protocol and explain what its purpose is (Section~\ref{sec:drb-def}). Direct RB has flexible, user-specified components, so we then provide guidance on how to choose these aspects of direct RB experiments (Sections~\ref{sec:stats}-\ref{sec:sampling}). We then explain why the direct RB protocol is defined the way it is, and we discuss how it differs from Clifford group RB (Sections~\ref{sec:circ-structure}-\ref{sec:randomized-out}). Finally, we compare direct RB to other RB protocols for directly benchmarking native gate sets (Section~\ref{sec:other-methods}).

\subsection{The direct RB protocol}\label{sec:drb-def}
We now define the $n$-qubit direct RB protocol, and explain what it aims to measure. The definition given here is more general than that in \cite{proctor2018direct}. For example, in \cite{proctor2018direct}, direct RB is defined for gate sets containing only Clifford gates. Our definition here permits non-Clifford gates. Note that our definition for direct RB applies to gates on a set of $n \geq 1$ qubits. Generalizing to $n\geq 1$ qudits of arbitrary dimension is conceptually simple, as is the case with Clifford group RB, but we do not pursue this here. In addition to parameters that set the number of samples (which exist in all RB protocols), direct RB on $n$ qubits has two flexible, user-specified inputs:
\begin{enumerate}
\item A set of $n$-qubit gates (a.k.a., layers or cycles) to benchmark $(\mathbb{G})$. 
\item A sampling distribution $\Omega$ over the gate set $\mathbb{G}$.
\end{enumerate}

We denote direct RB of the gate set $\mathbb{G}$ with the sampling distribution $\Omega$ by DRB$(\mathbb{G},\Omega)$. We call $\mathbb{G}$ a ``gate set'' for consistency with common RB terminology, but note that $\mathbb{G}$ contains $n$-qubit gates---i.e., ``circuit layers'' or ``cycles''---not one- and two-qubit gates. All $n$-qubit gate sets generate some $n$-qubit group, or a dense subset of some group. This group plays an important role in the direct RB protocol, and we denote it by $\mathbb{C}$ (for brevity, for a gate set $\mathbb{G}$ that only generates a dense subset of a group $\mathbb{C}$ we also refer to $\mathbb{G}$ as generating $\mathbb{C}$). In most of our examples this group will be the $n$-qubit Clifford group, but this is not required (the conditions required of $\mathbb{C}$ are stated in Section~\ref{sec:gate-set}, which include that it is a unitary 2-design).
  
Having introduced $\mathbb{G}$ and $\Omega$, we are now ready to define the direct RB protocol. DRB$(\mathbb{G},\Omega)$ is a protocol for estimating $\epsilon_{\Omega}$ [see Eq.~\eqref{eq:epsilon-omega-def}] and it consists of the following procedure:
\begin{enumerate}
\item {\bf Sample the circuits.} For each $d$ in some set of user-specified \emph{benchmark depths} all satisfying $d \geq 0$, and for $k = 1,2,\dots, K_d$ for some user-specified $K_d\geq 1$:

\begin{enumerate}
\renewcommand{\labelenumii}{\labelenumi\arabic{enumii}.}
\item Choose a target output $n$-bit string $s_{d,k}$ for this circuit. This can be set to the all-zeros bit string (as in the conventional implementation of Clifford group RB \cite{magesan2011scalable, magesan2012characterizing}), or it can be chosen uniformly at random (our recommendation).

\item Sample a group element $F_{\textrm{sp}}$ uniformly from $\mathbb{C}$ (the group generated by $\mathbb{G}$),  and then find a circuit $C_{\text{sp}}$ that creates the state 
\begin{equation}
\ket{\psi} =U(F_{\text{sp}})\ket{0}^{\otimes n},
\end{equation}
from $\ket{0}^{\otimes n}$ [$U(\cdot)$ maps a gate or circuit to the unitary it implements---see Section~\ref{sec:definitions}]. That is, find a circuit $C_{\text{sp}}$ that 
satisfies
\begin{align}
U(C_{\text{sp}})\ket{0}^{\otimes n} &=U(F_{\text{sp}})\ket{0}^{\otimes n},
\end{align}
meaning that $C_{\textrm{sp}}$ is only required to faithfully implement $U(F_{\textrm{sp}})$'s action on $\ket{0}^{\otimes n}$. The circuit $C_{\text{sp}}$ can contain any gates, including gates that are not in $\mathbb{G}$, and it is ideally chosen to maximize the fidelity with which $\psi$ is produced. The circuit $C_{\text{sp}}$ is the first part of the sampled direct RB circuit, and we refer to it as the circuit's state preparation subcircuit.
 
\item Independently sample $d$ gates, $G_1$, $G_2$, $\dots$, $G_d$, from $\Omega$ (the user-specified distribution over $\mathbb{G}$). The circuit $G_d \cdots G_2G_1$ is the next part of the sampled direct RB circuit, and we refer to it as the circuit's ``core''.

\item Find a circuit $C_{\text{mp}}$ that, when applied after the two parts of the circuit sampled so far, maps the qubits to $|s_{d,k}\rangle$, and append it to the circuit sampled so far. That is, $C_{\text{mp}}$ is a circuit that satisfies
\begin{equation}
U(C_{\rm{mp}}\,G_d\cdots G_2\,G_{1}\,C_{\rm{sp}}) \ket{0}^{\otimes n} =|s_{d,k}\rangle.
\end{equation}
The circuit $C_{\rm{mp}}$ is the final part of the sampled direct RB circuit. As with the circuit $C_{\textrm{sp}}$, $C_{\text{mp}}$ can contain any gates, and we refer to it as the circuit's measurement preparation subcircuit.

\end{enumerate}
The sampling procedure of 1.1-1.4 creates the circuit
\begin{equation}
C_{d,k} = C_{\rm{mp}}G_{d} \cdots  G_2G_1C_{\rm{sp}},
\end{equation}
that, if run on a perfect quantum computer, always outputs the bit string $s_{d,k}$. That is, $|\langle s_{d,k}|U(C_{d,k})\ket{0}^{\otimes n}|^2 = 1$.

\item {\bf Run the circuits.} Run each of the sampled circuits $N$ times, for some user-specified $N \geq 1$. Estimate each circuit's success probability ($S_{d,k}$), as the frequency that the circuit $C_{d,k}$ returns its success bit string $s_{d,k}$ (we denote this estimate by $\hat{S}_{d,k}$). Note that the direct RB protocol does not specify the order that the circuits are run, but this is important if there could be significant drift in the system over the time period of the entire experiment \cite{mavadia2017experimental,fogarty2015nonexponential,fong2017randomized}. We recommend ``rastering'' \cite{proctor2019detecting} if possible---meaning looping through all of the circuits $N$ times, running each circuit once in each loop---as this facilitates detecting the presence of drift and a time-resolved RB analysis \cite{proctor2019detecting} (these analyses are not discussed further herein). If rastering is not possible then the order that circuits are run should be randomized, as this will typically reduce the impact of drift on the results.

\item {\bf Analyze the data.} Fit the estimated average probability of success, $\hat{S}_{d} = \sum_{k} \hat{S}_{d,k}/K_{d}$, versus benchmark depth ($d$) to the exponential decay function
\begin{equation}
S_d = A + B p^d, \label{eq:fitfunc}
\end{equation}
 where $A$, $B$ and $p$ are fit parameters (this is the same as the fit function used in Clifford group RB). If the success bit strings were sampled uniformly then fix $A=\nicefrac{1}{2^n}$. The estimate of the DRB$(\mathbb{G},\Omega)$ error rate of the gates ($\hat{r}_{\Omega}$) is then defined to be
\begin{equation}
\hat{r}_{\Omega} =\frac{(4^n-1)}{4^n}(1-\hat{p}).
\label{eq:r-def}
\end{equation}
where $\hat{p}$ is the fit value of $p$.
\end{enumerate}
 
Our definition for the direct RB procedure specifies fitting the data to the standard exponential decay function of Eq.~\eqref{eq:fitfunc}, and so DRB$(\mathbb{G}, \Omega)$ is only well-behaved if the direct RB data has approximately this form. However, it is not clear \emph{a priori} that the direct RB average success probability will decay exponentially. Moreover, even if the decay is an exponential it is not obvious what the direct RB error rate ($r_{\Omega}$) measures. Proving that $\hat{S}_d \approx A + Bp^d$ under broad conditions, providing an intuitive explanation for why $\hat{S}_d$ decays exponentially, and then explaining what $r_{\Omega}$ quantifies are three of the main aims of this paper. In particular, we will show that $r_{\Omega} \approx \epsilon_{\Omega}$.

\subsection{A simple numerical demonstration of direct RB}
We now demonstrate that direct RB works correctly---the decay is an exponential and $r_{\Omega} \approx \epsilon_{\Omega}$---in two simple scenarios. First, consider the case of gate-independent global depolarizing noise, and assume perfect state preparation and measurement sub-circuits. In this model, after each $n$-qubit gate the $n$ qubits are mapped to the completely mixed state with some probability $1-\lambda$.
This is arguably the simplest possible error model, and we would expect the error rate measured by any well-formed RB protocol to be the infidelity of this global depolarizing channel, which is $\epsilon =(4^n-1)(1 - \lambda)/4^n$. An explicit calculation confirms that this is the case here: 
\begin{equation}
S_d = \nicefrac{1}{2^n} + (1 -\nicefrac{1}{2^n})\lambda^d,
\end{equation}
which implies that $p = \lambda$  and $r_\Omega = (4^n-1)(1 - \lambda)/4^n$. 

\begin{figure}[t!]
\begin{center}
\includegraphics[width=9cm]{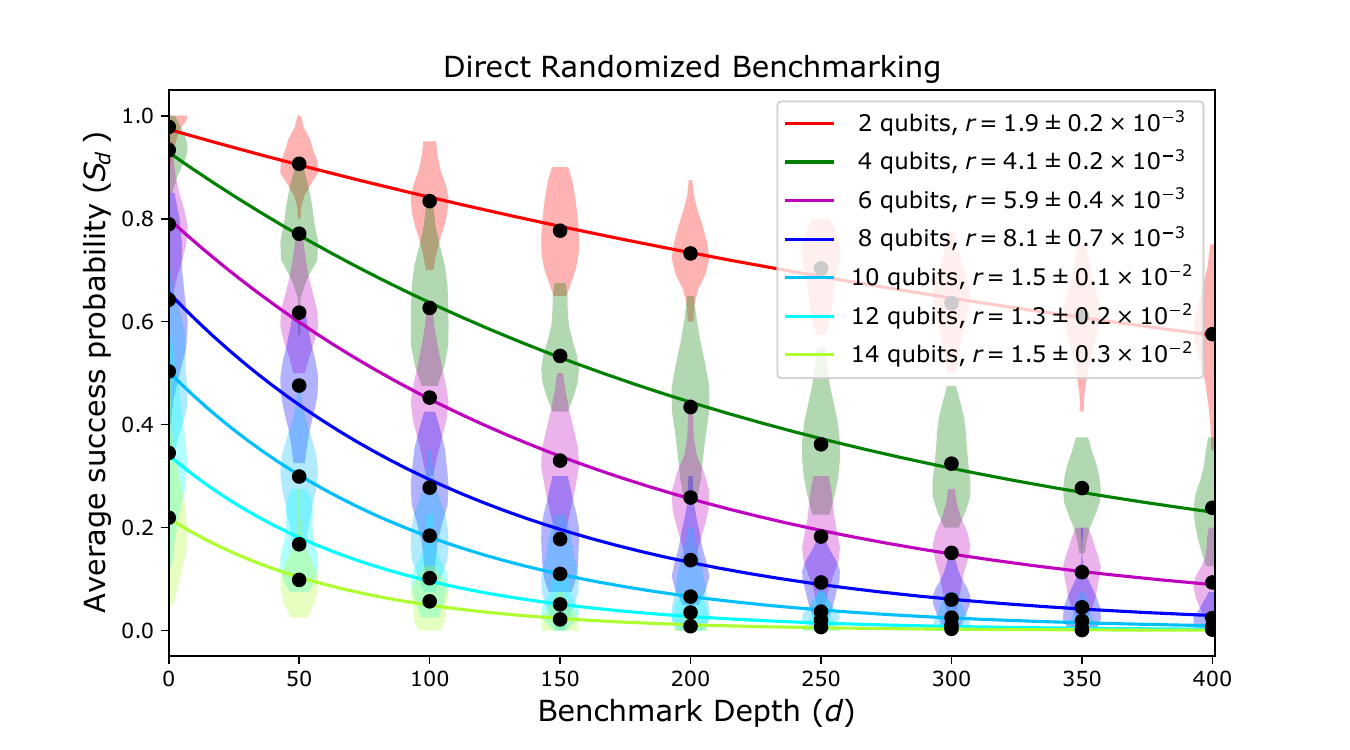}
\includegraphics[width=9cm]{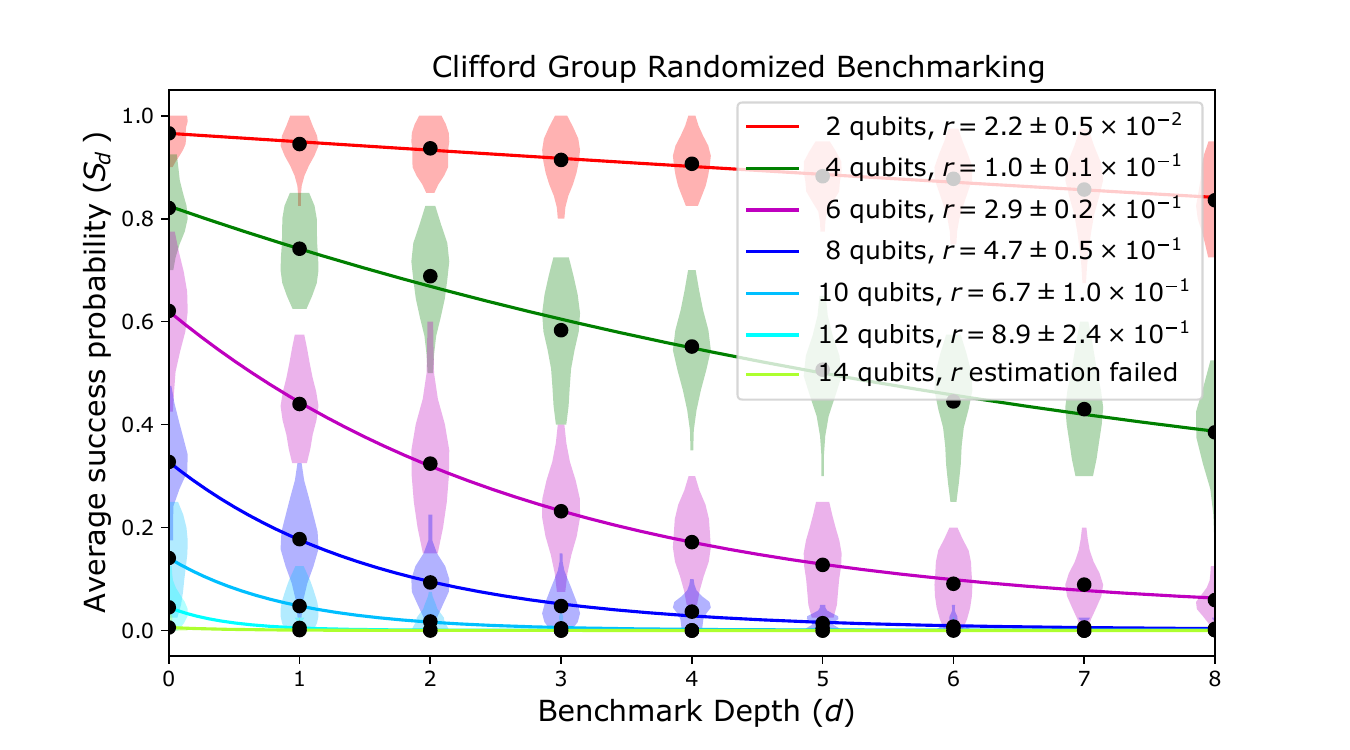}
\end{center}
\caption{Simulated $n$-qubit direct RB and Clifford group RB for $n=2,4,6,\dots,14$. Here, direct RB is benchmarking $n$-qubit gates that consist of a single layer of parallel one- and two-qubit gates. Clifford group RB is benchmarking the $n$-qubit Clifford group generated out of these layers. In this simulation, an imperfect $n$-qubit layer is modeled by the perfect unitary followed by independent uniform depolarization on each qubit at a rate of 0.1\%. The points and violin plots are the means and the distributions of the estimated circuit success probabilities versus benchmark depth ($d$), respectively, and the lines are fits of the means to $S_{d} = A + Bp^d$. The estimated RB error rates ($\hat{r}$), which are reported in the legends, are obtained from the fit decay rates, using Eq.~\eqref{eq:r-def}. We observe that the direct RB average circuit success probabilities ($S_{d}$) decay exponentially for all $n$, and that the direct RB error rate is given by $\hat{r} \approx 1 - (1 - 0.001)^{n} \approx n \times 0.001$. Direct RB is therefore accurately estimating the error rate of the $n$-qubit gates it is benchmarking. The Clifford group RB error rate grows very quickly with $n$---as it is benchmarking the $n$-qubit Clifford group, whose elements must be compiled into the available layers of parallel one- and two-qubit gates. The Clifford group RB $d=0$ intercept quickly decays to approximately $\nicefrac{1}{2^n}$ as $n$ increases, at which point $r$ is close to 100\% and it cannot be estimated to reasonable precision with a practical amount of data. This demonstrates that direct RB is feasible on substantially more qubits than standard Clifford group RB.}
\label{fig:simple-simulation}
\end{figure}

Global depolarizing errors are physically unrealistic, so we also consider another error model that is more realistic but that is still simple to understand: gate-independent \emph{local} depolarization. We simulated $n$-qubit direct RB for a hypothetical $n$-qubit processor (with $n=2,4,6,\dots,14$) whereby after each $n$-qubit gate is applied every qubit experiences independent uniform depolarization with an error rate of 0.1\%, i.e., the infidelity of each single-qubit depolarizing channel is $\epsilon = 0.001$. This error model has the convenient property that we can easily compute $\epsilon_{\Omega}$. The (entanglement) fidelity of a tensor product of $n$ error channels $\mathcal{E}_1, \dots, \mathcal{E}_n$ is the product of $\mathcal{E}_1$ to $\mathcal{E}_n$'s fidelities, i.e.,
\begin{equation}
F(\mathcal{E}_1 \otimes \cdots \otimes \mathcal{E}_n, I\otimes \cdots \otimes I) = F(\mathcal{E}_1,I)  \cdots F(\mathcal{E}_n,I),\label{eq:tensor-F}
\end{equation}
so $\epsilon_{\Omega}$ is given by $\epsilon_{\Omega} = 1 - (1 - 0.001)^{n} \approx n \times0.001$. Note that, because the error rates are gate-independent, $\epsilon_{\Omega}$ is again independent of the sampling distribution $\Omega$ in this example. Figure~\ref{fig:simple-simulation} (upper plot) shows $S_d$ versus $d$ (black circles) as well as the fit exponential (solid lines) and the estimates for $r_{\Omega}$ (in the legend), for each $n$. We observe that $S_{d}$ decays exponentially, and that $\hat{r}_{\Omega} \approx \epsilon_{\Omega}$. We specify the gate set and sampling distribution used in this simulation in Example 2 of Section~\ref{sec:gate-set}.

These numerical results also enable a heuristic assessment of how many qubits can feasibly be benchmarked using direct RB, given the approximate size of current gate error rates (and our suboptimal compilers for creating stabilizer states and measurements). We observe that, in these simulations, $S_d$'s
value at $d=0$ is significantly non-zero for even 14 qubits. Readout errors (which are not included in this simulation) would suppress this intercept further, but readout error of a similar size to the gate errors here ($0.1\%$) would have negligible effect on $S_0$.  Higher error rates would limit direct RB to fewer qubits.  We can estimate the largest feasible number of qubits using the numerical results in Section~\ref{sec:scaling}, where we compute the number of two-qubit gates in the circuits created by our (suboptimal) algorithm for generating random $n$-qubit stabilizer states. These results suggest that with, e.g., a two-qubit gate error rate of $0.5\%$ it would be feasible to benchmark around 12 qubits using direct RB. In summary, our assessment is that around 10-15 qubit could be benchmarked using direct RB for gate and readout errors in the range of $0.1\%$-$1\%$. Furthermore, values for a given set of error rates can be easily computed using simulations. We note, however, that more scalable versions of direct RB \cite{proctor2021scalable, hines2022demonstrating, Hines2024-qe} are, in our experience, likely to be more useful in practice beyond around 8-10 qubits.

\subsection{Direct RB's standard sampling parameters}\label{sec:stats}
The direct RB protocol has three user-specified parameters in common with all RB protocols, which we now briefly discuss. These are the user-specified benchmark depths (the values for $d$), the number of repetitions of each circuit ($N$), and the number of randomly sampled circuits at each benchmark depth $(K_d)$. These parameters predominantly control purely statistical aspects of the experiment---specifically, the precision with which the direct RB experiment estimates the underlying $N,K_d\to \infty$ direct RB error rate. How these parameters control the statistical uncertainty in an estimated RB error rate, and how to choose these parameters, has been studied in some detail for Clifford group RB \cite{magesan2012characterizing, wallman2014randomized, epstein2014investigating, granade2015accelerated, helsen2017multiqubit, hinks2018bayesian}. Much of this work can be applied to direct RB. We will therefore not discuss these parameters in detail herein. Instead, we will provide some simple numerical evidence that the direct RB error rate can be estimated from reasonable amounts of data. The simulated direct RB experiments of Fig.~\ref{fig:simple-simulation} use a practical amount of data: we used 9 benchmarking depths, $K_d=30$, and $N=40$, for a total of approximately $10^4$ samples for the direct RB instance at each value of $n$. Furthermore, the estimated direct RB error rates have low uncertainty (the uncertainties reported in the legend are $2\sigma$ and they are estimated using a standard bootstrap). Throughout the remainder of this paper we are predominantly interested in studying the behaviour of direct RB in the limit of $N,K_d\to \infty$; we denote the average circuit success probability and the direct RB error rate in this limit by $S_{d}$ and $r_{\Omega}$, respectively.

\subsection{Selecting the gate set to benchmark}\label{sec:gate-set}
One of the defining characteristics of direct RB is that the user specifies the $n$-qubit gate set ($\mathbb{G}$) that is to be benchmarked. This is in contrast with Clifford group RB, where the benchmarked gate set is necessarily the $n$-qubit Clifford group---or another group that is a unitary 2-design. However there is not total freedom in choosing $\mathbb{G}$ in direct RB. The direct RB gate set has to satisfy the following conditions, some of which depend on the performance of the processor that is to be benchmarked:
 \begin{itemize} 
\item The group $\mathbb{C} \subseteq \text{SU}(2^n)$ generated by $\mathbb{G}$ must be a unitary 2-design over $\text{SU}(2^n)$. The requirement that $\mathbb{C}$ is a unitary 2-design can likely be relaxed (with appropriate adaptions to the direct RB experiments and/or data analysis) using techniques from RB for general groups \cite{carignan2015characterizing, cross2016scalable, brown2018randomized, hashagen2018real, helsen2018new, Helsen2020-it, helsen2020general, Claes2020-cy}. This is an interesting open area of research, as these generalizations have the potential to enable direct RB on any gate set $\mathbb{G}$. Our formal theory for direct RB (Section~\ref{sec:general-theory}) relies on a slightly stronger condition on $\mathbb{G}$ than simply generating a unitary 2-design---it requires that $\mathbb{G}$ induces a \emph{sequence-asymptotic unitary 2-design}. The concept of a sequence-asymptotic unitary 2-design is introduced in Section~\ref{sec:sequads}, where we state necessary and sufficient conditions for a gate set to induce a sequence-asymptotic unitary 2-design. Here we state a simple condition that is sufficient (but not necessary) for a gate set $\mathbb{G}$ to induce a sequence-asymptotic unitary 2-design: the gate set $\mathbb{G}$ generates a unitary 2-design and contains the identity operation.

\item The $n$ qubits to be benchmarked must have sufficiently high fidelity operations for there to exist a circuit to prepare any random state from $\{ \ket{\psi} = U(C)\ket{0}^{\otimes n} \mid C \in \mathbb{C}\}$ with non-negligible fidelity. To actually run direct RB, the user also needs an explicit algorithm for finding these circuits. This is required so that errors in the state and measurement preparation subcircuits do not mean that $S_d \approx \nicefrac{1}{2^n}$ at $d=0$ (how far above $\nicefrac{1}{2^n}$ it is necessary for $S_0$ to be depends on how much data the user is willing to collect).

\item It must be feasible to sample the direct RB circuits using a classical computer, which requires, e.g., multiplying together arbitrary elements in $\mathbb{C}$, and sampling uniformly from $\mathbb{C}$. This condition is also required to implement standard RB over $\mathbb{C}$, and it is satisfied if $\mathbb{C}$ is the Clifford group.
\end{itemize}

None of the three conditions on $\mathbb{G}$, above, require any efficient scaling with the number of qubits $n$. Instead, it suffices that each condition is satisfied at the values for $n$ of interest. For example, perhaps only benchmarking one- or two-qubit gate sets is of interest (note that most RB experiments are one- or two-qubit Clifford group RB). The conditions also do not require that $\mathbb{G}$ is finite. This is illustrated by our first example, below, of a practically interesting type of gate set that can be benchmarked with direct RB. 

\vspace{0.2cm}
\noindent
\textbf{Example gate set 1 [$\mathbb{G}_{1}(n)$]:} An $n$-qubit gate set constructed from all possible combinations of parallel applications of \textsc{cnot} gates on connected qubits and the one-qubit gates $X(\nicefrac{\pi}{2})$ and $Z(\theta)$ for $\theta \in 
(-\pi,\pi]$ [where $X(\theta)$ and $Z(\theta)$ denote rotations by $\theta$ around $\sigma_x$ and $\sigma_z$, respectively].
\vspace{0.2cm}

This gate set is universal---i.e., it generates $\text{SU}(2^n)$---and so $\mathbb{G}_{1}(n)$ does not satisfy all our conditions for $n\gg 1$ [generating a uniformly (Haar) random pure state is not efficient in $n$]. However, this gate set does satisfy our conditions for small $n$. For example, generating any unitary in $\text{SU}(4)$ requires only three \textsc{cnot} gates \cite{Vidal2004-qe, Shende2004-yf}. The particular $n$-qubit universal gate set given above is just an example, and this reasoning holds for (almost) all few-qubit universal gate sets. Direct RB can therefore be used to benchmark almost any universal gate set over a few qubits.\footnote{It is only almost any universal gate set rather than every universal gate set as, e.g., it is possible to construct universal gate sets that require arbitrarily long sequences of gates to implement some one-qubit gates---but such gate sets are of no practical relevance.} This is arguably substantially simpler than the RB method used by Garion \emph{et al.}~\cite{Garion2020-gi} to benchmark a two-qubit gate set containing a controlled $Z(\nicefrac{\pi}{4})$ gate (which is not a Clifford gate). Note that, recently, direct RB of one- and two-qubit universal gate sets has been experimentally demonstrated in \cite{hines2022demonstrating}.
\vspace{0.2cm}

\noindent
\textbf{Example gate set 2 [$\mathbb{G}_{2}(n)$]:} An $n$-qubit gate set constructed from \textsc{cnot} gates and any set of generators for the single-qubit Clifford group (such as the Hadamard and phase gates).
\vspace{0.2cm}
 
This is the type of gate set used in the simulations of Fig.~\ref{fig:simple-simulation} (and shown in the schematic of Fig.~\ref{Fig:schematic}). In that simulation the gate set consisted of the Hadamard and phase gates as well as \textsc{cnot} gates between any pair of qubits. That is, this simulation uses all-to-all connectivity. However, note that direct RB with this gate set can be applied to a processor with any connectivity, as direct RB does not have to be applied to a processor's native gate set (applying direct RB to a standardized set of $n$-qubit layers could be useful for comparing different processors). In that case, \textsc{cnot} gates between distant qubits would be synthesized via \textsc{swap} gate chains. 

\vspace{0.2cm}
\noindent
\textbf{Example gate set 3 [$\mathbb{G}_{3}(n)$]:} An $n$-qubit gate set constructed from a maximally entangling two-qubit Clifford gate (e.g., \textsc{cnot} or \textsc{cphase}) and the full single-qubit Clifford group.
\vspace{0.2cm}

Direct RB of a gate set with this form is particularly robust and simple to understand from a theoretical perspective (this is the type of gate set that was used in the direct RB experiments of \cite{proctor2018direct}). As with $\mathbb{G}_{2}(n)$, $\mathbb{G}_{3}(n)$ specifies a family of gate sets. We now specify a particular convenient gate set within this family (which is entirely specified given a processor's connectivity).

\vspace{0.2cm}
\noindent
\textbf{Example gate set 4 [$\mathbb{G}_{4}(n)$]:} An $n$-qubit gate set consisting of all $n$-qubit gates composed from applying a layer $L_1$ and then a layer $L_2$ where these layers have the following forms: $L_1$ is a layer containing one of the 24 single-qubit Clifford gates on each qubit, and $L_2$ is a layer containing non-overlapping \textsc{cnot} gates on connected pairs of qubits. 
\vspace{0.2cm}

A random gate from $\mathbb{G}_{4}(n)$ locally randomizes the basis of each qubit (and applies some random arrangement of two-qubit gates)---if the marginal distribution over $L_1$ layers is uniform it implements \emph{local} 2-design randomization (composed with a more complex multi-qubit randomization step). A gate set of this form is assumed in parts of Section~\ref{sec:stochastic-theory}, in order to simplify the theory presented there. Note that direct RB circuits for $\mathbb{G}_4(n)$ have much in common with the random circuits of Google's quantum supremacy experiments \cite{arute2019quantum}, which are used in cross-entropy benchmarking \cite{boixo2018characterizing}. However, these direct RB circuits contain only Clifford gates (making them efficient to simulate), and they contains additional structure (the state preparation and measurement preparation subcircuits).

\subsection{Selecting the sampling distribution}\label{sec:sampling}
There are two main customizable aspects of a direct RB experiment: the gate set ($\mathbb{G}$) and the sampling distribution ($\Omega$). We now discuss the role of $\Omega$, and how to choose it. The direct RB sampling distribution $\Omega$ can be any probability distribution over $\mathbb{G}$ that has support on a subset $\mathbb{G}'$ of $\mathbb{G}$ that also satisfies all the requirements for a direct RB gate set (e.g., $\mathbb{G}'$ also generates $\mathbb{C}$). The sampling distribution $\Omega$ and the gate set $\mathbb{G}$ control what direct RB measures---direct RB estimates the mean infidelity of an $n$-qubit gate sampled from $\Omega$. Therefore, the primary consideration when selecting $\Omega$ is to choose a distribution that defines an error rate $\epsilon_{\Omega}$ of interest. There are infinite valid choices for $\Omega$, and so we do not attempt to discuss all interesting choices for $\Omega$ here. One category of 
sampling distributions we have found useful in experiments is one-parameter families of distributions $\Omega_{\xi}$ in which $\xi$ sets the expected two-qubit gate density of the sampled gate. There are many possible such families of distributions---in Appendix~\ref{app:samplers} we briefly describe a family of probability distributions that has been used in direct RB experiments (the ``edge grab'' sampler, introduced in \cite{proctor2020measuring}).

The sampling distribution $\Omega$ defines the error rate that direct RB measures, but it also impacts the reliability of direct RB, i.e., whether $S_d$ decays exponentially and how close $r_{\Omega}$ is to $\epsilon_{\Omega}$. So a sampling distribution should be chosen for which direct RB will be reliable. For every sampling distribution satisfying the above requirements, direct RB will be reliable for \emph{sufficiently low error} gates. This is because, for any $(\mathbb{G},\Omega)$ satisfying the criteria of direct RB, any Pauli error will be randomized by a depth $l$ $\Omega$-random circuit for sufficiently large $l$ (see the theory in Section~\ref{sec:stochastic-theory}). But this randomization can be very slow, i.e., the required depth $l$ can be very large, and so the error randomization can be much slower than the rate of errors---which, in  Section~\ref{sec:stochastic-theory}, we explain can result in unreliable direct RB (e.g., multi-exponential decays). For example, we could have $\Omega(G) = 1 - \delta$ for one gate $G$ and some $\delta \ll 1$. In this case, a length $l \ll \nicefrac{1}{\delta}$ sequence of gates sampled from $\Omega$ is almost certainly just $l$ repetitions of $G$---so sequences of this length do not randomize errors at all. The theory presented in Section~\ref{sec:stochastic-theory} will help to explain how to choose a sampling distribution that guarantees reliable direct RB. 

\subsection{The reason for randomized 2-design state preparation and measurement in direct RB circuits}\label{sec:circ-structure}
Direct RB is built on the simple idea of directly benchmarking a gate set $\mathbb{G}$ by running varied-depth circuits whose layers are sampled independently from a distribution over $\mathbb{G}$---a class of circuits that have been termed $\Omega$-\emph{distributed random circuits} \cite{hines2022demonstrating}. However, the depth $d$ direct RB circuits do not just consist of a depth $d$ $\Omega$-distributed random circuits. They surround the core direct RB circuit---the $\Omega$-distributed random circuit---with additional structure (see Fig.~\ref{Fig:schematic}). We now explain the purpose of this additional structure. The direct RB circuits begin with a randomized state preparation sub-circuit ($C_{\rm sp}$) and end with a measurement preparation sub-circuit ($C_{\rm mp}$). The purpose of $C_{\rm mp}$ is simply to return the qubits to the computational basis, maximizing error visibility and simplifying the data analysis. The subcircuit has the same purpose as the group inversion element in Clifford group RB.

The reasons for beginning a direct RB circuit with a randomized state preparation circuit ($C_{\rm sp}$) are now explained. First, together $C_{\rm sp}$ and $C_{\rm mp}$ implement an approximate (state) 2-design twirl on the error in the core circuit. This is because the state preparation subcircuit generates a sample from a (state) 2-design, which (if implemented perfectly) twirls the overall error channel of the core circuit, so that it behaves as though it is a global depolarizing channel. This will be shown in each of our complementary theories for direct RB (Section~\ref{sec:stochastic-theory} and Section~\ref{sec:general-theory}). An alternative interpretation of $C_{\rm sp}$ is that it seeds the random walk over the group $\mathbb{C}$ induced by a random sequence of group generators (gates from $\mathbb{G}$) with an initial uniformly random group element (that has been compiled into the state preparation for efficiency).

There is also a more practical but mundane reason for including the state preparation subcircuit $C_{\rm sp}$. If $C_{\rm sp}$ is \emph{not} included in a direct RB circuit, the contribution of errors in $C_{\rm mp}$ to a direct RB circuit's total failure probability can be strongly dependent on the depth of the core circuit ($d$). This would pollute $r_{\Omega}$, i.e., $r_{\Omega}$ would not accurately quantify the error per layer in the core of the direct RB circuit ($\epsilon_{\Omega}$). To see this, consider direct RB without the randomized state preparation, and assume an efficient compiler for creating $C_{\rm mp}$. Further, assume that $\mathbb{G}$ is small (compared to $\mathbb{C}$), so that the distribution over unitaries produced by a short sequence of gates sampled from $\Omega$ must be far from the uniform distribution over $\mathbb{C}$. When $d$ is small, the average depth of $C_{\rm mp}$ will grow approximately linearly with $d$ under many circumstances. This is because $C_{\rm mp}$ is likely to look very much like the depth-$d$ core circuit that it inverts, with the order reversed and individual gates inverted. However, as $d$ gets large, the core circuit converges to a random group element. So the mean depth (and all other properties) of the $C_{\rm mp}$ circuits asymptote to a fixed value that does not depend on $d$. Critically, this means that the depth of $C_{\rm mp}$ would have a nontrivial dependence on $d$. This variation will pollute the decay rate (and so $r_{\Omega}$), and may even cause non-exponential decays---its impact is uncontrolled, as the direct RB protocol is agnostic to how $C_{\rm mp}$ is implemented (just as the Clifford group RB is agnostic about the Clifford compiler). This effect could instead be mitigated by carefully designing the compiler for $C_{\rm mp}$, but this is a complicated undertaking. Instead, the problem is simply solved by the inclusion of $C_{\rm sp}$. With $C_{\rm sp}$ included, the average depth of $C_{\rm mp}$ is independent of $d$. The only residual $d$ dependence outside of the core circuit is in \emph{correlations} between the state that is prepared by $C_{\rm sp}$ and the state that must be mapped to the computational basis by $C_{\rm mp}$ (they are perfectly correlated at $d=0$ and uncorrelated as $d\to\infty$). These correlations can have an effect on $r_{\Omega}$, in principle, but in realistic settings they appear to have no observable effect on $r_{\Omega}$ (see Section~\ref{sec:stochastic-theory}).

\subsection{The reason for conditional compilation: improved scalability}\label{sec:scaling}
The state preparation and measurement sub-circuits within a direct RB circuit are very similar to the first and last gates in a standard Clifford group RB circuit (see Fig.~\ref{Fig:schematic})---and they play essentially the same roles (see above). However, they differ in a practically important way. Both Clifford group RB and direct RB begin with a circuit that implements a group element $F_{\rm sp}$ sampled uniformly from $\mathbb{C}$, but whereas Clifford group RB demands that this circuit implements $U(F_{\rm sp})$ on any input state, direct RB only requires that the circuit generates the same state as $U(F_{\rm sp})$ when applied to $\ket{0}^{\otimes n}$. We call this \emph{unconditional} and \emph{conditional} compilation, respectively. The same distinction holds for the final inversion step, used in direct and Clifford group RB.

\begin{figure}[t!]
\begin{center}
\includegraphics[width=9cm]{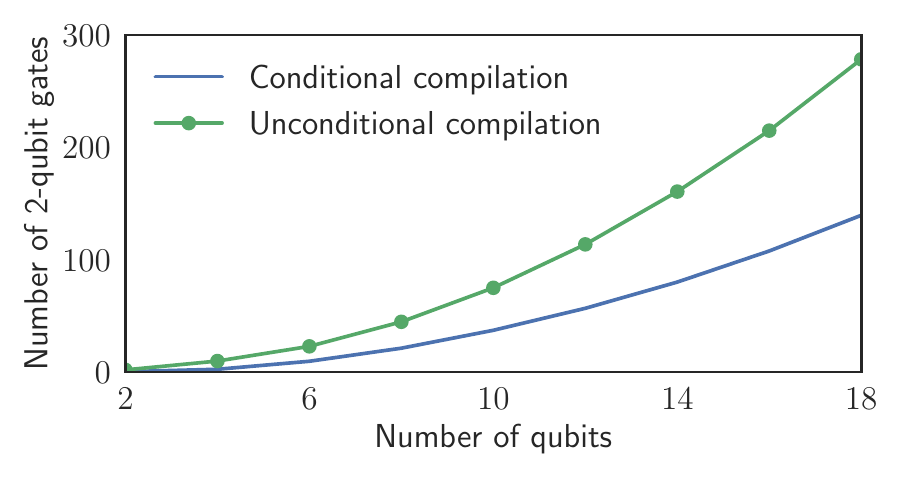}
\end{center}
\caption{A comparison of the mean number of two-qubit gates in circuits for implementing a uniformly random $n$-qubit Clifford gate either unconditionally (green connected points) or conditionally on the input of $\ket{0}^{\otimes n}$ (blue line). In the latter case, this is a circuit mapping $\ket{0}^{\otimes n}$ to a uniformly random $n$-qubit stabilizer state. This demonstrates the significant increase in the number of qubits on which direct RB is feasible---for a fixed two-qubit gate error rate---compared to Clifford group RB. This data was generated using the same open-source compilers that we use for all our direct RB and Clifford group RB simulations \cite{pygstiversion0.9.6, Nielsen2020-rd}, and here we compiled into one-qubit gates and $\textsc{cnot}$ gates between any pair of qubits (i.e., all-to-all connectivity).}
\label{Fig:compiler}
\end{figure}

We choose to use conditional compilation in direct RB circuits because it substantially increases the number of qubits on which direct RB is feasible---as illustrated by the Clifford and direct RB simulations shown in Fig.~\ref{fig:simple-simulation} (this difference between direct and Clifford group RB is the only reason for the difference in their $S_0$ values, which is the primary factor setting the number of qubits that it is feasible to benchmark). This is because conditional compilation results in circuits that are shallower and contain fewer two-qubit gates \cite{aaronson2004improved}. This is demonstrated in Fig.~\ref{Fig:compiler}, which compares the mean number of \textsc{cnot} gates in the circuits generated by a conditional and unconditional $n$-qubit Clifford compiler. To generate this plot we used open-source compilation algorithms \cite{pygstiversion0.9.6, Nielsen2020-rd} (that we also used for all the simulations herein). These algorithms are unlikely to generate circuits with minimal two-qubit gate counts (see \cite{aaronson2004improved, patel2003efficient, bravyi2020hadamard} for work on optimal Clifford compilation) although we have found that they perform reasonably well.\footnote{The algorithm we use is based on Gaussian elimination; the compiled circuits contain $O(n^2)$ two-qubit gates for both the state and unitary compilations. This scaling is not optimal: there are algorithms that generate circuits containing $O(n^2/\log(n))$ two-qubit gates \cite{aaronson2004improved}. But in the $n=O(10)$ regime we have found that our compilations contain many fewer two-qubit gates than the $O(n^2/\log(n))$ algorithms of~\cite{aaronson2004improved}. We have not compared our compilation algorithms to more recent work on Clifford compilation \cite{bravyi2020hadamard}.}

Note that the error rate measured by direct RB is (approximately) independent of the details of the compiler used to generate these states. The property of the compilation algorithm that is of importance to direct RB is the efficiency with which it generates these states---the higher the fidelity of these states the more qubits direct RB will be feasible on. This is convenient, as algorithms for generating many-qubit states/unitaries are typically ``black-box'' and the properties of the circuits they generate cannot be easily controlled. In contrast, with Clifford group RB the compiler entirely defines the physical meaning of the RB error rate---a Clifford group RB error rate cannot be used to quantify \emph{native gate} performance without a detailed understanding of the compiled circuits.

\subsection{The reason for randomizing the success outcome}\label{sec:randomized-out}
Direct RB specifically allows for uniform randomization of the success bit string (this is also possible with Clifford group RB \cite{harper2019statistical}, but it is not standard practice). This bit string randomization is not essential, but in our view it is preferable. This is because it guarantees that the $d\to \infty$ asymptote of the average RB success probability $S_d$ is $\nicefrac{1}{2^n}$ (for all Markovian errors), so we can fix $A=\nicefrac{1}{2^n}$ in $S_d = A + Bp^d$. As Harper \emph{et al.}~\cite{harper2019statistical} discuss the motivation and statistical impact of this bit string randomization (in the context of Clifford group RB), we do not do so further here.

\subsection{The error rate convention: choosing the decay rate scaling factor}\label{sec:scaling-factor}
For both direct and Clifford group RB, data are analyzed by fitting the average success probabilities to $S_d = A + Bp^d$ and then mapping $p$ to an error rate $r$ (or a fidelity $1-r$). In Clifford group RB (and other RB protocols), $p$ is conventionally mapped to an error rate defined by $r' = (2^n-1)(1-p)/2^n$ \cite{magesan2011scalable, magesan2012characterizing}. This is not the definition that we use. As specified in Eq.~\eqref{eq:r-def}, we define the direct RB error rate to be $r =(4^n-1)(1-p)/4^n$. For $n\gg 1$ the difference between $r$ and $r'$ is negligible, but it is substantial for $n \sim 1$. Our decision to use Eq.~\eqref{eq:r-def} to define $r$ is not specific to direct RB. The RB error rate $r'$ corresponds to the gate set's mean \emph{average gate infidelity}, whereas $r$ corresponds to the mean \emph{entanglement infidelity}. We choose to use a definition for the RB error rate that correspond to entanglement infidelity because of entanglement infidelities convenient properties. For example, for Pauli stochastic errors the entanglement infidelity corresponds to the probability of any Pauli error occurring (see Section~\ref{sec:definitions}) and the entanglement fidelity of multiple gates used in parallel is equal to the product of their entanglement fidelities (assuming no additional errors when gates are parallelized)---see Eq.~\eqref{eq:tensor-F}. Due to these convenient properties, this choice has also been made with other scalable benchmarking techniques (e.g., cycle benchmarking \cite{erhard2019characterizing}).

\subsection{Comparison to other methods for native gate randomized benchmarking}\label{sec:other-methods}
Direct RB is not the first or only proposal for RB directly on a gate set that generates a group. Below we explain the relationship between RB and each of these protocols:
\begin{itemize}
\item The method of Knill \emph{et al.}~\cite{knill2008randomized} benchmarks a set of gates that generate the one-qubit Clifford group, and this has been widely used as an alternative to Clifford group RB (e.g., see the references within \cite{boone2018randomized}). That method consists of uniform sampling over a specific set of one-qubit Clifford group generators. So it is a particular example of direct RB (as pointed out by Boone \emph{et al.}~\cite{boone2018randomized}) \emph{except} that it does not include the initial stabilizer state preparation step (which is of little importance in the single-qubit setting). So our theory for direct RB is further evidence that the protocol of \cite{knill2008randomized} is just as reliable and arguably as well-motivated as Clifford group RB, although it measures a different error rate.

\item The extensions of the method of Knill \emph{et al.}~up to three qubits \cite{ryan2009randomized} also fit within the framework of direct RB except that, again, in that method there is no stabilizer state initialization.

\item Independently of the development of direct RB, Fran{\c{c}}a and Hashagen proposed ``generator RB''  \cite{francca2018approximate}, which is also direct RB without the stabilizer state preparation step and without user-configurable sampling (but note that some of the theory within~\cite{francca2018approximate} also applies to direct RB).

\item  Since the development of direct RB, a protocol called mirror RB \cite{proctor2021scalable, hines2022demonstrating} has been introduced that adapts direct RB to improve its scalability. Mirror RB methods replace the stabilizer state preparation and measurement sub-circuits from direct RB with a mirror circuit reflection structure, meaning that they contain no large subroutines. Mirror RB techniques are more scalable than direct RB, and they are designed to measure the same error rate as direct RB ($\epsilon_{\Omega}$). However, the error rate measured by existing mirror RB methods is a less reliable estimate of $\epsilon_{\Omega}$ (it is typically a slight underestimate---see \cite{proctor2021scalable, hines2022demonstrating} for further details), so mirror RB is not a strict improvement on direct RB. Much of the theory presented in this paper is also applicable, or adaptable, to mirror RB.

\item  After the initial version of this paper appeared, a protocol called binary RB \cite{Hines2024-qe} was introduced that (like mirror RB) adapts direct RB to improve its scalability. Binary RB removes the stabilizer state preparation and measurement subcircuits from direct RB, replacing each with a single layer of single-qubit gates, and changes the data analysis using the ideas from direct fidelity estimation \cite{Flammia2011-qj}. This makes binary RB more scalable than direct RB, and it is designed to measure the same error rate as direct RB ($\epsilon_{\Omega}$). Much of the theory presented here is also applicable (or adaptable) to binary RB.
\end{itemize}

\section{Understanding direct RB using error scrambling}\label{sec:stochastic-theory}
In this section we present a simple approximate theory of direct RB. The theory in this section is designed to (1) highlight the key properties of a gate set, and a sampling distribution, that guarantee reliable direct RB, and (2) to explain why direct RB works when these conditions are satisfied. In this theory we make a range of simplifying assumptions that allow us to present simple conditions under which direct RB reliably estimates the average infidelity of the benchmarked gates. The main assumptions we make in this section are as follow (see also Table~\ref{table:theories}):
\begin{enumerate}
\item The gate set contains only Clifford gates (and generates the Clifford group).
\item All errors are stochastic Pauli errors.
\item Many ($n \gg 1$) qubits are involved. Around $n \sim 5$ suffices.
\item Each gate's error rate is low ($\epsilon \ll 1$).
\end{enumerate}
The first two assumptions are foundational for our arguments (although it is plausible that a conceptually similar, but likely more mathematically complex, theory could be created for general gates and general stochastic errors). This is because our arguments are based on how Pauli errors propagate through layers of Clifford gates. The third assumption---many qubits---is not necessary, but simplifies notation. This assumption enables us to drop $O(\nicefrac{1}{4^n})$ terms that appear because two random non-identity Pauli operators (errors) compose to the identity with probability $\nicefrac{1}{4^n-1}$. This effect is not negligible when $n \sim 1$. It can be accounted for in all of the following theory, but it causes notational complications that, in our opinion, obfuscate the key intuitions. As our theory in Section~\ref{sec:general-theory} is applicable to the few-qubit setting, we therefore ignore the $n \sim 1$ setting here. The final assumption is simply that the rate of error in a single layer of gates is small, which we require because otherwise we cannot guarantee any error scrambling occurring between two errors within a circuit. As we discuss in more detail later in this section, we expect that this assumption can be relaxed to the condition that the \emph{marginal} rate of errors on each qubit is small (which is a physically well-justified assumption for any $n$), but we do not rigorously show this herein.

\subsection{Clifford gates and stochastic Pauli errors}\label{ssec:stochastic-assumptions}
We now introduce our notation for this section, expand on some of our assumptions, and summarize the results of this section. In this section we are studying direct RB of an $n$-qubit gate set $\mathbb{G}$ that generates the Clifford group, with a sampling distribution $\Omega$ that has support on all of $\mathbb{G}$. Each $G \in \mathbb{G}$ is modeled by the ideal unitary $U(G)$ followed by a $G$-dependent Pauli stochastic channel (assumption 2). This Pauli stochastic channel is described by $4^n$ rates $\{\epsilon_{G,P}\}_{P \in \mathbb{P}}$ where $\epsilon_{G,P}$ is the probability that $U(G)$ is followed by the Pauli $P$ and $\mathbb{P}$ is the $n$-qubit Pauli group. So 
\begin{equation}
\sum_{P\in\mathbb{P}}\epsilon_{G,P} = 1,
\end{equation}
and $\epsilon_{G,I}$ is the probability of no error where $I$ is the identity Pauli operator. The probability of any Pauli error is
\begin{equation}
\sum_{P \in\mathbb{P}, P \neq I}\epsilon_{G,P} = \epsilon_{G},
\end{equation}
where $\epsilon_G$ is the entanglement infidelity of $G$. In this section, we will show that for a broad range of $\mathbb{G}$, $\Omega$, and stochastic Pauli error models (defined by $\{\epsilon_{G,P} \}_{G \in \mathbb{G}, P \in \mathbb{P}}$), direct RB is reliable. That is, the direct RB average success probability ($S_d$) decays exponentially ($S_d \approx A + Bp^d$) and the direct RB error rate ($r_{\Omega}$) is approximately equal to the mean entanglement infidelity of a gate sampled from $\Omega$ ($r_{\Omega} \approx \epsilon_{\Omega}$).

Under the assumptions introduced above, a benchmark depth $d$ direct RB circuit consists of:
\begin{enumerate}
\item A circuit preparing a uniformly random $n$-qubit stabilizer state $\psi$.
\item A circuit $C = G_{d}\cdots G_{1}$ of $d$ gates $G_{i}$ sampled from $\Omega$, which is called the direct RB circuit's ``core''.
\item A circuit and measurement that simulates the POVM $M_{\varphi} = \{\ket{\varphi}\bra{\varphi}, 1 -\ket{\varphi}\bra{\varphi}\}$, where $\varphi$ is the stabilizer state 
\begin{equation}
\ket{\varphi} = U(C)\ket{\psi}.
\end{equation}

We refer to this as a stabilizer state measurement.
\end{enumerate}

Until stated later in this section, we will assume that the stabilizer state preparation and measurement (SSPAM) are perfect, i.e., we exactly prepare $\psi$ and exactly measure $M_{\varphi}$. This is not a realistic assumption, but, as we argue below, the primary effect of SSPAM error is on the initial value of the success probability decay (i.e., $A+B$ in $S_d = A + Bp^d$). Therefore, SSPAM error has negligible impact on $r_{\Omega}$, and only impacts the number of qubits on which direct RB is feasible.

\subsection{Modelling direct RB circuits with a stochastic unravelling}\label{ssec:stochastic-unravelling}
\begin{figure}[t!]
\begin{center}
\includegraphics[width=9cm]{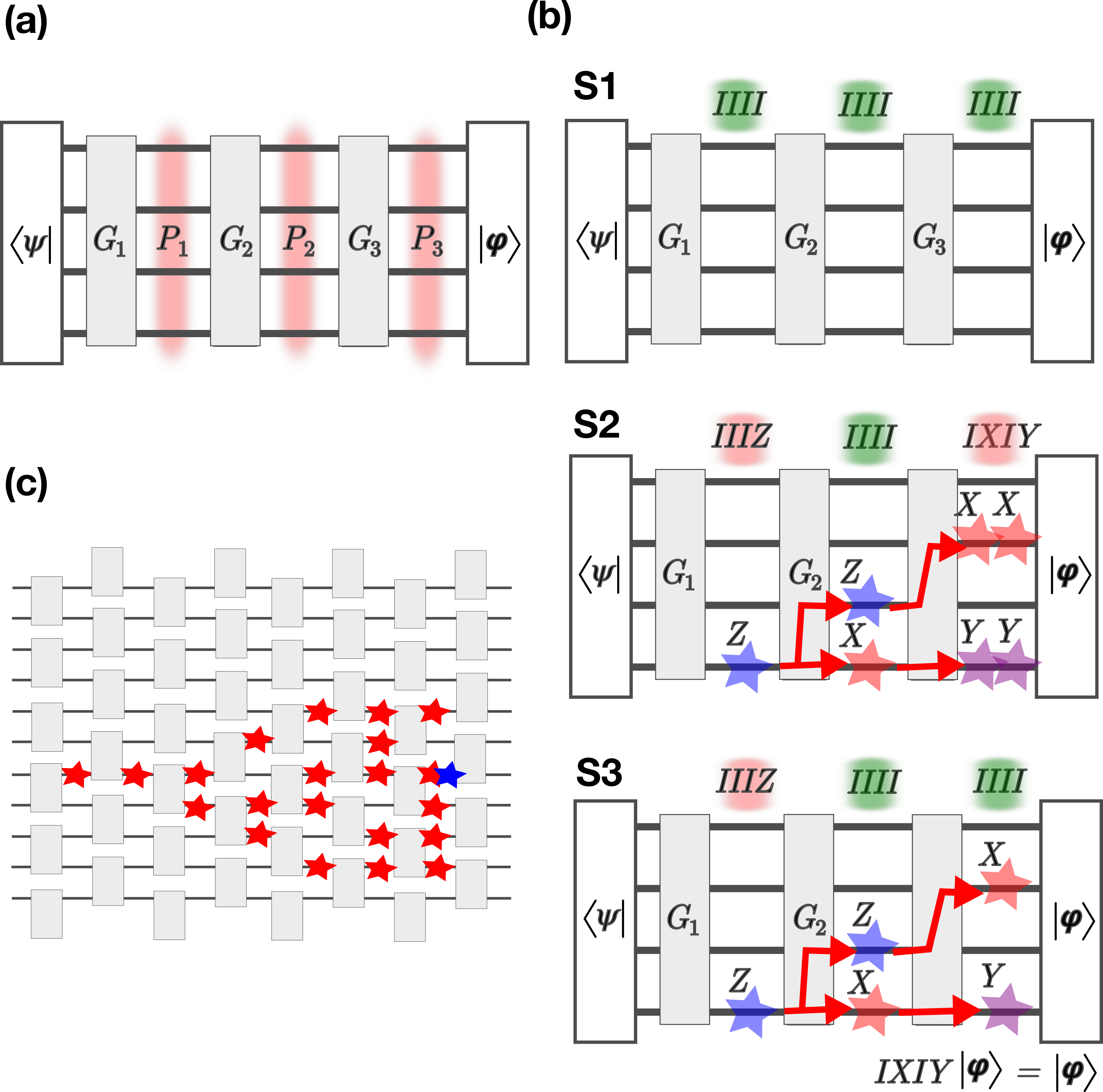}
\end{center}
\caption{Our theory for direct RB with stochastic Pauli errors uses (a) a stochastic unravelling of a Pauli stochastic error model [see Sections~\ref{ssec:stochastic-assumptions}--\ref{ssec:stochastic-unravelling}], illustrated here for a depth $d=3$ circuit. To prove that direct RB is reliable we (b) consider the probabilities of three mutually exclusive events [S1-S3; see Section~\ref{ssec:S1S2S3}] within that unravelling that each contribute to the probability that a depth $d$ direct RB circuit returns the ``success'' bit string ($S_d$). Our theory shows that if the probability of two errors cancelling within a direct RB circuit (the event S2) is negligible then $S_d = A + B(1-\epsilon_{\Omega})^d$ [see Sections~\ref{ssec:s1}--\ref{ssec:if-s2-neg}]. (c) The probability of two or more errors cancelling is negligible, because even a single-qubit Pauli error (leftmost red star) is almost certainly spread onto many qubits by a small number of random circuit layers (red stars illustrate a possible error spreading trajectory), so that it cannot cancel with a few-qubit error later in the circuit (blue star) [see Section~\ref{ssec:s2-is-neg}].}
\label{Fig:stochastic-theory}
\end{figure}

Throughout this section we use a \emph{stochastic unraveling} of the stochastic Pauli error model, and we treat a direct RB circuit, and the gates it contains, as random variables. This is illustrated in Fig.~\ref{Fig:stochastic-theory} (a) [for $d=3$]. In particular, a random depth $d$ direct RB circuit's output ``success'' bit ($B_d$, with $B_d \in \{0, 1\}$) is given by 
\begin{equation}
    B_d = |\bra{\varphi} \tilde{U}(C) \ket{\psi}|^2, \label{eq:Bd}
\end{equation}
where
\begin{equation}
\tilde{U}(C) = P_{d}U(G_{d})\cdots P_{2}U(G_{2})P_{1}U(G_{1}). \label{eq:unraveling}
\end{equation}
Here the $U(G_i)$ are unitary-valued random variables, that are mutually independent and $\Omega$-distributed, the $P_i$ are Pauli-operator-valued random variables where $P_i$ is equal to the Pauli operator $P$ with probability $\epsilon_{G_i,P}$ (so $P_i$ depends on $G_i$), and $\psi$ is a stabilizer-state valued random variable that is independent and uniformly distributed (and $\bra{\varphi} = \bra{\psi}U^{-1}(C)$). The $P_i$ correspond to the potential errors in the direct RB circuit. The average success probability of a depth $d$ direct RB circuit ($S_d$) is the expected value of $B_d$, i.e.,
\begin{equation}
    S_d =\mathbb{E}[B_d].
\end{equation}

In our theory below, we will show how $S_d$ can be calculated by considering the probabilities of different mutually exclusive ``patterns'' of Pauli errors within Eq.~\eqref{eq:unraveling}.

\subsection{Three routes to success: no errors, multiple errors cancel, or the measurement misses an error}\label{ssec:S1S2S3}
The first step in our theory is to show that the direct RB average success probability ($S_d$) can be decomposed into the sum of three contributions [see Fig.~\ref{Fig:stochastic-theory} (b)]: circuit execution instances in which (1) no Pauli errors occurred, (2) at least two Pauli errors occur but they cancel out, and (3) Pauli errors occur and they do not cancel, but they are unobserved due to the particular SSPAM in that circuit. We begin by noting that Eq.~\eqref{eq:Bd} can be rewritten as
\begin{equation}
B_d = \vert\langle \varphi \vert \tilde{U}(C) \vert\psi\rangle\vert^2 =   \vert\langle \psi \vert  U^{\dagger}(C)\tilde{U}(C) \vert\psi\rangle\vert^2.
\label{eq:S}
\end{equation}
Now, because the gates are all Clifford operators, we have that 
\begin{equation}
\tilde{U}(C) = PU(C), 
\end{equation}
where $P$, which is a Pauli-operator-valued random variable, is calculated by commuting each of the $P_i$ past the subsequent gates, and then multiplying them together. So
\begin{equation}
B_d = \vert\langle\psi\vert U(C)^{\dagger} P U(C) \vert\psi\rangle\vert^2 = \vert\langle\psi\vert \tilde{P} \vert\psi\rangle\vert^2,
\label{eq:S}
\end{equation}
where $\tilde{P}$ is the Pauli-operator-valued random variable given by 
\begin{equation}
\tilde{P}=U(C)^{\dagger} P U(C).
\end{equation}

Equation~\eqref{eq:S} implies that a direct RB circuit's outcome is the ``success'' bit string (i.e., $B_d=1$ rather than $B_d=0$) if and only if one of the following \emph{mutually exclusive} events occurs:
\begin{enumerate}
\item[S1.] All $P_{i} = I$, meaning that no errors occurred in $C$. This implies that $P = I$, so $\tilde{P} = I$. [See Fig.~\ref{Fig:stochastic-theory} (b) S1].
\item[S2.] Two or more of the $P_i$ operators are errors, i.e., they are not the identity, but we still have that $P = I$ up to a phase,  so $\tilde{P} = I$ up to a phase. This means that two or more errors occurred, but---after they are propagated past the circuit layers separating them---they compose to the identity. [See Fig.~\ref{Fig:stochastic-theory} (b) S2].
\item[S3.] One or more of the $P_i$ operators are errors \emph{and} they do not cancel ($P \neq I$), but they are nonetheless unobserved by the stabilizer measurement. This means that $\vert\psi\rangle$ is an eigenstate of $\tilde{P}$ (equivalently $\ket{\varphi}$ is an eigenstate of $P$). This means that $\tilde{P}$ is a stabilizer or anti-stabilizer of the prepared stabilizer state. [See Fig.~\ref{Fig:stochastic-theory} (b) S3].
\end{enumerate}

The benchmark depth $d$ average success probability ($S_d = \mathbb{E}[B_d]$) is the probability that any one of these mutually exclusive events (S1, S2, and S3) occurs in a random depth $d$ direct RB circuit. That is, we can write 
\begin{equation}
S_d = \text{Pr}_d(\text{S1}) +  \text{Pr}_d(\text{S2}) +  \text{Pr}_d(\text{S3}), \label{eq:PmS1S2S3}
\end{equation}
where $\text{Pr}_d(E)$ denotes the probability of the event $E$ occurring in a benchmark depth $d$ direct RB circuit. Because S1, S2, and S3 are exclusive events, by applying basic probability theory we can rewrite $S_d$ as
\begin{equation}
S_d = s_1 + (1 - s_1)s_2 + (1 - s_1)(1 - s_2)s_3, \label{eq:Pms1s2s3}
\end{equation}
where $s_1$ is the probability that S1 occurs, $s_2$ is the probability that S2 occurs given that S1 does not, and $s_3$ is the probability that S3 occurs given that S1 and S2 do not. That is, 
\begin{align}
s_1 &= \text{Pr}_d(\text{S1}),\label{eq:defS1}\\
s_2 &= \text{Pr}_d(\text{S2} \mid \bar{\text{S1}}), \label{eq:defS2}\\
s_3 &= \text{Pr}_d(\text{S3} \mid \bar{\text{S1}} \cap \bar{\text{S2}}). \label{eq:defS3}
\end{align}

In the next two subsection we show that $s_1$ and $s_3$ have very simple forms, before then turning to $s_2$.

\subsection{A formula for the probability of no errors}\label{ssec:s1}
We have shown that the direct RB decay ($S_d$) can be expressed in terms of three quantities---$s_1$, $s_2$, and $s_3$ [Eq.~\eqref{eq:Pms1s2s3}]. We now derive a formula for $s_1$, defined in Eq.~\eqref{eq:defS1}. This is the probability that no errors occur in a depth $d$ $\Omega$-distributed random circuit [see Fig.~\ref{Fig:stochastic-theory} (b) S1]. A depth $d$ $\Omega$-distributed random circuit consists of $d$ gates from $\mathbb{G}$ that are sampled independently from $\Omega$. Because $\epsilon_\Omega$ is the probability that a gate sampled from $\Omega$ experiences a Pauli error, $(1 - \epsilon_{\Omega})$ is the probability that no error occurs for a gate sampled from $\Omega$. As the $d$ gates in our circuit are independently sampled, we therefore simply have
 \begin{equation}
 s_1 = (1 - \epsilon_{\Omega})^d. \label{eq:s1}
\end{equation} 

\subsection{A formula for the probability that an error is not observed}\label{ssec:s2}
We now derive a formula for the probability that an error in a direct RB circuit is not observed by the measurement (i.e., $s_3$), which is one of the three quantities in our formula for $S_d$ in Eq.~\eqref{eq:Pms1s2s3}. Defined in Eq.~\eqref{eq:defS3}, $s_3$ is the probability that any errors in the core of a benchmark depth $d$ direct RB circuit do not cause the circuit to return the wrong bit string given that (i) at least one error occurs, and (ii) the errors do not cancel [see Fig.~\ref{Fig:stochastic-theory} (b) S3]. Together these conditions imply that $\tilde{P}$ in Eq.~\eqref{eq:S} is not the identity. So $s_3$ is the probability that $|\bra{\psi} \tilde{P} \ket{\psi}|^2 = 1$ given that $\tilde{P} \neq I$. The state $\psi$ is a uniformly random stabilizer state (that is independent of $\tilde{P}$), so it is an eigenstate of \emph{any} fixed Pauli error (i.e., a non-identity Pauli operator) with probability $(2^n - 1)/(4^n - 1)=1/(2^n+1)$. This is because any stabilizer state $\psi$ is an eigenstate of $2^n-1$ non-identity Pauli operators, and there are $4^n-1$ non-identity Pauli operators \cite{aaronson2004improved}. So 
 \begin{equation}
 s_3=\frac{1}{2^n + 1} = \frac{1}{2^n} + O(\nicefrac{1}{4^n}). \label{eq:s3}
 \end{equation}
 
This simple $d$-independent formula for $s_3$ is a consequence of the random state preparation step in direct RB---it guarantees that $s_3$ depends only on whether $\tilde{P}$ is an error, not on its unknown (and $d$-dependent) distribution over the $4^n-1$ Pauli errors. 

\subsection{Direct RB is reliable if the probability of error cancellation is negligible}\label{ssec:if-s2-neg}
We now explain why direct RB is reliable whenever the probability of multiple errors canceling in a direct RB circuit is negligible. Above, we derived simple and $d$-independent equations for $s_1$ [Eq.~\eqref{eq:s1}] and $s_3$ [Eq.~\eqref{eq:s3}]. Substituting these equations into Eq.~\eqref{eq:Pms1s2s3} we obtain
\begin{equation}
S_d = \frac{1}{2^{n}}+ \left( 1 - \frac{1}{2^{n}}\right) (1 - \epsilon_{\Omega})^d (1 - s_2) + s_2 + O(\nicefrac{1}{4^{n}}). \label{eq:Pm-with-O4n}
\end{equation}
This equation for $S_d$ depends on $n$, $\epsilon_{\Omega}$, and $s_2$, where $s_2$ is the probability that all the errors within a direct RB circuit's core cancel out given that at least one error occurs. If we assume that $s_2 \approx 0$ [and that $n \gg 1$, so that $\nicefrac{1}{4^{n}} \approx 0$], then Eq.~\eqref{eq:Pm-with-O4n} becomes
\begin{equation}
S_d \approx \frac{1}{2^{n}}+ \left( 1 - \frac{1}{2^{n}}\right) (1 - \epsilon_{\Omega})^d,
\end{equation}
i.e., $S_d \approx A+ B p^d$ with $A=\nicefrac{1}{2^{n}}$, $B= 1-\nicefrac{1}{2^{n}}$, and $p = 1 - \epsilon_{\Omega}$, so
\begin{equation}
r_{\Omega} \approx \epsilon_{\Omega}.
\end{equation}
This implies that, when the $s_2$ is small, direct RB is guaranteed to be reliable. The remainder of our theory consisting of showing that $s_2$ is negligible under very broad conditions.

\subsection{The probability of error cancellation in direct RB circuits is negligible}\label{ssec:s2-is-neg}
 We now explain why the probability of errors cancelling in direct RB circuits is typically negligible (i.e., $s_2 \approx 0$). The core of the argument is:
 \begin{itemize}
     \item[(i)] random circuit layers quickly turn an error that occurs on few qubits into an error impacting many qubits \cite{Hunter-Jones2018-zh,Von_Keyserlingk2018-ts,Nahum2018-zs}, and 
     \item[(ii)] the probability that a many-qubit error cancels with another error is negligible, so
     \item[(iii)] if errors are delocalized faster than they occur then $s_2 \approx 0$, resulting in reliable direct RB. 
 \end{itemize}
Therefore, direct RB will be reliable as long as $ \epsilon_{\Omega} < \nicefrac{1}{k_{\textrm{delocal}}}$ where $k_{\textrm{delocal}}$ is the number of layers required to delocalize any error and $\epsilon_{\Omega}$ is the infidelity of a gate sampled from $\Omega$. Importantly, $k_{\textrm{delocal}}$ is a small constant---it is independent of $n$. This argument is illustrated in Fig.~\ref{Fig:stochastic-theory} (c), and we expand upon it below.
 
Our argument will use some simplifying assumptions, one of which uses the concept of a Pauli $P$'s \emph{weight} $W(P)$. The weight $W(P)$ of the Pauli $P$ is the number of qubits on which $P$ acts non-trivially, i.e., $W(P)=w$ if $P$ has a non-identity action ($X$, $Y$ or $Z$) on $w$ qubits and it is an identity on the remaining $n-w$ qubits. We assume:
\begin{enumerate}
    \item An even number of qubits $n$ with an $\alpha$-regular connectivity graph and an $\alpha$-edge-coloring, e.g., a ring ($\alpha=2$), torus ($\alpha=4$), or fully connected graph ($\alpha=n-1$).
    \item  An $n$-qubit gate set $\mathbb{G}=\{G\}$ consisting of all $n$-qubit gates of the form $G=L_2L_1$ where:
    \begin{enumerate}
        \item $L_1$ is a layer containing one of the 24 single-qubit Clifford gates on each qubit, and 
        \item $L_2$ is a layer of \textsc{cnot} gates that contains all $\nicefrac{n}{2}$ \textsc{cnot} gates corresponding to one of the $\alpha$ edge colors on the connectivity graph (so there are $\alpha$ different two-qubit gate layers). 
    \end{enumerate}
    \item $\Omega$ is the uniform distribution over $\mathbb{G}$.
    \item The probability of high-weight errors is negligible, i.e., for every gate $G$, $\sum_{P \in \mathbb{P}_{4}} \epsilon_{G,P} \approx 0$ where $P_{w}$ is the set of all Pauli operators of weight $w$ or greater. 
    \item $n \gg 1$, as assumed throughout this section.
    \end{enumerate}
The first three of these assumptions constitute a type of spatial homogeneity: these assumptions imply that when a Pauli error occurs on a qubit $q$, $\Omega$-distributed random layers cause it to experience the same randomization process independent of the basis of the error ($X$, $Y$ or $Z$) and on which qubit it occurs. This simplifies the following argument, but it is not essential to it.\footnote{The following derivation can be reformulated in terms of the Pauli error that is randomized slowest by $\Omega$ distributed layers.} The fourth assumption is also not essential, but it is physically reasonable and it also simplifies the argument. The designation of weight-4 and above errors as high-weight, and weight-3 and below errors as low-weight is somewhat arbitrary---any other weight $w \ll n$ can be used to define ``low-weight'' in the following derivation.

We aim to show that $s_2$ is negligible, where $s_2$ [defined in Eq.~\eqref{eq:defS2}] is the probability that (i) two or more errors occur in $\tilde{U}(C)$ [Eq.~\eqref{eq:unraveling}] and (ii) these errors cancel, so that $\tilde{U}(C)=U(C)$. We consider the case of (at least) two errors occurring in Eq.~\eqref{eq:unraveling}, by conditioning on errors occurring after layers $i$ and $i+k+1$, in Eq.~\eqref{eq:unraveling}, for some $i,k \geq 1$.
This means that we are conditioning on the Pauli $P_i$ and $P_{i+k+1}$ being not equal to the identity---so they are  now random variables that are distributed over the $4^n-1$ possible Pauli errors---and we will denote these conditional random variables by $P$ and $Q'$, respectively. Now, because $Q'$ is a random variable that models an error occurring on gate $G_{i+k+1}$ it is not independent of $U(G_{i+k+1})$. So we commute $Q'$ in front of $U(G_{i+k+1})$, by defining $Q = U(G_{i+k+1})^{-1}Q'U(G_{i+k+1})$ [$Q$ is essentially just a pre-gate Pauli error, rather than a post-gate Pauli error]. Therefore $P$ and $Q$ are independent Pauli-operator-valued random variables with $k$ independent random unitaries $U_{i+k:i} \equiv U(G_{i+k}) \cdots U(G_i)$ separating them [see the illustration in Fig.~\ref{Fig:stochastic-theory} (c)].

\begin{figure}[t!]
\begin{center}
\includegraphics[width=9cm]{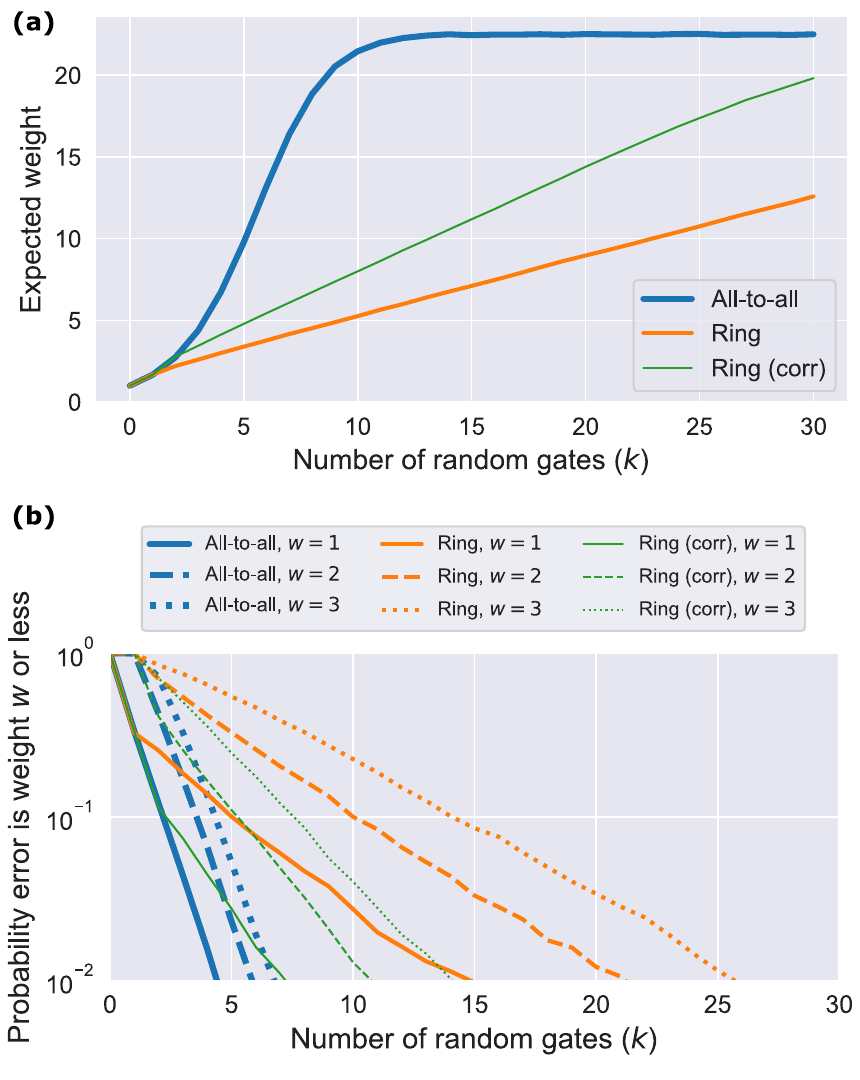}
\end{center}
\caption{In the main text we prove that direct RB is guaranteed to be reliable if direct RB's random gates convert a weight-1 Pauli error into a high-weight Pauli error (a weight-4 or above error) before there is a high probability that another error occurs. To show that weight-1 errors are quickly converted into high-weight errors by random gates from practically relevant gate sets we simulated this randomization process. (a) The expected weight of a weight-1 Pauli error after it has been propagated through $k$ $n$-qubit gates $G=L_2L_1$, where $L_1$ is a layer of uniformly random single-qubit Clifford gates, and $L_2$ is a random layer containing $\nicefrac{n}{2}$ \textsc{cnot} gates between pairs of connected qubits (see Section~\ref{ssec:s2-is-neg}). We show the expected weighted versus $k$ for all-to-all connectivity (blue), ring connectivity (orange), and ring connectivity with \emph{correlated} sampling of the \textsc{cnot} layers (green) in which the $L_2$ layers alternate between coupling a qubit with its neighbour to its left or right (this speeds up error spreading). These results do not depend on the location of the initial Pauli error, because we assume a form of spatial homegeniety (an $\alpha$-regular connectivity graph that is $\alpha$-edge-colorable). (b) The probability that a weight-1 Pauli error is still a weight-$w$ or less error after it has been propagated through $k$ $n$-qubit gates, for $w=1,2,3$. As expected (see main text), we find that these probabilities drop off quickly with $k$. These results are for $n=30$ qubits [the probabilities in (b) upper-bound their values for $n > 30$].}
\label{fig:error-scrambling}
\end{figure}

We now show that the probability that $P$ and $Q$ cancel is negligible---i.e., the probability of $Q U_{i+k:i} P = U_{i+k:i}$ (up to a phase) is negligible---when $k$ is sufficiently large. This is the condition
\begin{equation}
  \textrm{Pr}[ P^{(k)} = i^aQ] < \delta_{\textrm{cancel}},
\end{equation}
 for some small $\delta_{\textrm{cancel}} \geq 0$ (and where $i^a$ is any phase), where 
 \begin{equation}
 P^{(k)} = U_{i+k:i} P U^{-1}_{i+k:i}.
 \end{equation}
Now, by assumption [see (4) above] the probability that $Q$ is a high-weight error (weight 4 or higher) is negligible, and so 
\begin{equation}
  \textrm{Pr}[ P^{(k)} = i^aQ] \leq    \textrm{Pr}[ W(P^{(k)}) = W(Q)]  \leq \kappa(3,k)
\end{equation}
where $\kappa(w,k)$ is the probability that $P^{(k)}$ is a weight-$w$ or fewer error:
\begin{equation}
\kappa(w,k) \equiv \sum_{w'=1}^{w} \textrm{Pr}[W(P^{(k)}) = w'].
\end{equation}
So, the probability that $P$ and $Q$ cancel is bounded by the probability that $P^{(k)}$ is a weight-3 or fewer error.

The probability that $P^{(k)}$ is a weight-3 or fewer error decreases rapidly with increasing $k$, independent of $P$'s distribution over Pauli errors. The probability distribution for $P$ that minimizes the expected weight of $P^{(k)}$ is a distribution that only has support on weight-1 errors. But a weight-1 Pauli operator rapidly increases in weight as it is commuted through $k$ $\Omega$-distributed random gates, as illustrated in Fig.~\ref{Fig:stochastic-theory} (c). To see why this is, consider the effect of commuting some Pauli operator $P_{\textrm{in}}$ through a random gate $G=L_2L_1$ [$L_1$ and $L_2$ are as defined in assumption (2), above].
\begin{itemize}
\item \emph{Propagation through $L_1$}: The Pauli operator $P_{\textrm{in}}$ is first commuted through a layer of independent and uniformly random single-qubit Clifford gates ($L_1$). This has a local homogenising effect on $P_{\textrm{in}}$. Specifically, if we let $E(P_{\textrm{in}})$ denote the set of qubits that $P_{\textrm{in}}$ acts non-trivially on, then this layer transforms $P_{\textrm{in}}$ into a \emph{uniformly random} non-identity Pauli on that set of qubits. Therefore, $P_{\textrm{in}} \to P_{\textrm{int}}$ where $P_{\textrm{int}} $ is uniformly distributed over the $3^{w}$ different Pauli errors on $E(P_{\textrm{in}})$ where $w = W(P_{\textrm{in}})$. 
\item \emph{Propagation through $L_2$}: The randomized Pauli operator $P_{\textrm{int}}$ is then propagated through a random layer of $\textsc{cnot}$ gates ($L_2$). In expectation, this will typically (but not always) spread the error to more qubits as long as $W(P_{\textrm{int}}) < \nicefrac{3n}{4}$.
\begin{itemize}
\item  If none of the qubits in $E(P_{\textrm{int}})$ are connected to each other, $L_2$ cannot couple pairs of qubits within $E(P_{\textrm{int}})$. In this case, $L_2$ contains \textsc{cnot} gates that couple each qubit within $E(P_{\textrm{in}})$ with a qubit outside of $E(P_{\textrm{in}})$, spreading the error onto that other qubit with a probability of $\nicefrac{2}{3}$. This is because a \textsc{cnot} gate maps 4 of the 6 weight-1 Pauli operators to weight-2 Pauli operators. Therefore, the effect of $L_2$ is to spread the error to additional qubits: it maps $P_{\textrm{int}} \to P_{\textrm{out}}$ where the expected weight of $P_{\textrm{out}}$ is 
\begin{equation}
\mathbb{E}[W(P_{\textrm{out}})] = W[P_{\textrm{in}}] + 2W[P_{\textrm{in}}]/3 = 5W[P_{\textrm{in}}]/3.
\end{equation}
\item If some of the qubits in $E(P_{\textrm{int}})$ are connected to each other, there are two effects that compete to decrease or increase the weight of $P_{\textrm{int}}$. Those $\textsc{cnot}$ gates in $L_2$ that interact two qubits within $E(P_{\textrm{int}})$ will correct (i.e., remove) the error on one of the qubits with probability $\nicefrac{4}{9}$. This is because a \textsc{cnot} gate maps 4 of the 9 weight-2 Pauli operators to weight-1 Pauli operators. In contrast, those $\textsc{cnot}$ gates in $L_2$ that interact a qubit from $E(P_{\textrm{int}})$ with one outside of $E(P_{\textrm{int}})$ spread an error to that other qubit with probability $\nicefrac{2}{3}$.
\end{itemize}
\end{itemize}

The distribution of $P^{(k)}$ can be calculated by $k$ applications of the above randomization process to $P$. The Pauli-operator-valued random variable $P$ is (by assumption) distributed over only low-weight errors, but the process of propagating $P$ through $j$ gates ($G=L_2L_1$) causes the expected weight of the error $P^{(j)}$ to quickly increase with $j$. The rate of increase depends on the probability at step $j$ that a $\textsc{cnot}$ in $L_2$ couples two qubits within $E(P^{(j)})$ instead of coupling one qubit within $E(P^{(j)})$ with one outside of $E(P^{(j)})$. This probability depends on the qubits' connectivity, as well as the initial distribution of $P$. The result is that $\mathbb{E}[W(P^{(j)})]$ initially grows linearly with $j$ if $\alpha = 2$ (ring connectivity) and exponentially if $\alpha=n-1$ (all-to-all connectivity) \cite{Hunter-Jones2018-zh,Von_Keyserlingk2018-ts,Nahum2018-zs} [see Fig.~\ref{fig:error-scrambling} (a), and note that in all cases $\mathbb{E}[W(P^{(j)})] \to 3n/4$ as $j \to \infty$]. Most importantly for our purposes, for \emph{any} connectivity $\alpha$ the number of layers $k$ required to satisfy $\kappa(3,k) \approx 0$ is a \emph{constant}, i.e., it is independent of $n$. [see Fig.~\ref{fig:error-scrambling} (b) for $\kappa(3,k)$ when $n=30$---this figure is discussed further below]. 

We are now ready to state a sufficient condition for reliable direct RB. A condition that is independent of the underlying Pauli error model is preferable (as this is unknown in practice), so we define:
\begin{equation}
K(w,k) = \max_{P} \kappa(w,k) \equiv \max_P\left\{\sum_{w'=1}^{w} \textrm{Pr}[W(P^{(k)}) = w']\right\}.
\end{equation}
This is $\kappa(w,k)$ maximized over all possible distributions for $P$---the worst-case is when $P$ has support only on weight-1 errors. $K(w,k)$ quantifies the probability that a Pauli-operator-valued random variable $P$ is a weight-$w$ or less error after it is propagated through $k$ random layers, maximized over all possible distributions for $P$. We now define $k_{\textrm{delocal}}$ to be the smallest $k$ such that $K(3,k) < \delta_{\textrm{delocal}}$ where $ \delta_{\textrm{delocal}} > 0$ is some small constant. The above argument has shown that:
\begin{itemize}
    \item[(i)] two or more errors that are separated by at least $k_{\textrm{delocal}}$ layers have negligible probability to cancel, and
    \item[(ii)] $k_{\textrm{delocal}}$ is independent of $n$. 
\end{itemize}
Direct RB will therefore be reliable as long as the probability that two or more errors occur within $k_{\textrm{delocal}}$ gates of each other is negligible. We can guarantee this with the following condition:
 \begin{equation}\label{eq:delocal}
\epsilon_{\Omega}  < \nicefrac{1}{k_{\textrm{delocal}}}.
\end{equation}
This is a simple and easy-to-verify condition that is sufficient for reliable direct RB. Note that it is easy-to-verify as (i) $\epsilon_{\Omega}$ can be estimated by direct RB,\footnote{Even when direct RB is not reliable, it will provide a reasonable order-of-magnitude estimate for $\epsilon_{\Omega}$. Alternatively, an approximate value for $\epsilon_{\Omega}$ may already be known.} and (ii) $k_{\textrm{delocal}}$ can be efficiently calculated using simulations, for a particular Clifford gate set $\mathbb{G}$ and sampling distribution $\Omega$.

Equation~\eqref{eq:delocal} shows that direct RB will be reliable whenever $\epsilon_{\Omega}$ is smaller than some constant ($\nicefrac{1}{k_{\textrm{delocal}}}$). However, $\epsilon_{\Omega}$ will typically grow with $n$. Therefore, Eq.~\eqref{eq:delocal} will not be satisfied for all $n$ for any realistic $n$-parameterized error models (e.g., constant single- and two-qubit gate error rates). So it is interesting to consider whether there is a weaker condition that guarantees reliable direct RB.\footnote{Even though direct RB will not be feasible when $n\gg 1$ \emph{and} $\epsilon_{\Omega}$ is large because the SSPAM subcircuits of direct RB circuit will not be implementable with significantly non-zero fidelity.} Two errors can cancel only if, when propagated past the intervening layers, they occur on the same set of qubits (as implied by the illustration of Fig.~\ref{Fig:stochastic-theory} (c)). So, the probability of error cancellation will be minimal even if two errors often occur within $k_{\textrm{delocal}}$ layers of each other as long as those errors almost certainly occur on distant qubits. This suggests that direct RB will be reliable if $\epsilon_{\Omega, \textrm{perQ}} < \nicefrac{1}{k_{\textrm{delocal}}}$ where $\epsilon_{\Omega, \textrm{perQ}}$ is some quantification of the error rate per-qubit. This can be formalized if we assume a local Pauli error model, i.e., every gate's error channel is a tensor products of local Pauli stochastic channels. Then our argument (above) implies that direct RB will be reliable if
\begin{equation}
  \epsilon_{\Omega, \textrm{perQ},\max} < \nicefrac{1}{k_{\textrm{delocal}}},
\end{equation}
where $  \epsilon_{\Omega, \textrm{perQ},\max}$ is the maximum of the average infidelities of the $n$ qubits. Note that, in this error model, $  \epsilon_{\Omega, \textrm{perQ},\max}$ is well-defined because the tensor-product structure of the error maps means that each qubit has a well-defined infidelity per $n$-qubit gate.

Our sufficient condition for reliable direct RB [Eq.~\eqref{eq:delocal}] depends on the number of random layers needed to spread a low-weight error onto approximately 4 or more qubits ($k_{\textrm{delocal}}$). We have argued that $k_{\textrm{delocal}}$ is independent of $n$ (for $n\gg 1$) and that it decreases with increasing qubit connectivity ($\alpha$). To quantify $k_{\textrm{delocal}}$ and verify its dependence on $\alpha$, we used simulations to calculate the probability $K(w,k)$ that a weight-1 error is a weight $w$ or less error after it has been commuted past $k$ random gates. Figure~\ref{fig:error-scrambling} (b) shows the results, for $w=1,2,3$ and two extremal connectivities: all-to-all (blue) and ring (orange) connectivity [we also show $K(w,k)$ for a ring connectivity with \emph{correlated} sampling of $L_2$ layers, so that the $L_2$ layers alternate between coupling a qubit with its neighbour on its left or right, which significantly speeds up error propagation]. We find that $K(w,k)$ for $w=1,2,3$ decreases rapidly for both linear and all-to-all connectivity. These results are for $n=30$ qubits, but note that there is only a weak $n$ dependence and these $K(w,k)$ upper-bound the values of $K(w,k)$ for $n > 30$ qubits.

\begin{figure}[t!]
\begin{center}
\includegraphics[width=7cm]{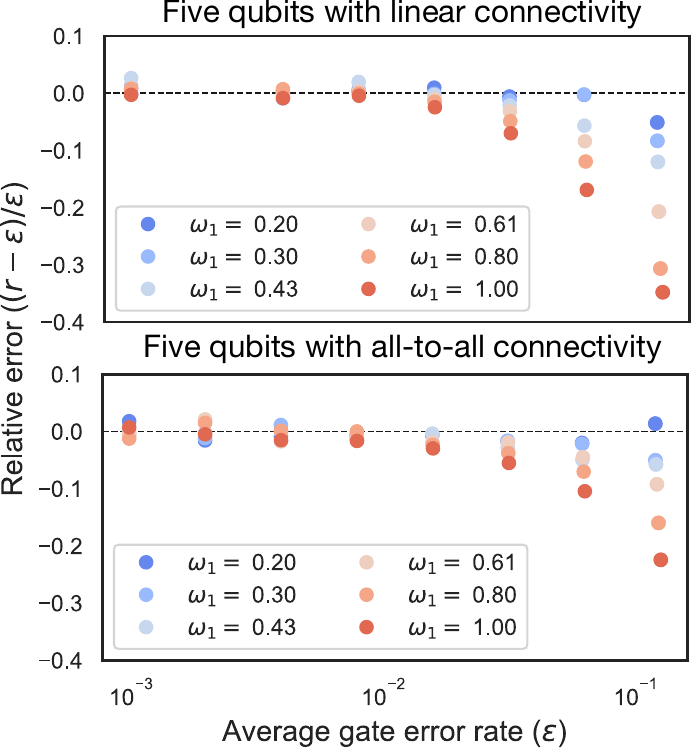}
\end{center}
\caption{Simulations investigating the reliability of direct RB in the slow scrambling regime. The discrepancy between the estimated direct RB error rate ($r$) and the actual average gate error rate ($\epsilon$) in simulations, as quantified by the relative error $\delta = (r - \epsilon)/\epsilon$. The simulations are of a 5-qubit device with linear connectivity (upper plot) and all-to-all connectivity (lower plot). By varying the average gate error ($\epsilon$) and the in-homogeneity of the error rates across the qubits (related to $\omega_1$) we see that direct RB is reliable under a wide range of different connectivities and stochastic error models. Details of the simulation, including definitions of $\omega_1$ and $\epsilon$ are given in the main text.}
\label{Fig:varied-error-simulation}
\end{figure}

\subsection{Numerical simulations demonstrating negligible error cancellation}\label{sec:spreading}
Our theory implies that direct RB is reliable whenever the probability of error cancellation is negligible, and that this error cancellation probability is negligible under broad conditions. It also implies that the error cancellation probability decreases as
\begin{enumerate}
\item[(1)] the error rates of the gates decreases, 
\item[(2)] the homogeneity of the error probabilities increases (reduced bias), and 
\item[(3)] the speed of error delocalization and scrambling increases.
\end{enumerate}
We investigated these claims using numerical simulations of direct RB. Our simulations are of direct RB on five qubits with a gate set containing gates constructed from parallel applications of (1) \textsc{cnot} gates between connected qubits, and (2) all 24 single qubit Clifford gates. 

A processor's connectivity strongly effects the maximum speed with which errors can be spread. With all-to-all connectivity, a single random layer can spread a localized error so that, for every qubit, there is an $O(\nicefrac{1}{n})$ probability that the error has been been spread to that qubit. In contrast, with linear connectivity, a single random layer can only spread an error to one of at most two adjacent qubits---meaning that there is a significant probability that an error that is delocalized is then relocalized by a subsequent random layer. So, to investigate effect (3), above, we simulated direct RB for two extremal processor connectivities: all-to-all and linear connectivity. We investigate effects (1) and (2), above, by varying the strength of the errors and by interpolating from (i) equal error probabilities on all qubits, and (ii) errors occurring only on a single qubit. In the simulated error model, qubit $i$ (with $i=1,2,\dots,5$) had an error rate of $\epsilon_i=\tilde{\epsilon} \omega_i$ for weights given by 
\begin{equation}
\omega_i =\frac{\alpha^{i}}{\sum_{j=1}^{5}\alpha^j},
\end{equation}
for some $\alpha$. On each qubit, each gate experiences the same error map, and each Pauli occurs with equal probability, i.e., local depolarization. The value of $\alpha$ controls the homogeneity of the error rates on the different qubits. The value of $\tilde{\epsilon}$ sets the error rate: 
\begin{equation}
\epsilon_{\Omega} = 1 - \prod_i(1-\tilde{\epsilon}\omega_i) \approx \tilde{\epsilon}.
\end{equation}

We simulated direct RB under the above error model, for linear and all-to-all connectivity, with both $\alpha$ and $\epsilon$ varied.\footnote{For each connectivity and each value of $\tilde{\epsilon}$ and $\alpha$ we independently simulated a direct RB experiment 50 times, and each direct RB experiment consisted of 30 circuits at benchmark depths $d=0,1,2,4,8 \dots, 128$. In each simulated direct RB experiment we compute the estimate of $r_{\Omega}$, and then we take the mean of these 50 values.} We estimated $r_{\Omega}$ from the data, and compared it to $\epsilon_{\Omega}$. The results of these simulations are summarized in Fig.~\ref{Fig:varied-error-simulation}, which quantifies the accuracy of the direct RB error rate $r_{\Omega}$ by the relative error $\delta = (r_{\Omega}-\epsilon_{\Omega})/\epsilon_{\Omega}$. 
For $\epsilon_{\Omega} \leq 1\%$ we see that direct RB is accurate (i.e., $\delta \approx 0$) for both qubit connectivities and even when all the error is on a single qubit ($w_1 = 1$). This is because even when the error is all on one qubit, the random circuit layers almost certainly delocalize any error before another error can occur (if $\epsilon_{\Omega} = 1\%$ an error is expected to occur approximately every 100 gates, and an error is delocalized in many fewer than 100 gates, as shown in Fig.~\ref{fig:error-scrambling}). As the error rate increases up to $\epsilon_{\Omega} = 10\%$, we observe that $r_{\Omega}$ begins to underestimate $\epsilon_{\Omega}$, and this underestimate is worse with linear connectivity and with greater bias in the error (increasing $w_1$). This is because, as $w_1$ increases, sequential errors in a direct RB circuit are more likely to occur on the same qubit, and when $\epsilon_{\Omega}$ is large enough an error is not delocalized with high probability before another error occurs on the same qubit. These results therefore validate our theory of direct RB.

\subsection{The impact of stabilizer state preparation and measurement errors}
The theory in this section has so far assumed perfect SSPAM (although the simulations, above, do not). We now briefly consider the effect of imperfect SSPAM---which is unavoidable in practice. The main impact of errors in the SSPAM is to decrease the $d=0$ success probability ($S_0$). Preparing a typical stabilizer state from a computational basis state requires $O(\nicefrac{n^2}{\log n})$ one- and two-qubit gates \cite{aaronson2004improved,patel2003efficient,bravyi2020hadamard}, so errors in the SSPAM operations are the main limitation on the number of qubits that direct RB (on Clifford gates) can be applied to, as shown by the simulations in Fig.~\ref{fig:simple-simulation}. To confirm that SSPAM errors do not have a significant impact on the estimated direct RB error rate, we simulated direct RB under the same error model, but with or without SSPAM error. Specifically, SSPAM was either implemented by compiling them into circuits containing imperfect native gates, or SSPAM was implemented without error.

\begin{figure}[t!]
\begin{center}
\includegraphics[width=10cm]{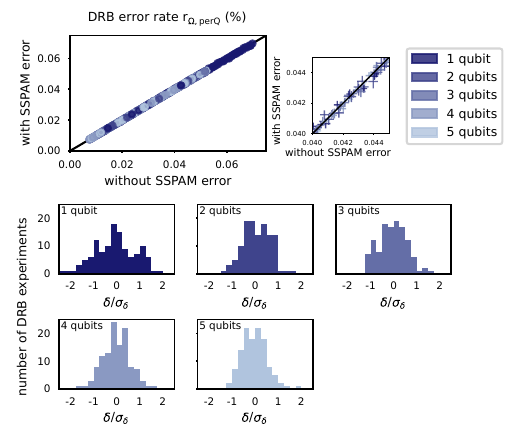}
\end{center}
\caption{Numerical simulations demonstrate that the direct RB error rate is unaffected by errors in the stabilizer state preparation and measurement (SSPAM) subcircuits, including the error in the intrinsic state preparation and measurement (SPAM). (Top) Each point shows the estimates for the direct RB error rate per qubit $r_{\Omega, \textrm{perQ}}=1-(1-r_{\Omega})^{\nicefrac{1}{n}}$ obtained when direct RB is simulated under the same error model with imperfect SSPAM (horizontal axis) or with artificially perfect SSPAM (vertical axis). Each data point corresponds to a different stochastic Pauli error model with randomly sampled error rates (details in main text). (Bottom) The relative error $\delta$ in the direct RB error rate with respect to the direct RB error rate with perfect SSPAM circuits, divided by its standard deviation $\sigma_{\delta}$ (the standard deviation is calculated using a nonparametric bootstrap). These histograms exhibit no statistically significant evidence for any difference between $r_{\Omega}$ with SSPAM error and $r_{\Omega}$ without SSPAM error.}
\label{fig:sspam}
\end{figure}

We simulated direct RB on $n$ qubits with $n= 0,1,\dots,5$ and with 120 randomly-generated stochastic Pauli error models for each $n$. We used a gate set constructed from parallel applications of $X(\nicefrac{\pi}{2})$,$Y(\nicefrac{\pi}{2})$, $I$, and \textsc{cnot} gates. For error models with $n \geq 2$, the Pauli error rates in the model are sampled such that the expected two-qubit gate infidelity is $q$ for 120 evenly-spaced values $q \in [0.004, 0.024]$, and each single-qubit gate has an infidelity of $0.1q$. For error models with $n = 1$, the Pauli error rates on each single-qubit gate are sampled so that the expected infidelity of each gate is $q$. Figure~\ref{fig:sspam} shows the results of these simulations. We observe no systematic difference in the direct RB error rate with and without SSPAM error (see upper plots in Fig.~\ref{fig:sspam}). Moreover, there is no statistically significant evidence of any difference between $r_{\Omega}$ with or without SSPAM error (see the histograms in Fig.~\ref{fig:sspam}). This is evidence that $r_{\Omega}$ is not significantly affected by physically realistic SSPAM errors. This is consistent with the theory for direct RB that we present in Section~\ref{sec:general-theory}.

\subsection{A heuristic theory for direct RB with general errors}
In this section we have presented a theory that explains how and why direct RB works when applied to gates that experience stochastic Pauli errors. We now explain how the above theory can be extended to a \emph{descriptive} theory for direct RB with general Markovian errors. The fundamental idea is that layers of random gates decohere errors: coherent errors can systematically add or cancel across short distances in a random circuit, but (on average) they cannot systematically add or cancel across many layers of a random circuit. This can be most easily formalized for circuits containing layers of random single qubit gates, as assumed throughout this section. This is because a single layer of uniformly random single-qubit Clifford gates (an $L_1$ layer) is sufficient to project \emph{any} error map (a CPTP superoperator) onto the space spanned by tensor-products of local depolarizing channels \cite{gambetta2012characterization}. So, for direct RB circuits that contain layers of uniformly random single-qubit Clifford gates (or Pauli gates), the behavior of direct RB under any Markovian error model can be modeled using its ``equivalent'' stochastic Pauli error model---i.e., the model in which each error map is replaced with its Pauli-twirl. Therefore, the theory in this section also arguably proves that direct RB is reliable for general Markovian errors, although it does not make precise mathematical predictions in that case. This complements the \emph{predictive} theory for direct RB of gates that experience general Markovian errors that we present in Section~\ref{sec:general-theory}. That theory does not require the above assumption about the distribution of the random layers in the direct RB circuits (i.e., uniformly random $L_1$ layers), but (unlike the theory in this section) it does not prove that direct RB is reliable when $n\gg 1$ except with very stringent conditions on the gate's error rates.

\section{Sequence-asymptotic unitary designs}\label{sec:sequads}
Our theory for direct RB with general Markovian errors (Section~\ref{sec:general-theory}) relies on the concept of \emph{sequence-asymptotic unitary 2-designs}. In this section we define sequence-asymptotic unitary 2-designs and we prove some results about their properties. Informally, a sequence-asymptotic unitary 2-design is a gate set $\mathbb{G}$ and a sampling distribution $\Omega$ over $\mathbb{G}$ such that the set of unitaries created by length $l$ sequences of gates sampled from $\Omega$ converges to a unitary 2-design as $l \to \infty$. The concept of sequence-asymptotic unitary 2-designs is related to approximate unitary 2-designs \cite{dankert2009exact, gross2007evenly}, and the broader theory of convergence to unitary $t$-designs and other scrambling properties of random circuits \cite{brown2015decoupling, sekino2008fast, brandao2016local, Oszmaniec2022-gs, Hunter-Jones2018-zh,Von_Keyserlingk2018-ts,Nahum2018-zs}. But, because we only use sequence-asymptotic unitary 2-designs as a tool for studying direct RB, we do not discuss the relationship with this prior work.

A plausibly sufficient condition for $(\mathbb{G},\Omega)$ to constitute a sequence-asymptotic unitary 2-design is that $\mathbb{G}$ generates a unitary 2-design (assuming that $\Omega$ has support on all of $\mathbb{G}$). This is because, for a wide class of generating sets and probability distributions, a random sequence of group generators converges to a random group element with increasing length sequences \cite{diaconis1988group, diaconis1995random, saloff2004random}. However, this convergence does not happen for all probability distributions over all generating sets: the long-length distribution can oscillate between the uniform distribution on different parts of the group (an example is given towards the end of this section, and this is only possible if $\mathbb{G}$ is contained within a coset of a proper subgroup of $\mathbb{C}$). In this section, we will show the following: given a set of generators for any group that is a unitary 2-design, length-$l$ sequences of $\Omega$-distributed and independently sampled generators converge to a unitary 2-design as $l \rightarrow \infty$, if and only if the set of length-$l$ sequences also generate unitary 2-designs for all $l$.

\subsection{Notation}\label{sec:notation}
First we need to define some notation. In this section we consider a quantum system of dimension $d$, without the restriction to $d=2^{n}$ used throughout the rest of this paper. Let $\mathfrak{L}\left(H_{d}\right)$ denote the space of linear operators on the $d$-dimensional Hilbert space $H_{d}$. Let $\mathbb{T}\left(H_{d}\right)$ denote the space of superoperators on $H_{d}$, which is the set of all linear operators $\mathcal{S}$ whereby $\mathcal{S}$ : $\mathfrak{L}\left(H_{d}\right) \rightarrow \mathfrak{L}\left(H_{d}\right)$.  In the following proofs, we will use Hilbert-Schmidt space, and superoperator ``stacking''. Let
\begin{equation}
\mathrm{B}=\left\{B_{0}, B_{1}, \ldots, B_{d^{2}-1}\right\}
\end{equation}
be an orthonormal basis for $\mathfrak{L}\left(H_{d}\right)$, with respect to the Hilbert-Schmidt inner product 
\begin{equation}
\langle X, Y\rangle=\operatorname{Tr}\left(X^{\dagger} Y\right).
\end{equation}
We can always choose this basis such that $B_{0}=\mathbb{1} / \sqrt{d}$ and $\operatorname{Tr}(B_{j})=0$ for $j>0$, and this will be assumed herein. For example, with $d=2^{n}$ the elements of this basis can be the $n$-qubit Pauli operators multiplied by $1 / \sqrt{2^{n}}$.

Any $\rho \in \mathfrak{L}\left(H_{d}\right)$ can be expanded with respect to the basis $\mathrm{B}$ as
\begin{equation}
\rho=\sum_{j}\langle B_{j},\ \rho\rangle B_{j}.
\end{equation}
As such, $\rho$ may be represented by the column vector
\begin{equation}
\sket{\rho}=\left(\left\langle B_{0}, \rho\right\rangle,\left\langle B_{1}, \rho\right\rangle, \ldots,\left\langle B_{d^{2}-1}, \rho\right\rangle\right)^{T} .
\end{equation}
Similarly, a superoperator $\mathcal{X}$ may be represented as the matrix $\mathcal{X}^{\mathrm{HS}}$ with elements $\mathcal{X}_{i j}^{\mathrm{HS}}=\left\langle B_{i}, \mathcal{X} B_{j}\right\rangle$. In the following we will drop the superscript; we will not use a distinct notation for a superoperator and the matrix representation of the superoperator acting on Hilbert-Schmidt space.

In this and the next section we will be considering linear maps from superoperators to superoperators, which we will refer to as ``superchannels'' and denote using the script font (e.g., $\mathscr{L}$ ). When considering superchannels it is often convenient to represent matrices acting on Hilbert-Schmidt space (i.e., superoperators) as vectors in a vector space, and superchannels as matrices acting on that vector space. To achieve this we use the invertible linear ``stacking'' map
\begin{equation}
\operatorname{vec}\left(| B_{j} \rrangle \llangle B_{k} |\right) = |B_{k} \rrangle \otimes |B_{j}\rrangle,
\end{equation}
which stacks the columns of a matrix. We denote the inverse ``unstacking'' operation by
\begin{equation}
\operatorname{unvec} \left( | B_{k} \rrangle \otimes | B_{j} \rrangle \right) = | B_{j} \rrangle \llangle B_{k} |.
\end{equation}
From this definition, it follows that
\begin{equation}
\operatorname{vec}(\mathcal{A B C})= \left(\mathcal{C}^{T} \otimes \mathcal{A} \right) \operatorname{vec}(\mathcal{B})
\end{equation}
for Hilbert-Schmidt matrices $\mathcal{A}, \mathcal{B}$ and $\mathcal{C}$. Therefore, any superchannel $\mathscr{K}$ with the action $\mathscr{K}(\mathcal{B})=\sum_{i} \mathcal{A}_{i} \mathcal{B C}_{i}$ may be rewritten as a matrix
\begin{equation}
  \textrm{mat}[\mathscr{K}] = \sum_{i}\left(\mathcal{C}_i^{T} \otimes \mathcal{A}_i\right),
\end{equation}
that acts on vectors $|\mathcal{B}) \equiv \operatorname{vec}(\mathcal{B})$, where we will use the $|\cdot)$ notation to succinctly denote stacked superoperators while distinguishing them from pure states in Hilbert space (denoted $|\cdot\rangle)$ and Hilbert-Schmidt vectors (denoted $\sket{\cdot})$.

\subsection{Ordinary unitary designs}
Before introducing sequence-asymptotic unitary designs, we first review the definition of an ``ordinary'' unitary $t$-design \cite{gross2007evenly, dankert2009exact}. Although not the original definition \cite{dankert2009exact}, an equivalent definition for a unitary $t$-design is \cite{gross2007evenly}:

\begin{definition}
The gate set $\mathbb{G} \subseteq \mathbb{U}$ is a unitary t-design if
\begin{equation}
\frac{1}{|\mathbb{G}|} \sum_{\mathcal{G} \in \mathbb{G}} \mathcal{G}^{\otimes t}=\int_{\mathcal{U} \in \mathbb{U}} \mathcal{U}^{\otimes t} d \mu
\end{equation}
where $d \mu$ is the Haar measure on $\mathbb{U}$.
\end{definition} 

The class of designs important herein are the unitary 2-designs. The definition of a unitary $t$-design implies that an equivalent characterization of a unitary 2-design is the following: a gate set is a unitary 2-design if and only if, when it ``twirls'' any channel, it projects this channel onto the space of depolarizing channels. Specifically, define the $\mathbb{G}$-twirl superchannel $\mathscr{T}_{\mathbb{G}}: \mathbb{T}\left(H_{d}\right) \rightarrow \mathbb{T}\left(H_{d}\right)$ by
\begin{equation}
\mathscr{T}_{\mathbb{G}}(\mathcal{E})=\frac{1}{|\mathbb{G}|} \sum_{\mathcal{G} \in \mathbb{G}} \mathcal{G E} \mathcal{G}^{\dagger}.
\end{equation}
The twirl superchannel is a linear operator on superoperators. If $\mathbb{G}$ is a unitary 2-design then
\begin{equation}\label{eq:twirled_dep}
\mathscr{T}_{\mathbb{G}}(\mathcal{E})=\mathcal{D}_{\lambda(\mathcal{E})}, \quad \lambda(\mathcal{E})=\frac{\operatorname{Tr}[\mathcal{E}]-1}{d^{2}-1},
\end{equation}
for any trace-preserving superoperator $\mathcal{E} \in \mathbb{T}\left(H_{d}\right)$ \cite{carignan2015characterizing}, where $\mathcal{D}_{\lambda(\mathcal{E})}$ is a ``global'' depolarizing channel (i.e, uniform depolarization on the $d$-dimensional space).

Eq.~\eqref{eq:twirled_dep} implies that the twirl superchannel for any unitary 2-design is a rank 2 projector, with support on the depolarizing subspace $\mathbb{S}_{\textrm{dep}}$ spanned by the identity channel and any non-identity depolarizing channel. Explicitly,
\begin{equation} \label{eq:dep-subspace}
\mathbb{S}_{\textrm{dep}}=\operatorname{span}\left\{\mathcal{D}_{0}, \mathbb{1}\right\},
\end{equation}
where $\mathbb{1}$ denotes the identity superoperator and $\mathcal{D}_{0}$ is the completely depolarizing channel.

\subsection{Sequence-asymptotic unitary designs}\label{subsec:seq-asy}
We now introduce the notion of a sequence-asymptotic unitary 2-design. This concept is distinct from the ``asymptotic unitary 2-designs'' of Gross \emph{et al.}~\cite{gross2007evenly}, which are asymptotically unitary 2-designs with respect to scaling of the dimension $d$. This definition will be in terms of a set $\mathbb{G}$ of unitaries and a probability distribution $\Omega$ over $\mathbb{G}$. Note that we do not assume that $\mathbb{G}$ is finite. Instead, we assume the weaker condition that $\mathbb{G}$ contains a finite number of ``gates'' that are each either fixed (i.e., a constant matrix) or continuously parameterized. For example, a one-qubit gate set of this sort $\mathbb{G}$ is the Hadamard gate and a continuous family of rotations around $\sigma_{z}$. The probability distribution $\Omega$ should be interpreted as an arbitrary weighted mixture of probability distributions over these fixed or parameterized gates. That is, sampling from $\Omega$ consists of first selecting one of the gates, and if this selected gate is continuously parameterized we then further sample a value for the continuous parameter(s) according to some distribution. We also assume that all of the probability density functions over any continuous parameters are continuous functions. So when we use the notation $\sum_{\mathcal{G} \in \mathbb{G}}$ this should be interpreted as a sum over the finite number of (perhaps parameterized) gates and an integral over any continuous parameters in each gate.

\begin{definition}\label{def:seq-asym}
    Let $\mathbb{G} \subseteq \mathbb{U}$ be a unitary gate set and $\Omega$ : $\mathbb{G} \rightarrow[0,1]$ be a probability distribution over $\mathbb{G}$. The tuple $(\mathbb{G}, \Omega)$ is a sequence-asymptotic unitary $t$-design if
\begin{equation}
\lim _{m \rightarrow \infty}\left(\sum_{\mathcal{G} \in \mathbb{G}} \Omega(\mathcal{G}) \mathcal{G}^{\otimes t}\right)^{m}=\int_{\mathcal{U} \in \mathbb{U}} \mathcal{U}^{\otimes t} d \mu.
\end{equation}
\end{definition}
If $\mathbb{G}$ is a unitary $t$-design then $(\mathbb{G},\Pi)$ is also a sequence-asymptotic unitary $t$-design, where $\Pi$ is the uniform distribution (this can be shown by induction).

\begin{definition}
    The $(\mathbb{G}, \Omega)$-twirl superchannel is the linear map $\mathscr{T}_{\mathbb{G}, \Omega}: \mathbb{T}\left(H_{d}\right) \rightarrow \mathbb{T}\left(H_{d}\right)$ with the action
\begin{equation}
\mathscr{T}_{\mathbb{G}, \Omega}(\mathcal{E})=\sum_{\mathcal{G} \in \mathbb{G}} \Omega(\mathcal{G}) \mathcal{G} \mathcal{E} \mathcal{G}^{\dagger} .
\end{equation}
\end{definition} 
If $(\mathbb{G}, \Omega)$ is a sequence-asymptotic unitary 2-design, Definition~\ref{def:seq-asym} implies that the $m^{\text {th }}$ power of the $(\mathbb{G}, \Omega)$ twirl superchannel $\mathscr{T}_{\mathbb{G}, \Omega}$ converges to a rank 2 projector as $m \rightarrow \infty$. As such, the absolute values of the eigenvalues of $\mathscr{T}_{\mathbb{G}, \Omega}$ outside of this subspace must be strictly bounded above by $1$. The converse is also true: this condition on the eigenvalues of $\mathscr{T}_{\mathbb{G}, \Omega}$ is sufficient for any $(\mathbb{G}, \Omega)$ to be a sequence-asymptotic unitary 2-design. We state this more formally in the following proposition, for which we will require the notion of the fixed points of a linear map. The fixed points of a linear map $L: V \rightarrow V$, for some vector space $V$, are those $v \in V$ such that $L(v)=v$. That is, they are the eigenvectors with eigenvalue 1. For example, the space of all fixed points of the twirling map for any unitary 2-design is $\mathbb{S}_{\rm {dep}}$ [as defined in Eq.~\eqref{eq:dep-subspace}]. The following are formal requirements on properties of the map $\mathscr{T}_{\mathbb{G}, \Omega}$, that will allow us to characterize the sequence-asymptotic unitary 2-designs. 

\begin{proposition}\label{prop:seqasymreq}
Let $\mathbb{G} \subseteq \mathbb{U}$ be a unitary gate set and $\Omega: \mathbb{G} \rightarrow[0,1]$ be a probability distribution over $\mathbb{G}$. The tuple $(\mathbb{G}, \Omega)$ is a sequence-asymptotic unitary 2-design if and only if the following two conditions hold:

\begin{enumerate}
\item The space of all fixed points of $\mathscr{T}_{\mathbb{G}, \Omega}$ is the subspace of depolarizing channels $\mathbb{S}_{\rm {dep }}$.

\item If $\mathscr{T}_{\mathbb{G}, \Omega}(\mathcal{E})=\lambda \mathcal{E}$ with $\mathcal{E} \notin \mathbb{S}_{\rm {dep }}$ then $|\lambda|<1$.

\end{enumerate}
\end{proposition}
\begin{proof} We first prove that if $(\mathbb{G}, \Omega)$ is a sequence-asymptotic unitary 2-design then (1) and (2) are implied. If $(\mathbb{G}, \Omega)$ is a sequence-asymptotic unitary 2-design then the fixed points of $\mathscr{T}_{\mathbb{G}, \Omega}^{m}$ converge to $\mathbb{S}_{\textrm{dep}}$ as $m \rightarrow \infty$. The superchannel $\mathscr{T}_{\mathbb{G}, \Omega}$ has complex eigenvalues of modulus at most $1$. One way to see this is by observing that $\operatorname{mat}\left[\mathscr{T}_{\mathbb{G}, \Omega}\right]:=\sum_{\mathcal{G} \in \mathbb{G}} \Omega(\mathcal{G}) \mathcal{G}^* \otimes \mathcal{G}$, which has the same spectrum as $\mathscr{T}_{\mathbb{G}, \Omega}$, is a convex sum of unitaries, and so it is a CPTP map. Hence, any fixed point of $\mathscr{T}_{\mathbb{G}, \Omega}^{m}$ is an eigenchannel of $\mathscr{T}_{\mathbb{G}, \Omega}$ with an eigenvalue of unit modulus, and so it is also a fixed point of $\mathscr{T}_{\mathbb{G}, \Omega}$ if this eigenvalue is $1$. Moreover, all of the unit modulus eigenvalues of $\mathscr{T}_{\mathbb{G}, \Omega}$ must be equal to $1$, because if this was not the case then $\mathscr{T}_{\mathbb{G}, \Omega}^{m}$ would not have a convergent spectrum as $m \rightarrow \infty$, which would imply that $(\mathbb{G}, \Omega)$ was not a sequence-asymptotic unitary 2-design. Hence, the space of fixed points of $\mathscr{T}_{\mathbb{G}, \Omega}$ is $\mathbb{S}_{\textrm{dep}}$, and the eigenvalues for all eigenchannels outside of this subspace have absolute value strictly less than $1$.

It remains to prove that if (1) and (2) hold then $(\mathbb{G}, \Omega)$ is a sequence-asymptotic unitary 2-design. If (1) and (2) hold this immediately implies that $\mathscr{T}_{\mathbb{G}, \Omega}^{m}$ converges to a rank-2 projector onto $\mathbb{S}_{\rm {dep }}$ as $m \rightarrow \infty$. Projection onto this subspace is the action of the right hand side of Definition~\ref{def:seq-asym} for $t=2$ \cite{carignan2015characterizing}, so (1) and (2) imply that Definition~\ref{def:seq-asym} holds for $t=2$. 
\end{proof}

We now prove a lemma that will be of use later. Here and later, we let $\mathbb{G}^k = \{g_k \dots g_1 \mid g_i \in \mathbb{G}\}$ be the gate set consisting of the set of all length $k$ sequences of gates from $\mathbb{G}$ and $\Omega^k$ be the distribution over $\mathbb{G}^k$ obtained by $k$ convolutions of $\Omega$, i.e., $\Omega^k(g_k\dots g_1) = \Omega(g_k)\cdots\Omega(g_1)$.

\begin{lemma}\label{lemma:alldesign}
    The tuple $(\mathbb{G}, \Omega)$ is a sequence-asymptotic unitary 2-design if and only if, for all $k>0$, $(\mathbb{G}^k, \Omega^k)$ is a sequence-asymptotic unitary 2-design.
    \begin{proof}
        Suppose $(\mathbb{G}, \Omega)$ is a sequence-asymptotic unitary 2-design. The composite of twirling maps is a twirling map. Thus by taking the composite map, (1) and (2) of Proposition~\ref{prop:seqasymreq} hold. The reverse direction is trivial.
    \end{proof}
\end{lemma}

Proposition~\ref{prop:seqasymreq} allows us to test whether or not a given $(\mathbb{G}, \Omega)$ tuple is a sequence-asymptotic unitary 2-design, by explicitly calculating the spectrum of $\mathscr{T}_{\mathbb{G}, \Omega}$. However, for an $n$-qubit gate set, this brute force calculation becomes quickly intractable as $n$ increases---using the stacked Hilbert-Schmidt representation, $\mathscr{T}_{\mathbb{G}, \Omega}$ is a $16^{n} \times 16^{n}$ dimensional matrix. Moreover, this proposition is not particularly instructive for the task of finding sequence-asymptotic unitary 2-designs. In the following proposition and theorem, we give an alternative characterization of sequence-asymptotic unitary 2-designs.

\begin{proposition}\label{prop:preservedsubspace}
Let $\mathbb{G} \subseteq \mathbb{U}$ be a unitary gate set and $\Omega:\mathbb{G} \rightarrow[0,1]$ be a probability distribution over $\mathbb{G}$ with support on all of $\mathbb{G}$. Then $(\mathbb{G}, \Omega)$ is a sequence-asymptotic unitary 2-design if and only if the subspace of $\mathbb{T}\left(H_{d}\right)$ containing all superoperators that are preserved up to a (possibly superoperator-dependent) phase under conjugation by all elements of $\mathbb{G}$ is the space of depolarizing channels $\mathbb{S}_{\rm {dep}}$.
\end{proposition}

\begin{proof} Assume that $(\mathbb{G},\Omega)$ is a sequence-asymptotic unitary 2-design. Let $\mathbb{S}$ denote the set of all superoperators preserved under conjugation by all elements of $\mathbb{G}$ up to a phase, meaning the set of all $\mathcal{S} \in \mathbb{T}\left(H_{d}\right)$ satisfying $\mathcal{G S G}{ }^{-1}=e^{i\theta_{\mathcal{S}}}\mathcal{S}$ for all $\mathcal{G} \in \mathbb{G}$, where $\theta_{\mathcal{S}}$ is a phase that depends on the superoperator being conjugated. If the superoperator $\mathcal{S}$ is preserved under conjugation up to a phase by all elements of $\mathbb{G}$ then it is a unit-modulus eigenvector of the $(\mathbb{G}, \Omega)$-twirl superchannel $\mathscr{T}_{\mathbb{G}, \Omega}$. Thus, by $(2)$ in Proposition \ref{prop:seqasymreq}, it must have $\theta_{\mathcal{S}}=0$ because $(\mathbb{G}, \Omega)$ is a sequence-asymptotic unitary 2-design by assumption. Hence, $\mathbb{S}$ is a subspace of the set of all fixed points of $\mathscr{T}_{\mathbb{G}, \Omega}$. If $\mathbb{G}_{\Omega}$ is a sequence-asymptotic unitary 2-design then Proposition~\ref{prop:seqasymreq} states that the space of all fixed points of $\mathscr{T}_{\mathbb{G}, \Omega}$ is $\mathbb{S}_{\rm{dep}}$. Therefore, as all depolarizing maps are preserved under conjugation by any unitary gate, if $(\mathbb{G}, \Omega)$ is a sequence-asymptotic unitary 2-design then $\mathbb{S} = \mathbb{S}_{\rm{dep}}$.

Next, we must show that $\mathbb{S} = \mathbb{S}_{\rm{dep}}$ implies that $(\mathbb{G}, \Omega)$ is a sequence-asymptotic unitary 2-design. Assume that $\mathbb{S}=\mathbb{S}_{\rm {dep}}$, and consider $\mathcal{V} \in \mathbb{S}$ such that $\mathbb{G}\mathcal{V}\mathbb{G}^{-1} = e^{i\theta_{\mathcal{V}}}\mathcal{V}$. By definition, $\mathcal{V} \in \mathcal{S}$ and hence by assumption, $\mathcal{V} \in \mathbb{S}_{\rm {dep}}$. If $\mathcal{V}$ has support on the identity superoperator, we find that $\theta_{\mathcal{V}} = 0$ since the identity commutes with all elements of $\mathbb{G}$. The only element in $\mathbb{S}_{\rm{dep}}$ without support on the identity superoperator is the completely depolarizing map. However, this as well cannot have a nontrivial phase, as that would imply that channels that are mixtures of the identity superoperator and the completely depolarizing channel would be required to have the same nontrivial phase, which we have just concluded is impossible. Hence, if $\mathcal{V}$ is preserved up to phase under conjugation by all elements of $\mathbb{G}$, it is preserved with a phase of $1$.

Now, consider the $(\mathbb{G}, \Omega)$-twirl superchannel $\mathscr{T}_{\mathbb{G}, \Omega}$, and consider any $\mathcal{V}$ satisfying $\mathscr{T}_{\mathbb{G}, \Omega}(\mathcal{V})=e^{i \theta} \mathcal{V}$ for some $e^{i \theta}$. For clarity, let us express this as a matrix acting on a vector, by using the ``stack'' operation. We then have
\begin{equation}
\sum_{\mathcal{G} \in \mathbb{G}} \Omega(\mathcal{G})[\mathcal{G}^* \otimes \mathcal{G}] \operatorname{vec}(\mathcal{V})=e^{i \theta} \operatorname{vec}(\mathcal{V})
\end{equation}
where the matrix acting on $\operatorname{vec}(\mathcal{V})$ is a convex sum of unitaries, or a weighted sum and integral over unitaries with the total ``weighting'' integrating to $1$. Because unitaries preserve the length of vectors, this can only hold if $[\mathcal{G} \otimes \mathcal{G}] \operatorname{vec}(\mathcal{V})=e^{i \theta} \operatorname{vec}(\mathcal{V})$ for all $\mathcal{G} \in \mathbb{G}$ for which $\Omega(\mathcal{G})>0$, and this is all $\mathcal{G} \in \mathbb{G}$ by the conditions of the proposition. Hence, returning to the unstacked representation, we have that
\begin{equation}
\mathcal{G} \mathcal{V} \mathcal{G}^{-1}=e^{i \theta} \mathcal{V}
\end{equation}
for all $\mathcal{G} \in \mathbb{G}$. But we have shown above that, if $\mathbb{S}=\mathbb{S}_{\rm {dep }}$, this holds if and only if $\mathcal{V} \in \mathbb{S}_{\rm {dep}}$ and $e^{i \theta}=1$. As such, any eigen-operator of $\mathscr{T}_{\mathbb{G}, \Omega}$ outside of the depolarizing subspace has an eigenvalue with absolute value strictly less than $1$ in modulus. Proposition~\ref{prop:seqasymreq} then implies that $(\mathbb{G}, \Omega)$ is a sequence-asymptotic unitary $2$-design.
\end{proof}

Proposition~\ref{prop:preservedsubspace} implies the following theorem:
\begin{theorem}\label{theorem:generateddesign}
Let $\mathbb{G} \subseteq \mathbb{U}$ be a unitary gate set that either generates a group or generates a dense subset of a group, $\mathbb{C}$, and let $\Omega: \mathbb{G} \rightarrow[0,1]$ be a probability distribution over $\mathbb{G}$ with support on all of $\mathbb{G}$. Let $\mathbb{G}^k = \{g_1\dots g_k \mid g_i \in \mathbb{G}\}$, which generates a group, or a dense subset of a group, $\mathbb{C}^k$.  Then $(\mathbb{G}, \Omega)$ is a sequence-asymptotic unitary 2-design if and only if, for all $k$, $\mathbb{C}^k$ is a unitary 2-design.
\end{theorem}

\begin{proof} 
First we show that if each $\mathbb{C}^k$ is a unitary 2-design, then $\mathbb{G}$ is a sequence-asymptotic unitary 2-design. If $\mathbb{G}$ generates a unitary 2-design, then the depolarizing subspace is a subspace of the space preserved up to a phase by $\mathbb{G}$. Suppose there is some element in the preserved space that is not depolarizing. There is a $k$-string (a length $k$ sequence of $G \in \mathbb{G}$) that brings this phase arbitrarily close to one. Then, in the group generated by these $k$-strings, this element is also preserved by $\mathbb{C}^k$ with a phase arbitrarily close to $1$. However, we also have that $\mathbb{C}^k$ is a unitary 2-design and thus element must also belong to the depolarizing subspace, and thus we have a contradiction.

Next, we prove that if there is some $\mathbb{C}^k$ that is not a unitary 2-design, then $\mathbb{G}$ is not a sequence-asymptotic unitary 2-design. Let $\mathbb{S}_{\mathbb{G}^k}$ and $\mathbb{S}_{\mathbb{C}^k}$ denote the space of all superoperators preserved under conjugation up to a phase by all elements of $\mathbb{G}^k$ and $\mathbb{C}^k$, respectively. Because $\mathbb{G}^k$ generates $\mathbb{C}^k$, or generates a dense subset of $\mathbb{C}^k$, then $\mathbb{S}_{\mathbb{G}^k}=\mathbb{S}_{\mathbb{C}^k}$. If $\mathbb{C}^k$ is not a unitary 2-design then since $\mathbb{S}_{\mathbb{C}^k} = \mathbb{S}_{\mathbb{G}^k}$, by Proposition~\ref{prop:preservedsubspace} the elements of $\mathbb{S}_{\mathbb{G}^k}$ preserve some superoperator up to a phase that is not a depolarizing channel and thus $\mathbb{S}_{\mathbb{G}^k} \neq \mathbb{S}_{\rm {dep}}$. But if $\mathbb{S}_{\mathbb{G}^k} \neq \mathbb{S}_{\rm {dep}}$, then $\mathbb{S}_{\mathbb{C}^k} \neq \mathbb{S}_{\rm {dep}}$. Therefore, $(\mathbb{G}^k, \Omega^k)$ is not a sequence-asymptotic unitary 2-design, and hence by Lemma~\ref{lemma:alldesign},  $(\mathbb{G}, \Omega)$ is not a sequence-asymptotic unitary 2-design.
\end{proof}

Theorem~\ref{theorem:generateddesign} characterizes the class of all gate sets, and probability distributions over those gate sets, that form sequence-asymptotic unitary 2-designs. Perhaps surprisingly, a necessary and sufficient condition is that the generated group by $\mathbb{G}^k$ is a unitary $2$-design for every $k$. As an example of a consequence of this theorem, consider the gate set $\mathbb{G}=\{\sqrt{Z} H, Z \sqrt{Z} H\}$ for which both elements have a period of $3$. $\mathbb{G}$ generates a subgroup of the Clifford group that is a unitary $2$-design. However, $\mathbb{G}^{3}=\{I, X, Y, Z\}$ is the Pauli group, which does not form a 2-design. Hence, $\mathbb{G}$ does not induce a sequence-asymptotic unitary 2-design.

In any instance where a gate set $\mathbb{G}$ does generate a unitary 2-design but does not induce a sequence-asymptotic unitary 2-design (for any $\Omega$), $\mathbb{G}$ can be altered so that it can induce a sequence-asymptotic unitary 2-design by adding a single element to $\mathbb{G}$. In particular, any gate set $\mathbb{G}$ that generates a unitary 2-design and contains the identity gate is a sequence-asymptotic unitary 2-design. This is a practically reasonable and easy-to-verify condition on a gate set.

\section{A theory of Direct RB with gate-dependent errors}\label{sec:general-theory}
In this section we present two closely-related theories for direct RB under general Markovian errors that build on (and are of similar rigor to) the most accurate theories of Clifford group RB under general gate-dependent errors \cite{proctor2017randomized, wallman2017randomized, merkel2018randomized, carignan2018randomized}. These theories can be understood in terms of Fourier transforms on functions over groups (as in Ref.~\cite{merkel2018randomized}), but we chose not to explicitly use this particular mathematical framework. The theories in this section show that the direct RB decay is an exponential ($S_d \approx A + Bp^d$) for sufficiently small gate errors, and that the direct RB error rate ($r_{\Omega}$) is equal to a gauge-invariant definition for the $\Omega$-weighted average infidelity ($\epsilon_{\Omega}$) of the benchmarked gate set. The main assumptions we make in this section (see also  Table~\ref{table:theories}) are as follows:
\begin{enumerate}
    \item All errors are Markovian, i.e., a gate's error can be represented by a CPTP map on $n$ qubits.
    \item All errors are small ($\epsilon \ll 1$). 
\end{enumerate}
As we will see, the maximum value of $\epsilon$ for which this theory holds (and proves reliability of direct RB) is a subtle function of the gate set, the number of qubits, and the sampling distribution. However, for few-qubit gate sets (up to $n \sim 4$) and practically-relevant choices for the gate set and sampling distribution (e.g., the uniform distribution over a gate set consisting of $\pi/2$ rotations around the $X$ and $Y$ axes), the conditions on $\epsilon$ are physically reasonable.

This section begins with some preliminaries: in Section~\ref{ssec:gen-theory-assumptions} we introduce some notation and three additional more technical and less important assumptions that are used in this section but not elsewhere in the paper; and in Section~\ref{ssec:gauge} we review the concept of gauge, and gauge-invariant metrics. In Section~\ref{ssec:exact-theory} we introduce the first of the two closely-related theories for direct RB that we present in this section. This theory---which we refer to as the \emph{$\mathscr{R}$-matrix theory} for direct RB---combines the regular representation of the group $\mathbb{C}$ with the superoperators for the imperfect gates to construct a matrix $\mathscr{R}$ that can be used to exactly compute the direct RB decay ($S_d$).

In Sections~\ref{ssec:lmatrix-theory}-\ref{ssec:r-infidelity} we then present our second theory for direct RB with general gate-dependent errors---which we refer to as the \emph{$\mathscr{L}$-matrix theory} for direct RB---that enables us to show that $r_{\Omega} \approx \epsilon_{\Omega}$. This theory combines the superoperator representation of the group $\mathbb{C}$ with the superoperators for the imperfect gates to construct a matrix $\mathscr{L}$ that is an imperfect version of the twirling superchannel that appears in our theory of sequence-asymptotic unitary 2-designs (Section~\ref{sec:sequads}). In Section~\ref{ssec:lmatrix-theory} we show how $S_d$ can be approximated by a function of the $d^{\rm th}$ power of this $\mathscr{L}$ matrix. In Section~\ref{ssec:spectral-decomp} we then show how $S_d$ can be expanded into a linear combination of many exponentials, using the spectral decomposition of $\mathscr{L}$. In Section~\ref{ssec:approximate-exponential} we use the properties of sequence-asymptotic unitary 2-designs (Section~\ref{sec:sequads}) to show that all but two terms in this linear combination are negligible, i.e., $S_d \approx A + B\gamma^d$ where $\gamma$ is an eigenvalue of $\mathscr{L}$ that is close to 1. This implies that direct RB approximately measures $r_{\gamma} = (4^n-1)(1-\gamma)/4^n$, i.e., $r_{\Omega} \approx r_{\gamma}$. In Section~\ref{ssec:r-infidelity} we conclude our theory, by building on a proof of Wallman's~\cite{wallman2017randomized} to show that $r_{\gamma}$ is equal to a gauge-invariant definition for the $\Omega$-weighted average infidelity ($\epsilon_{\Omega}$) of the benchmarked gate set. Finally, in Section~\ref{ssec:gen-theory-sims} we use the exact $\mathscr{R}$-matrix theory for direct RB, and numerical simulations, to validate our approximate $\mathscr{L}$-matrix theory for direct RB---by demonstrating that $|(r_{\Omega}-r_{\gamma})/r_{\gamma}|$ is small for a selection of randomly generated error models on a single-qubit gate set. The theories presented in this section have a close relationship to the Fourier transforms over a group, as is the case for Clifford group RB \cite{merkel2018randomized}.

\subsection{Additional assumptions and notation}\label{ssec:gen-theory-assumptions}
In addition to the two key assumptions stated above, we will assume that:
\begin{enumerate}
\item[3.] The state preparation and measurement components of direct RB circuits contain only gates from the benchmarked gate set ($\mathbb{G}$). This is without loss of generality, as the sampling distribution $\Omega$ need not have support on all of $\mathbb{G}$. 
\item[4.] Unconditional compilation of the initial group element and final inversion group element.
\item[5.] No randomization of the output bit string.
\end{enumerate}
These assumptions mean that the state preparation circuit implements a random element of $\mathbb{C}$, and the measurement preparation subcircuit implements the unique element in $\mathbb{C}$ that inverts the preceding circuit (as in Clifford group RB). These assumptions are not necessary for much of theory in this section, and we suspect (although have not proven) that they could be discarded. However, they greatly simplify notation.

Because this section is heavy on notation, we simplify some of the notation used elsewhere in this paper. Unlike elsewhere in this paper, we will use $G \in \mathbb{G}$ and $C \in \mathbb{C}$ to represent group elements [elsewhere we use, e.g., $U(G)$] rather than instructions to implement logic operations. Throughout this paper, we denote the superoperator representing the imperfect [perfect] implementation of $G \in \mathbb{G}$ by $\tilde{\mathcal{G}}(G)$ [$\mathcal{G}(G)$]. Similarly, in this section we will denote the superoperator representing the imperfect [perfect] implementation of $C \in \mathbb{C}$ (which are used in the state preparation and measurement stages of direct RB circuits) by $\tilde{\mathcal{C}}(G)$ [$\mathcal{C}(G)$]. Note that $\mathcal{C}(\cdot)$ is a representation of the group $\mathbb{C}$. We will assume a low-error gate set, so that each $\tilde{\mathcal{G}}(G)$ is a small deviation from $\mathcal{G}(G)$ (i.e., $\tilde{\mathcal{G}}(G) \approx \mathcal{G}(G)$) for all $G \in \mathbb{G}$ and similarly $\tilde{\mathcal{C}}(C) \approx \mathcal{C}(C)$ for all $C \in \mathbb{C}$. We will quantify this assumption when used. 

\subsection{Background: gauge-invariant metrics}\label{ssec:gauge}
In this section we will show that the direct RB error rate ($r_{\Omega}$) is related to a gauge-invariant version of the average infidelity of an $n$-qubit gate sampled from $\Omega$ ($\epsilon_{\Omega}$). We will therefore first review the concept of gauge in gate sets, as well as gauge-variant and gauge-invariant properties of a gate set. The sets of all as-implemented and ideal operations in direct RB circuits can be represented by
\begin{align}
\mathbb{G}_{\text {all }}&= \left\{\sbra{x}\right\}_{x \in \mathbb{B}_n} \cup \{\mathcal{G}(G)\}_{G \in \mathbb{G}} \cup \{\mathcal{C}(C)\}_{C \in \mathbb{C}}  \cup \left\{\sket{0^{n}}\right\}, \\
\tilde{\mathbb{G}}_{\text {all }}&= \left\{\sbra{E_x}\right\}_{x \in \mathbb{B}_n} \cup  \{\tilde{\mathcal{G}}(G)\}_{G \in \mathbb{G}} \cup \{\tilde{\mathcal{C}}(C)\}_{C \in \mathbb{C}}  \cup \left\{\sket{\rho}\right\},
\end{align}
respectively, where $\mathbb{B}_n$ is the set of all length $n$ bit strings, $E_x$ is a measurement effect representing the $x$ measurement outcome (so $\sbra{E_x} \approx \sbra{x}$), and $\rho$ represents imperfect state initialization (so $\sket{\rho} \approx \sket{0^n}$). (Furthermore, due to our assumption in this section that the target bit string is always the all-zeros bit string, we only need the $x=0^n$ effect).

All ideal and actual outcomes of circuits in direct RB can be computed from $\mathbb{G}_{\textrm{all}}$ and $\tilde{\mathbb{G}}_{\textrm{all}}$, respectively. However, the predictions of these models are unchanged under a gauge transformation---all operation sets of the form
\begin{equation}\label{eq:gaugetransformations}
\tilde{\mathbb{G}}_{\text {all }}(\mathcal{M})= \left\{\sbra{E_x}\mathcal{M}^{-1}\right\}_{x \in \mathbb{B}_n} \cup \left\{\mathcal{M}\tilde{\mathcal{G}}_i\mathcal{M}^{-1}\right\} \cup \left\{\mathcal{M}\sket{\rho}\right\},
\end{equation}
make identical predictions for the outcome distributions of all circuits, where $\mathcal{M}$ is any invertible linear map, called a gauge transformation \cite{merkel2013self, blume2017demonstration}. If $f$ is a property of $\tilde{\mathbb{G}}_{\textrm{all}}$, it is only experimentally observable if 
\begin{equation}
f\left(\tilde{\mathbb{G}}_{\text {all }}\right)=f\left(\tilde{\mathbb{G}}_{\text {all }}(\mathcal{M})\right),
\end{equation}
for all $\mathcal{M}$, i.e., $f$ must be gauge-invariant. Many metrics of error for a gate or set of gates that are defined as functions of process matrices are not gauge-invariant \cite{merkel2013self, blume2017demonstration, proctor2017randomized}. This includes the [in]fidelity of a superoperator, so $\epsilon_{\Omega}$ is not gauge-invariant. This implies that an RB error rate (which is observable) cannot be equal to the standard, gauge-variant definition of the average infidelity of a gate set \cite{proctor2017randomized}, i.e., to connect RB error rates to infidelities we need a gauge-invariant definition for infidelity \cite{proctor2017randomized, wallman2017randomized, carignan2018randomized}.

\subsection{An exact theory of direct randomized benchmarking}\label{ssec:exact-theory}
Here we present a theory for direct RB that is exact, generalizing a theory for Clifford group RB presented in \cite{proctor2017randomized}. This theory, which we call the $\mathscr{R}$-matrix theory for direct RB, provides an exact formula for the average success probability ($S_d$), as a function of an imperfect gate set $\tilde{\mathbb{G}}$ and a sampling distribution $\Omega$. This formula is efficient in $d$ (but not $n$) to compute. We will find this theory useful for verifying the accuracy of the approximate theory (that shows that $r_{\Omega} \approx \epsilon_{\Omega}$) introduced later in this section. Our exact theory requires one additional assumptions: we require that $\mathbb{G}$ generates a finite group $\mathbb{C}$.

Our theory starts from the formula for $S_d$ that follows directly from the definition of $S_d$ and the direct RB circuits. Specifically,
\begin{equation}\label{eq:pm}
S_{d}= \sbra{E_0} \tilde{\mathcal{S}_d} \sket{\rho},
\end{equation}
where
\begin{multline}\label{eq:sup-sd}
\tilde{\mathcal{S}}_d=\sum_{G_d \in \mathbb{G}} \cdots \sum_{G_1 \in \mathbb{G}} \sum_{C \in \mathbb{C}}\large[\Omega(G_d)\cdots \Omega(G_1)\Pi(C) \tilde{\mathcal{C}}(C^{-1}G_1^{-1}\cdots G_d^{-1})\tilde{\mathcal{G}}(G_d)\cdots\tilde{\mathcal{G}}(G_1) \tilde{\mathcal{C}}(C) \large],
\end{multline}
where (as above) $\Pi$ is the uniform distribution. To expresses $\tilde{\mathcal{S}}_d$ (and so $S_d$) in an efficient-in-$d$ form, we use the regular representation [$R(\cdot)$] of the group $\mathbb{C}$. To specify these matrices explicitly, we index the elements of $\mathbb{C}$ by $i=1,2,\dots, |\mathbb{C}|$, with $i=1$ corresponding to the identity group element. Then $R(G)$ is the $\mathbb{C} \times \mathbb{C}$ matrix with elements given by $R(G)_{jk} = 1$ if $G C_k= C_j$ and $R(G)_{jk} = 0$ otherwise. So, $R(G)$ is a permutation matrix in which the $i^{\rm th}$ row encodes the product of $G$ and the $i^{\rm th}$ group element.
We now use the regular representation to define a matrix [$\mathscr{R}_{\mathbb{G},\tilde{\mathbb{G}}, \Omega}$] that can be used to model sequences of gates from $\tilde{\mathbb{G}}$ that are independent samples from $\Omega$. We define $\mathscr{R}_{\mathbb{G},\tilde{\mathbb{G}}, \Omega}$ by
\begin{equation}\label{eq:rmatrix}
\mathscr{R}_{\mathbb{G},\tilde{\mathbb{G}},\Omega} = \sum_{G \in \mathbb{G}} \Omega(G) R(G) \otimes \tilde{\mathcal{G}}(G).
\end{equation}

We now express $\tilde{\mathcal{S}}_d$ in terms of $\mathscr{R}$ matrices, and to do so we will need the vector 
\begin{equation}
\vec{v} = (1,0,0,\dots)^T \otimes I,
\end{equation}
where $I$ is the $4^n \times 4^n$ dimensional identity superoperator (and the first entry of $\vec{v}$ corresponds to the identity group element). We can express $\tilde{\mathcal{S}}_d$ in terms of $\mathscr{R}$ matrices and $\vec{v}$ as follows:
\begin{equation}\label{eq:rmatrix-sup}
\tilde{\mathcal{S}}_d = \vec{v}^T \mathscr{R}_{\mathbb{C},\tilde{\mathbb{C}},\Pi} \mathscr{R}_{\mathbb{G},\tilde{\mathbb{G}},\Omega}^d \mathscr{R}_{\mathbb{C},\tilde{\mathbb{C}},\Pi} \vec{v},
\end{equation}
noting that $\mathscr{R}_{\mathbb{C},\tilde{\mathbb{C}},\Pi}$ represents the application of a uniformly random element from $\tilde{\mathbb{C}}$. To understand why this equation holds, consider 
\begin{equation}
\vec{w} = \mathscr{R}_{\mathbb{C},\tilde{\mathbb{C}},\Pi}\mathscr{R}_{\mathbb{G},\tilde{\mathbb{G}},\Omega}^d \mathscr{R}_{\mathbb{C},\tilde{\mathbb{C}},\Pi} \vec{v}.
\end{equation}
This is a vector of superoperators where the $i^{\textrm{th}}$ element of $\vec{w}$ consists of an average over all sequences of superoperators consisting of (1) a uniformly random element of $\tilde{\mathbb{C}}$, (2) $d$ elements of $\tilde{\mathbb{G}}$ sampled from $\Omega$, and (3) a uniformly random element of $\tilde{\mathbb{C}}$, \emph{conditioned} on the overall ideal (error-free) action of the sequence being equal to the group element $\mathcal{C}(C_i)$. To average over all direct RB sequences of length $d$ we simply need to condition on the overall action of the sequence being the identity---which is $\vec{v}^T\vec{w}$.

To obtain our final result---a formula for $S_d$---we substitute Eq.~\eqref{eq:rmatrix-sup} into Eq.~\eqref{eq:pm} to obtain
\begin{equation}\label{eq:sd-r-matrix}
S_{d}=\sbra{E_0} \vec{v}^T \mathscr{R}_{\mathbb{C},\tilde{\mathbb{C}},\Pi} \mathscr{R}_{\mathbb{G},\tilde{\mathbb{G}},\Omega}^d \mathscr{R}_{\mathbb{C},\tilde{\mathbb{C}},\Pi} \vec{v} \,\sket{\rho}.
\end{equation}
This formula for $S_{d}$ is efficient in $d$ to calculate (simply via multiplication of $\mathscr{R}$ matrices, or via the spectral decomposition of the two $\mathscr{R}$ matrices). The $\mathscr{R}$ matrices grow quickly with both the size of the generated group (i.e., with $|\mathbb{C}|$) and $n$. However, the $\mathscr{R}$ matrices are sufficiently small when $\mathbb{C}$ is the single-qubit Clifford group (they are $96 \times 96$ matrices) to use this theory to numerically evaluate $S_{d}$ (without resorting to sampling) for any single-qubit gate set $\mathbb{G}$ that generates the Clifford group. As such, this theory is useful for numerically studying direct RB in the single-qubit setting---and this will allow us to validate our theory that shows that $r_{\Omega} \approx \epsilon_{\Omega}$ (see Section~\ref{ssec:gen-theory-sims}).

Equation~\eqref{eq:sd-r-matrix} can be rewritten in terms of the $d^{\rm th}$ power of the eigenvalues of $\mathscr{R}_{\mathbb{G},\tilde{\mathbb{G}},\Omega}$. However, this does not immediately imply that $S_d \approx A + Bp^d$ (each $\mathscr{R}$ matrix has $4^n \times |\mathbb{C}|$ eigenvalues).

\subsection{Modelling the direct RB decay with twirling superchannels}\label{ssec:lmatrix-theory}
In this and the subsequent subsections, we present our  \emph{$\mathscr{L}$-matrix theory} for direct RB. In this theory, we will show how the direct RB decay, and error rate, can be approximately expressed in terms of the following matrix
\begin{align}\label{eq:l-matrix}
\mathscr{L}_{\mathbb{G}, \tilde{\mathbb{G}}, \Omega} =\sum_{G \in \mathbb{G}} \Omega(G) \mathcal{G}(G) \otimes  \tilde{\mathcal{G}}(G).
\end{align}
Like the $\mathscr{R}$ matrix [Eq.~\eqref{eq:rmatrix}], this $\mathscr{L}$ matrix consists of summing over a tensor product of (1) a representation of the group $\mathbb{C}$ generated by $\mathbb{C}$ and (2) the as-implemented superoperators. In the case of the $\mathscr{R}$ matrix, this representation of the group was the regular representation [$R(\cdot)$], and in the case of the $\mathscr{L}$ matrix it is the standard superoperator representation [$\mathcal{G}(\cdot)$] of the group.

To derive our theory, we first show how the direct RB success probability ($S_d$) can be approximately expressed as a function of twirling superchannels $\mathscr{Q}$ whereby $\text{mat}[\mathscr{Q}] = \mathscr{L}$. In subsequent subsections, this will enable us to show that the direct RB decay is approximately exponential, and that $r_{\Omega} \approx \epsilon_{\Omega}$. Our theory will use \emph{error maps} for the as-implemented group elements $\tilde{\mathbb{C}}$. Let
\begin{equation}
\Lambda(C)= \mathcal{C}(C)^{\dagger}\tilde{\mathcal{C}}(C),
\end{equation}
for $C \in \mathbb{C}$, which is the ``pre-gate'' error map for the superoperator $\tilde{\mathcal{C}}(C)$. 
We also use the average error map ($\bar{\Lambda}$), defined by
\begin{equation}
\bar{\Lambda} = \sum_{C \in \mathbb{C}} \Pi(C) \Lambda(C).
\end{equation}

The first step to deriving our $\mathscr{L}$-matrix theory for direct RB is to approximate the imperfect initial group element using a gate-independent error channel. To do so, we first note that Eq.~\eqref{eq:sup-sd} can be rewritten as
\begin{multline}\label{eq:sup-sd-init}
\tilde{\mathcal{S}}_d=\sum_{C \in \mathbb{C}}\sum_{G_d \in \mathbb{G}} \cdots \sum_{G_1 \in \mathbb{G}} \large[\Pi(C) \Omega(G_d)\cdots \Omega(G_1) \tilde{\mathcal{C}}(C)\tilde{\mathcal{G}}(G_d)\cdots\tilde{\mathcal{G}}(G_1) \tilde{\mathcal{C}}(\{CG_d \cdots G_1\}^{-1}) \large],
\end{multline}
i.e., we have rewritten the first group element in terms of the $G_i$ and the inversion group element (which is now denoted by $C$). Now we have that
\begin{equation}
\tilde{\mathcal{C}}(\{CG_d \cdots G_1\}^{-1}) =
  \mathcal{G}^{\dagger}(G_1)\cdots \mathcal{G}^{\dagger}(G_d)  \mathcal{C}^{\dagger}(C)\Lambda(\{CG_d \cdots G_1\}^{-1}),
\end{equation}
and by substituting this into Eq.~\eqref{eq:sup-sd-init}, we obtain
\begin{multline}
\tilde{\mathcal{S}}_d=\sum_{C \in \mathbb{C}}\sum_{G_d \in \mathbb{G}} \cdots \sum_{G_1 \in \mathbb{G}} \large[\Pi(C) \Omega(G_d)\cdots \Omega(G_1)\tilde{\mathcal{C}}(C)\tilde{\mathcal{G}}(G_d)\cdots\tilde{\mathcal{G}}(G_1)  \,\,\,\,\times \\ \mathcal{G}^{\dagger}(G_1)\cdots  \mathcal{G}^{\dagger}(G_d) \mathcal{C}^{\dagger}(C) \Lambda((\{G_d\cdots G_1\}^{-1}) \large].
\end{multline}
By replacing the error maps in this equation (i.e., the error maps for each initial group element) with their average, we obtain
\begin{multline}
\tilde{\mathcal{S}}_d=\sum_{C \in \mathbb{C}}\sum_{G_d \in \mathbb{G}} \cdots \sum_{G_1 \in \mathbb{G}} \large[\Pi(C) \Omega(G_d)\cdots \Omega(G_1)\tilde{\mathcal{C}}(C)\tilde{\mathcal{G}}(G_d)\cdots\tilde{\mathcal{G}}(G_1)  \,\,\,\,\times \\ \mathcal{G}^{\dagger}(G_1)\cdots  \mathcal{G}^{\dagger}(G_d) \mathcal{C}^{\dagger}(C) \large]\bar{\Lambda} + \Delta_d, \label{eq:sd-mean-approx}
\end{multline}
where $\Delta_d$ absorbs the approximation error. This approximation breaks the dependencies, enabling independent averaging, as we can rearrange Eq.~\eqref{eq:sd-mean-approx} to
\begin{multline}
\tilde{\mathcal{S}}_d= \Bigg\{\sum_{C \in \mathbb{C}}\Pi(C) \tilde{\mathcal{C}}(C) \bigg\{\sum_{G_d \in \mathbb{G}}\Omega(G_d) \tilde{\mathcal{G}}(G_d)\cdots \\ \Big\{\sum_{G_1 \in \mathbb{G}}\Omega(G_1) \tilde{\mathcal{G}}(G_1)\mathcal{G}^{\dagger}(G_1) \Big\}  \cdots \mathcal{G}^{\dagger}(G_d)\bigg\} \mathcal{C}^{\dagger}(C) \Bigg\} \bar{\Lambda}  + \Delta_d.
\end{multline}
This can be more concisely expressed using two superchannels:
\begin{equation}\label{eq:sd-sup-lmatrix}
\tilde{\mathcal{S}}_d =  \left(\mathscr{Q}_{\mathbb{C}, \tilde{\mathbb{C}}, \Pi} \circ \mathscr{Q}_{\mathbb{G}, \tilde{\mathbb{G}}, \Omega}^{d}\right)[I] \bar{\Lambda} + \Delta_d,
\end{equation}
where $I$ is the identity superoperator, and $\mathscr{Q}_{\mathbb{G}, \tilde{\mathbb{G}}, \Omega}$ and $\mathscr{Q}_{\mathbb{C}, \tilde{\mathbb{C}}, \Pi}$ are imperfect twirling superchannels, defined by
\begin{align}
\mathscr{Q}_{\mathbb{G}, \tilde{\mathbb{G}}, \Omega}[\mathcal{E}] &=\sum_{G \in \mathbb{G}} \Omega(G) \tilde{\mathcal{G}}(G)  \mathcal{E} \mathcal{G}^{\dagger}(G),\\
\mathscr{Q}_{\mathbb{C}, \tilde{\mathbb{C}}, \Pi}[\mathcal{E}] &=\sum_{C \in \mathbb{C}} \Pi(C)\tilde{\mathcal{C}}(C)  \mathcal{E} \mathcal{C}^{\dagger}(C).
\end{align}
By substituting Eq.~\eqref{eq:sd-sup-lmatrix} into Eq.~\eqref{eq:pm} we obtain
\begin{equation}\label{eq:sd-qmatrices}
S_{d}=\sbra{E_0}\left(\mathscr{Q}_{\mathbb{C}, \tilde{\mathbb{C}}, \Pi} \circ \mathscr{Q}_{\mathbb{G}, \tilde{\mathbb{G}}, \Omega}^{d}\right)[I]\sket{\rho'}+\delta_{d},
\end{equation}
where $\sket{\rho^{\prime}}=\bar{\Lambda}\sket{\rho} $ and 
\begin{equation}
\delta_d= \sbra{E_0}\Delta_d \sket{\rho}.
\end{equation}
Equation~\eqref{eq:sd-qmatrices} represents the average direct RB success probability in terms of imperfect twirling superchannels, which builds on similar results for Clifford group RB \cite{proctor2017randomized, wallman2017randomized, chasseur2015complete}. 

Later we will apply the theory of sequence-asymptotic unitary 2-designs (Section~\ref{sec:sequads}) to understand $S_d$, and to do so we will find it useful to replace the $\mathscr{Q}$ superchannels in Eq.~\eqref{eq:sd-qmatrices} with the closely related $\mathscr{L}$ matrices [see Eq.~\eqref{eq:l-matrix}]. Consider
\begin{equation}
\textrm{vec}(\mathscr{Q}_{\mathbb{C}, \tilde{\mathbb{C}}, \Pi} \circ \mathscr{Q}_{\mathbb{G}, \tilde{\mathbb{G}}, \Omega}^{d}[I]) = \textrm{mat}(\mathscr{Q}_{\mathbb{C}, \tilde{\mathbb{C}}, \Pi}) \textrm{mat}(\mathscr{Q}_{\mathbb{G}, \tilde{\mathbb{G}}, \Omega})^{d}|I),
\end{equation}
where, as introduced in Section~\ref{sec:sequads}, $\textrm{vec}[\cdot]$ denotes  the ``stacking'' map, $\textrm{mat}[\cdot]$ turns a superchannel into a matrix and $\sdket{\cdot}$ turns a superoperator into a vector. So, e.g., we find that
\begin{align}
\textrm{mat}[\mathscr{Q}_{\mathbb{G}, \tilde{\mathbb{G}}, \Omega}]&=\sum_{G \in \mathbb{G}} \Omega(G) \mathcal{G}(G) \otimes \tilde{\mathcal{G}}(G) = \mathscr{L}_{\mathbb{G}, \tilde{\mathbb{G}}, \Omega},
\end{align}
where $\mathscr{L}_{\mathbb{G}, \tilde{\mathbb{G}}, \Omega}$ is the matrix introduced in Eq.~\eqref{eq:l-matrix} [here we have assumed a Hermitian basis for superoperators, so that $\mathcal{G}(G)$'s element are real numbers].
We can therefore rewrite Eq.~\eqref{eq:sd-qmatrices} in terms of two $\mathscr{L}$ matrices:
\begin{equation}\label{eq:sd-lmatrices}
S_{d}=\sbra{E_0}\textrm{unvec}\left[ \mathscr{L}_{\mathbb{C}, \tilde{\mathbb{C}}, \Pi}\mathscr{L}_{\mathbb{G}, \tilde{\mathbb{G}}, \Omega}^{d}\sdket{I} \right]\sket{\rho'}+\delta_{d}.
\end{equation}

Equation~\eqref{eq:sd-lmatrices} contains an approximation error term ($\delta_{d}$) and we now bound this term. Using basic properties of CPTP maps and the diamond norm it may be shown that
\begin{equation}
\left|\delta_{d}\right| \leq \left\|\Delta_d\right\|_{\diamond}  \leq \sum_{C \in \mathbb{C}} \Pi(C)\left\|\bar{\Lambda}-\Lambda(C)\right\|_{\diamond} \equiv \Delta,
\end{equation}
where $\|\cdot\|_{\diamond}$ is the diamond norm.\footnote{An explicit derivation that immediately implies this relation is given in \cite{proctor2017randomized} in the context of an equivalent relation in Clifford group RB (see the supplemental material in \cite{proctor2017randomized}, and also see \cite{magesan2012characterizing, chasseur2015complete} for similar work). Note that this upper-bound is $d$ independent, unlike the similar bound for approximating Clifford group RB decay curves in \cite{magesan2012characterizing} that grow with $d$ and which is therefore not able to provide useful guarantees on $S_d$ \cite{proctor2017randomized, wallman2017randomized}.} Error maps are not gauge-invariant, so the size of $\delta_{d}$ (and the size of our upper-bound on $\left|\delta_{d}\right|$) depends on the representation of $\tilde{\mathbb{G}}_{\rm all}$. Even for low-error gates---here meaning that every $\Lambda(C)$ is close to the identity in some representation---$\Delta$ and $\delta_{d}$ can always be made large by choosing a ``bad'' gauge in which to express $\tilde{\mathbb{G}}_{\rm all}$. But our bound holds for any CPTP representation, and so $\left|\delta_{d}\right| \leq \Delta_{\min }$ with $\Delta_{\min }$ the minimum of $\Delta$ over all CPTP representations of $\tilde{\mathbb{G}}_{\rm all}$. In the following, we will assume we are using a representation in which $\delta_{d}$ is small, which always exists with sufficiently low-error gates. Finally, we note that it is possible to improve our bound on $\delta_{d}$, using an analysis similar to Wallman's \cite{wallman2017randomized} treatment of Clifford group RB, which bounds a quantity that is closely related to $\delta_d$ from above by a function that decays exponentially in $d$.

\subsection{Modelling the direct RB decay using the spectrum of twirling superchannels}\label{ssec:spectral-decomp}
We have shown that the direct RB average success probability ($S_d$) can be approximated by a function of the $d^{\textrm{th}}$ power of $\mathscr{L}_{\mathbb{G}, \tilde{\mathbb{G}}, \Omega}$ [see Eq.~\eqref{eq:sd-lmatrices}]. This implies that $S_d$ can be approximated using the spectral decomposition of $\mathscr{L}_{\mathbb{G}, \tilde{\mathbb{G}}, \Omega}$. We now derive the functional form of this spectral decomposition, which in the subsequent subsection we will use to show that $S_d \approx A + Bp^d$.
Let $\tilde{\gamma}_{i}$ for $i=1,2, \ldots, 16^{n}$ denote the eigenvalues of $\mathscr{L}_{\mathbb{G}, \tilde{\mathbb{G}}, \Omega}$ ordered by descending absolute value, i.e., $\left|\tilde{\gamma}_{i}\right| \geq\left|\tilde{\gamma}_{i+1}\right|$ for all $i$. Then Eq.~\eqref{eq:sd-lmatrices} implies that $S_{d}$ may be written as
\begin{equation} \label{eq:sd-spectral}
S_{d}=\tilde{\omega}_{1} \tilde{\gamma}_{1}^{d}+\tilde{\omega}_{2} \tilde{\gamma}_{2}^{d}+\cdots+\tilde{\omega}_{16^{n}} \tilde{\gamma}_{16^{n}}^{d}+\delta_{d},
\end{equation}
where the $\tilde{\omega}_{k}$ are $d$-independent parameters. Note that the eigenvalues of an $\mathscr{L}$ matrix are gauge-invariant---because gauge transformations act on a $\mathscr{L}$ matrix by matrix conjugation, with a matrix of the form $I \otimes \mathcal{M}$---so the $\tilde{\gamma}_i$ are gauge-invariant.

We now derive a formula for $\tilde{\omega}_k$. It is convenient to rewrite Eq.~\eqref{eq:sd-lmatrices} as 
\begin{equation}\label{eq:sd-supsd-dash}
S_{d}=\sbra{E_0}\tilde{S}_d'\sket{\rho^{\prime}}+\delta_{d},
\end{equation}
where
\begin{equation}\label{eq:supsd-dash}
 \tilde{\mathcal{S}}_d' =\textrm{unvec}\left[\mathscr{L}_{\mathbb{C}, \tilde{\mathbb{C}}, \Pi}\mathscr{L}_{\mathbb{G}, \tilde{\mathbb{G}}, \Omega}^{d}\sdket{I} \right].
 \end{equation}
We now write $ \tilde{\mathcal{S}}_d'$ in terms of the eigenvectors and eigenvalues of the two $\mathscr{L}$ matrices, $\mathscr{L}_{\mathbb{G}, \tilde{\mathbb{G}}, \Omega}$  and $\mathscr{L}_{\mathbb{C}, \tilde{\mathbb{C}}, \Pi}$.
Let $\tilde{\eta}_{i}$ for $i=1,2, \ldots, 16^{n}$ denote the eigenvalues of $\mathscr{L}_{\mathbb{C}, \tilde{\mathbb{C}}, \Pi}$ ordered by descending absolute value, i.e., $\left|\tilde{\eta}_{i}\right| \geq\left|\tilde{\eta}_{i+1}\right|$ for all $i$ (noting that $\tilde{\gamma}_i$ denote the ordered eigenvalues of $\mathscr{L}_{\mathbb{G}\tilde{\mathbb{G},\Omega}}$). Moreover, let $\mathcal{L}_{\mathbb{C}, \tilde{\mathbb{C}}, i}$ and $\mathcal{R}_{\mathbb{C}, \tilde{\mathbb{C}}, i}$ ($\mathcal{L}_{\mathbb{G}, \tilde{\mathbb{G}}, i}$ and $\mathcal{R}_{\mathbb{G}, \tilde{\mathbb{G}}, i}$) be superoperators that, when stacked, are mutually orthonormal left and right eigenvectors of the matrix $\mathscr{L}_{\mathbb{C},\tilde{\mathbb{C}}, \Pi}$ (the matrix $\mathscr{L}_{\mathbb{G},\tilde{\mathbb{G}}, \Omega}$ ) with eigenvalue $\tilde{\eta}_{i}$ (eigenvalue $\tilde{\gamma}_{i}$). That is
\begin{align}
\mathscr{L}_{\mathbb{G},\tilde{\mathbb{G}}, \Omega} \sdket{\mathcal{R}_{\mathbb{G}, \tilde{\mathbb{G}}, i}} &= \tilde{\gamma}_{i} \sdket{\mathcal{R}_{\mathbb{G}, \tilde{\mathbb{G}}, i}}, \\
\sdbra{\mathcal{L}_{\mathbb{G},\tilde{\mathbb{G}}, i}} \mathscr{L}_{\mathbb{G}, \tilde{\mathbb{G}}, \Omega} &= \sdbra{\mathcal{L}_{\mathbb{G}, \tilde{\mathbb{G}}, i}}\tilde{\gamma}_{i}, \\
\mathscr{L}_{\mathbb{C}, \tilde{\mathbb{C}}, \Pi} \sdket{ \mathcal{R}_{\mathbb{C}, \tilde{\mathbb{C}}, i}} &= \tilde{\eta}_{i} \sdket{\mathcal{R}_{\mathbb{C}, \tilde{\mathbb{C}}, i}}, \\ 
\sdbra{\mathcal{L}_{\mathbb{C},\tilde{\mathbb{C}}, i}} \mathscr{L}_{\mathbb{C}, \tilde{\mathbb{C}}, \Pi} &= \sdbra{\mathcal{L}_{\mathbb{C}, \tilde{\mathbb{C}}, i}} \tilde{\eta}_{i},
\end{align}
with
\begin{equation}
\left(\mathcal{L}_{\mathbb{G}, \tilde{\mathbb{G}}, i} \mid \mathcal{R}_{\mathbb{G}, \tilde{\mathbb{G}}, j}\right)=\delta_{ij}, \quad\left(\mathcal{L}_{\mathbb{C}, \tilde{\mathbb{C}}, i} \mid \mathcal{R}_{\mathbb{C}, \tilde{\mathbb{C}}, j}\right)=\delta_{ij} .
\end{equation}
Writing the two $\mathscr{L}$ matrices in terms of their eigenvalues and their left and right eigenvectors, and substituting this decomposition of the $\mathscr{L}$ matrices into Eq.~\eqref{eq:supsd-dash}, we obtain
\begin{align}\label{eq:supds-dash-spectral}
 \tilde{\mathcal{S}}_d' = \sum_{jk}\tilde{\eta}_j\tilde{\gamma}_k^d \textrm{unvec}\left[ |\mathcal{R}_{\mathbb{C},\tilde{\mathbb{C}}, j})(\mathcal{L}_{\tilde{\mathbb{C}},\mathbb{C},j}|\mathcal{R}_{\mathbb{G},\tilde{\mathbb{G}}, k})(\mathcal{L}_{\mathbb{G},\tilde{\mathbb{G}},k}|  \sdket{I}\right].
\end{align}
By substituting Eq.~\eqref{eq:supds-dash-spectral} into Eq.~\eqref{eq:sd-supsd-dash} and comparing the resultant equation to Eq.~\eqref{eq:sd-spectral}, it follows that
\begin{equation}\label{eq:tilde-wk}
\tilde{\omega}_{k}=\sbra{E'_0} \mathcal{S}_{d,k}'\sket{\rho}
\end{equation}
where
\begin{equation}\label{eq:sup-sdk-dash}
   \mathcal{S}_{d,k}' =  \sum_{j} \tilde{\eta}_{j}\textrm{ unvec }\left[|\mathcal{R}_{\mathbb{C}, \tilde{\mathbb{C}}, j})(\mathcal{L}_{\mathbb{C}, \tilde{\mathbb{C}}, j} | \mathcal{R}_{\mathbb{G}, \tilde{\mathbb{G}}, k})(\mathcal{L}_{\mathbb{G}, \tilde{\mathbb{G}}, k} |\sdket{I}\right].
\end{equation}
This formula shows that we can quantify $\tilde{\omega}_k$ by studying the overlap in the eigenchannels of the two $\mathscr{L}$ matrices, as well as the spectrum of $\mathscr{L}_{\mathbb{C},\tilde{\mathbb{C}},\Pi}$.

\subsection{The direct RB decay is approximately exponential}\label{ssec:approximate-exponential}
We have now shown that the average success probability ($S_d)$ in direct RB can be written as a linear combination of the $d^{\textrm{th}}$ powers of the $16^n$ eigenvalues of the $\mathscr{L}$ matrix ($\tilde{\gamma}_k$), and we have derived formula for the coefficients in this sum ($\tilde{\omega}_k$). We now use the theory of sequence-asymptotic unitary 2-designs (Section~\ref{sec:sequads}), as well as the theory of ordinary unitary 2-designs, to show that $S_d \approx A + B\gamma^d$ where $\gamma \equiv \tilde{\gamma}_2$ is the second largest eigenvalue of $\mathscr{L}_{\mathbb{G},\tilde{\mathbb{G}},\Omega}$. Our theory relies on two properties of $\mathbb{G}$, $\Omega$, and $\tilde{\mathbb{G}}$: (1) $(\mathbb{G},
\Omega)$ is a sequence-asymptotic unitary 2-design, and (2) the gates are low error, i.e., $\tilde{\mathbb{G}}$ is close to $\mathbb{G}$ and $\tilde{\mathbb{C}}$ is close to $\mathbb{C}$. 

This second condition, above, implies that both the $\mathscr{L}$ matrices that appear in Eq.~\eqref{eq:supsd-dash} [$\mathscr{L}_{\mathbb{G},\tilde{\mathbb{G}},\Omega}$ and $\mathscr{L}_{\mathbb{C},\tilde{\mathbb{C}}, \Pi}$] will be small perturbations on the $\mathscr{L}$ matrices for perfect, error-free gate sets [i.e., $\mathscr{L}_{\mathbb{G},\mathbb{G},\Omega}$ and $\mathscr{L}_{\mathbb{C},\mathbb{C}, \Pi}$]. The group generated by $\mathbb{G}$, i.e, $\mathbb{C}$, is a unitary 2-design. Therefore
\begin{equation}
\mathscr{L}_{\mathbb{C},\mathbb{C} ,\Pi}=\sum_{C \in \mathbb{C}}  \Pi(C) \mathcal{C}(C) \otimes \mathcal{C}(C),
\end{equation}
is a projection onto the space spanned by a completely depolarizing channel [$\mathcal{D}_0$] and the identity superoperator [$I$] (see Section~\ref{sec:sequads}). Letting $\eta_i$ denote the ordered eigenvalues of $\mathscr{L}_{\mathbb{C},\mathbb{C} ,\Pi}$, we have that (1) $\eta_1=\eta_2=1$, with the two-dimensional eigen-space for this eigenvalue spanned by $\mathcal{D}_{0}$ and $I$, and (2) $\eta_i=0$ for $i \geq 3$. Similarly, Propopsition~\ref{prop:seqasymreq} implies that, because $(\mathbb{G},\Omega)$ is a sequence-asymptotic unitary 2-design,
\begin{equation}
\mathscr{L}_{\mathbb{G},\mathbb{G} ,\Omega}=\sum_{G \in \mathbb{G}}  \Omega(G) \mathcal{G}(G) \otimes \mathcal{G}(G),
\end{equation}
has two unit eigenvalues, which also correspond to the two-dimensional depolarizing subspace, and all other eigenvalues have absolute value strictly less than one. So, letting $\gamma_i$ denote the ordered eigenvalues of $\mathscr{L}_{\mathbb{G},\mathbb{G} ,\Omega}$, we have that (1) $\gamma_i$ $\gamma_1=\gamma_2=1$, with the two-dimensional eigen-space for this eigenvalue spanned by $\mathcal{D}_{0}$ and $I$, and (2) $|\gamma_i| \leq 1$ for all $i \geq 3$. Using Eqs.~\eqref{eq:tilde-wk}-\eqref{eq:sup-sdk-dash}, this then implies that if $\tilde{\mathbb{G}}=\mathbb{G}$ and $\tilde{\mathbb{C}}=\mathbb{C}$, i.e., the gates are perfect, then  $\tilde{\omega}_k=0$ for $k \geq 3$.

\begin{figure}
    \centering
    \includegraphics[scale=.7]{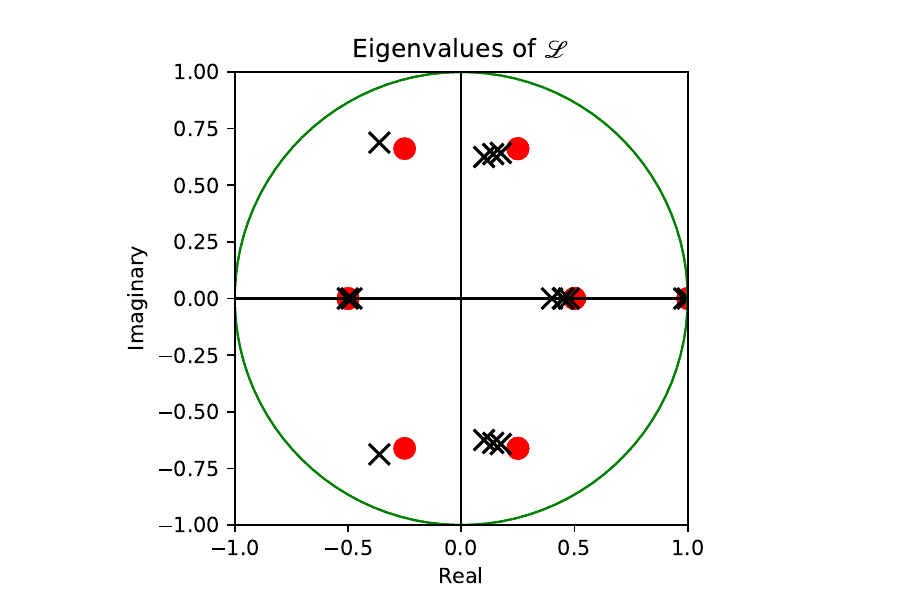}
    \caption{The eigenvalues of the $\mathscr{L}_{\tilde{\mathbb{G}},\mathbb{G},\Omega}$ matrix for the single qubit gate set $\mathbb{G}=\{X(\nicefrac{\pi}{2}), Y(\nicefrac{\pi}{2})\}$ without errors [red circles], i.e., $\tilde{\mathbb{G}} =\mathbb{G}$, and with a small coherent over-rotation error on each gate [black crosses], i.e. $\tilde{\mathbb{G}} = \{X(\nicefrac{\pi}{2}+\epsilon), Y(\nicefrac{\pi}{2}+\epsilon)\}$, with $\epsilon=0.1$. There are two unit eigenvalues for the error-free gate set, and all other eigenvalues have absolutely value strictly less than 1. Here the upper spectral gap---meaning the difference in the magnitudes of the second and third largest eigenvalues of $\mathscr{L}$---is \nicefrac{1}{2} for the error-free gate set, and it is still close to $\nicefrac{1}{2}$ for the imperfect gates.}
    \label{fig:l-argand}
\end{figure}

Consider now the case of imperfect but low-error gates, i.e., $\tilde{\mathbb{G}} \approx \mathbb{G}$ and $\tilde{\mathbb{C}} \approx \mathbb{C}$. As long as the errors are small, the eigenvalues of $\mathscr{L}_{\mathbb{G},\mathbb{G} ,\Omega}$ will be close to those of $\mathscr{L}_{\mathbb{G},\tilde{\mathbb{G}},\Omega}$. This is demonstrated with an example in Fig.~\ref{fig:l-argand}. Furthermore, as long as the errors are small enough so that the gap between the 2\textsuperscript{nd} largest eigenvalue and the smaller eigenvalues of $\mathscr{L}_{\mathbb{G},\mathbb{G} ,\Omega}$ does not close, the span of the eigenvectors of $\tilde{\gamma}_1$ and $\tilde{\gamma}_2$ is a small perturbation on the depolarizing subspace. Similarly, as long as the errors are small enough so that the gap between the 2\textsuperscript{nd} largest eigenvalue and the zero eigenvalues of $\mathscr{L}_{\mathbb{C},\mathbb{C} ,\Pi}$ does not close, the span of the eigenvectors of $\tilde{\eta}_1$ and $\tilde{\eta}_2$ is a small perturbation on the depolarizing subspace. 
This then implies that 
\begin{equation}
|\tilde{\omega}_k| \ll 1 \quad \forall k \geq 3.
\end{equation}
Substituting this into Eq.~\eqref{eq:sd-spectral}, and by noting that 1 is always an eigenvalue of an $\mathscr{L}$ matrix for trace-preserving superoperators (we are assuming CPTP superoperators throughout), we find that 
\begin{equation}
S_{d} = A+B \gamma^{d} + \delta_d + \delta_d'
\end{equation}
where $A=\omega_1$, $B=\omega_2$, $\gamma=\tilde{\gamma}_2$ and $\delta_d' = \sum_{k\geq 3}^{16^n} \tilde{\gamma}_k^d \omega_k$ consists of a sum of $16^n-2$ small terms. Therefore, for small $n$, we have that
\begin{equation}
S_{d} \approx A+B \gamma^{d},
\end{equation}
i.e., the direct RB average success probability decay is approximately an exponential (plus a constant).

Interestingly, note that the above theory does not show that $|\delta_d'|$ is small for $n\gg 1$. This is because we have only shown that $\delta_d'$ is the sum of $16^n-1$ small terms. Furthermore, the above theory has limited applicability in the $n\gg 1$ regime for another reason: for any $\mathbb{G}$ consisting of gates constructed from parallel one- and two-qubit gates, the upper spectral gap in the $\mathscr{L}_{\mathbb{G},\mathbb{G} ,\Omega}$ matrix must decrease in size as $n$ increases. This is because the spectral gap is closely related to convergence to a unitary 2-design, and the number of layers of one- and two-qubit gates needed to implement a good approximation to unitary 2-design increases with $n$. Yet our less formal theory for direct RB---presented in Section~\ref{sec:stochastic-theory}---shows that direct RB \emph{is} reliable in the $n \gg 1$ setting, i.e., we do have $S_d \approx A + Bp^d$ for some $A$, $B$ and $p$, even when $n \gg 1$. It is an interesting open question as to how to obtain this result from the $\mathscr{R}$ or $\mathscr{L}$ matrix theories presented in this section.

\subsection{The direct RB measures a version of infidelity}\label{ssec:r-infidelity}
We have now shown that the direct RB success probability ($S_d$) is approximately given by $S_d \approx A + B\gamma^d$ where $\gamma$ is the second largest eigenvalue of the superchannel $\mathscr{L}_{\mathbb{G}, \tilde{\mathbb{G}}, \Omega}$. The direct RB error rate ($r_{\Omega}$) is defined to be $r_{\Omega} = (4^n-1)(1-p)/4^n$, where $p$ is obtained by fitting $S_d$ to $S_d = A +Bp^d$. The theory presented so far therefore implies that $r_{\Omega} \approx r_{\gamma}$ where 
\begin{equation}
r_{\gamma} = (4^n-1)(1-\gamma)/4^n.
\end{equation}
The final part of our theory is to show that $r_{\gamma}$ is (exactly) equal to a gauge-invariant version of the mean infidelity of a gate sampled from $\Omega$ ($\epsilon_{\Omega}$). That is, we now show that direct RB measures the $\Omega$-weighted average infidelity of the gate set $\tilde{\mathbb{G}}$ it is used to benchmark, in the same sense that Clifford group RB measures the mean infidelity of an implementation of the Clifford group  \cite{wallman2017randomized, proctor2017randomized}.

First we introduce a term for a gate set $\tilde{\mathbb{G}}$ that has sufficiently small errors for the upper spectral gap in the $\mathscr{L}_{\mathbb{G}, \tilde{\mathbb{G}}, \Omega}$ to not close (which is a condition that we required, above, to show that $S_d \approx A + B\gamma^d$).
\begin{definition}\label{def:gap}
Let $(\mathbb{G}, \Omega)$ be a sequence-asymptotic unitary 2-design, and let $\tilde{\mathbb{G}}$ represent some implementation of $\mathbb{G}$. This implementation is $(\mathbb{G}, \Omega)$-benchmarking compatible if the upper spectral gap of $\mathscr{L}_{\mathbb{G}, \tilde{\mathbb{G}}, \Omega}$ does not close.
\end{definition}

\begin{figure}[t!]
    \centering
    \includegraphics[scale=.6]{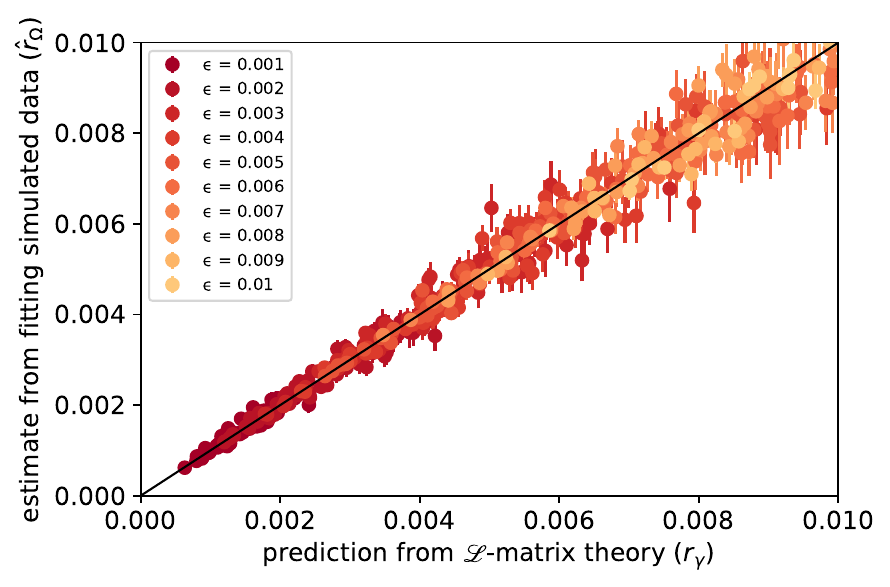}
    \caption{Verifying the accuracy of our $\mathscr{L}$ matrix theory for direct RB, using numerical simulations. We show the theory's prediction for the direct RB error rate ($r_{\gamma}$) versus an estimate $\hat{r}_{\Omega}$ of the true direct RB error rate ($r_{\Omega}$) obtained by simulating direct RB. Each data point shows $r_{\gamma}$ versus $\hat{r}_{\Omega}$ for a different randomly generated Markovian error models (see main text for error model details). We find that $r_{\gamma}$ and $\hat{r}_{\Omega}$ are consistent, but note that small systematic deviations between $r_{\gamma}$ and $r_{\Omega}$ would not be visible due to the finite sample error on $\hat{r}_{\Omega}$.}
    \label{fig:markovianscatter}
\end{figure}

Next, we present a rather technical proposition, that generalizes a result derived by Wallman \cite{wallman2017randomized} (see Theorem 2 therein) for Clifford group RB:
\begin{proposition}\label{prop:lmatdep}
Let $(\mathbb{G}, \Omega)$ be a sequence-asymptotic unitary 2-design, and let $\tilde{\mathbb{G}}$ represent an implementation of $\mathbb{G}$ that is sufficiently low error to be $(\mathbb{G}, \Omega)$-benchmarking compatible. Let $\gamma$ be the second largest eigenvalue of $\mathscr{L}_{\mathbb{G}, \tilde{\mathbb{G}}, \Omega}$ by absolute value, and let $\mathcal{E}_{1}$ and $\mathcal{E}_{\gamma}$ be eigenoperators of $\mathscr{L}_{\mathbb{G}, \tilde{\mathbb{G}}, \Omega}$ with eigenvalues 1 and $\gamma$, respectively. Then $\mathcal{E}=\mathcal{E}_{1}+\mathcal{E}_{\gamma}$ satisfies
\begin{equation}
\mathscr{L}_{\mathbb{G}, \tilde{\mathbb{G}}, \Omega}(\mathcal{E})=\mathcal{D}_{\gamma} \mathcal{E},
\end{equation}
where $\mathcal{D}_{\gamma}$ is the depolarization channel with decay constant $\gamma$.
\end{proposition} 
\begin{proof}
    See Appendix~\ref{app:proof}.
\end{proof}
We now use Proposition~\ref{prop:lmatdep} to prove that $r_{\gamma}=\epsilon_{\Omega}(\tilde{\mathbb{G}}(\mathcal{E}), \mathbb{G})$. Here 
\begin{equation}\label{eq:gi-inf}
    \epsilon_{\Omega}(\tilde{\mathbb{G}}(\mathcal{E}), \mathbb{G}) =  \sum_{G \in \mathbb{G}} \Omega(G)\epsilon(\mathcal{E}\tilde{\mathcal{G}}(G)\mathcal{E}^{-1}, \mathcal{G}(G)),
\end{equation}
where $\mathcal{E}$ is as defined in Proposition~\ref{prop:lmatdep}. Equation~\eqref{eq:gi-inf} is the $\Omega$-weighted average infidelity between the superoperators in $\mathbb{G}$ and $\tilde{\mathbb{G}}(\mathcal{E})$, where $\tilde{\mathbb{G}}(\mathcal{E})$ expresses the imperfect operations in a particular gauge:\footnote{Note that the $\mathcal{E}$ transformation is also applied to the SPAM whenever any probabilities are to be calculated, but for simplicity we have dropped the SPAM operations from the notation throughout this subsection.}
\begin{equation}
\tilde{\mathbb{G}}(\mathcal{E}):=\left\{\mathcal{E} \tilde{\mathcal{G}}(G) \mathcal{E}^{-1} \mid G \in \mathbb{G} \right\}.
\end{equation}
Note that the superoperators in $\tilde{\mathbb{G}}(\mathcal{E})$ are not necessarily completely positive (this follows because it is true in the case of Clifford group RB \cite{proctor2017randomized, wallman2017randomized}, and Clifford group RB is a special case of direct RB). Furthermore, note that $\tilde{\mathbb{G}}(\mathcal{E})$ is ill-defined if $\mathcal{E}$ is singular, and in any such case we take $\tilde{\mathbb{G}}(\mathcal{E})$ to instead be defined using a small perturbation on $\mathcal{E}$. In the following proposition, which is a generalization of a result for Clifford group RB shown in \cite{proctor2017randomized, wallman2017randomized}, we prove that $r_{\gamma}=\epsilon_{\Omega}(\tilde{\mathbb{G}}(\mathcal{E}), \mathbb{G})$.

\begin{proposition}\label{prop:rgamma}
Let $\mathbb{G}, \Omega, \tilde{\mathbb{G}}, \gamma$ and $\mathcal{E}$ be as defined in Proposition~\ref{prop:lmatdep}. Then $r_{\gamma}=(d^2-1)(1-\gamma) / d^2$ satisfies
\begin{equation}
r_{\gamma}=\epsilon_{\Omega}(\tilde{\mathbb{G}}(\mathcal{E}), \mathbb{G}).
\end{equation}
\end{proposition} 
\begin{proof} From Proposition~\ref{prop:lmatdep} we have that $\mathscr{L}_{\mathbb{G}, \tilde{\mathbb{G}}, \Omega}(\mathcal{E})=\mathcal{D}_{\gamma} \mathcal{E}$, and so $\mathscr{L}_{\mathbb{G}, \tilde{\mathbb{G}}, \Omega}(\mathcal{E}) \mathcal{E}^{-1}=\mathcal{D}_{\gamma}$. Taking the entanglement infidelity to the identity superoperator of both sides of this equality, and using the definition of $\mathscr{L}_{\mathbb{G}, \tilde{\mathbb{G}}, \Omega}$, we have that
\begin{equation}\label{eq:deptoid}
\epsilon\left(\sum_{G \in \mathbb{G}} \Omega(G) \mathcal{G}^{-1}(G) \mathcal{E} \tilde{\mathcal{G}}(G) \mathcal{E}^{-1}, \mathbb{1}\right)=\epsilon(\mathcal{D}_{\gamma}, \mathbb{1})
\end{equation}
Now using the well-known relation \cite{magesan2012characterizing}
\begin{equation}
\epsilon(\mathcal{D}_{\lambda}, \mathbb{1})=\frac{(4^n-1)(1-\lambda)}{4^n}
\end{equation}
and the definition of $r_{\gamma}$, it follows that $\epsilon(\mathcal{D}_{\gamma}, \mathbb{1})=r_{\gamma}$. Using this equality, the linearity of $\epsilon$, and the easily verified relation $\epsilon(B^{-1} A, \mathbb{1})=\epsilon(A, B)$ for any superoperators $A$ and $B$ with $B$ invertible, Eq.~\eqref{eq:deptoid} implies that
\begin{equation}
\sum_{G \in \mathbb{G}} \Omega(\mathcal{G}) \epsilon\left(\mathcal{E} \tilde{\mathcal{G}}(G) \mathcal{E}^{-1}, \mathcal{G}(G)\right)=r_{\gamma}
\end{equation}
By noting that $\tilde{\mathbb{G}}(\mathcal{E})=\{\mathcal{E} \tilde{\mathcal{G}}(G)\mathcal{E}^{-1} \mid G \in \mathbb{G}\}$, it follows immediately from the definition of $\epsilon_{\Omega}$, which is the average infidelity of a gate sampled from $\Omega$, that the left hand side of this equation is the $\epsilon_{\Omega}(\tilde{\mathbb{G}}(\mathcal{E}), \mathbb{G})$. As such $r_{\gamma}=\epsilon_{\Omega}(\tilde{\mathbb{G}}(\mathcal{E}), \mathbb{G})$, as required.
\end{proof}

We now summarize the results of our theory for direct RB in the following lemma:
\begin{lemma}
The direct RB error rate ($r_{\Omega}$) is approximately equal to the $\Omega$-weighted average infidelity of the benchmarked gate set: $r_{\Omega} \approx \epsilon_{\Omega}(\tilde{\mathbb{G}}(\mathcal{E}), \mathbb{G})$.
\end{lemma}

\begin{figure}[t!]
    \centering
    \includegraphics[scale=.6]{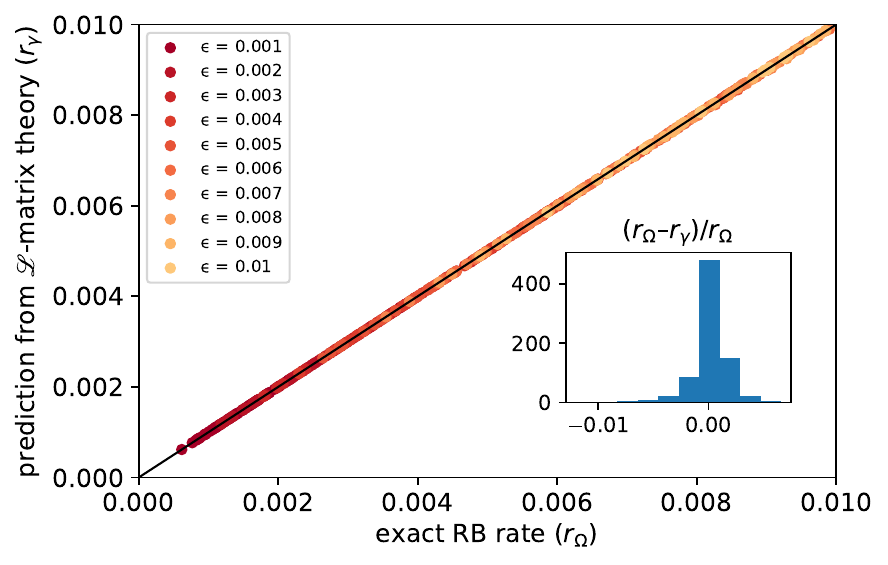}
    \caption{Verifying the accuracy of our $\mathscr{L}$ matrix theory for direct RB, using our exact $\mathscr{R}$ matrix theory. We show the $\mathscr{L}$ matrix theory's prediction for the direct RB error rate ($r_{\gamma}$) versus $r_{\Omega}$ computed from our exact $\mathscr{R}$ matrix theory (the $\mathscr{R}$ matrix theory enables computing $S_d$ exactly, and we then fit $S_d$ to $S_d=A+Bp^d$). Each data point shows $r_{\gamma}$ versus $r_{\Omega}$ for the same randomly generated Markovian error models as in Figure~\ref{fig:markovianscatter}. The inset shows a histogram of the relative error [$(r_{\Omega} - r_{\gamma})/r_{\Omega}$], demonstrating that our $\mathscr{L}$ matrix theory is extremely accurate. We observe extremely close agreement between $r_{\Omega}$ and $r_{\gamma}$, with no more than $\pm 1\%$ relative error (see inset histogram).}
    \label{fig:markovian}
\end{figure}

\subsection{Validating the theory using numerical simulation}\label{ssec:gen-theory-sims}
In this section we verify the correctness of our $\mathscr{L}$ matrix theory for direct RB, using numerical simulations. We simulated direct RB on a single qubit with a gate set consisting of $X(\nicefrac{\pi}{2})$ and $Y(\nicefrac{\pi}{2})$ gates. We choose an error model consisting of random Markovian errors with rates that contribute no more than $\varepsilon$ to the gate's infidelity, for some small $\varepsilon > 0$, using the error generator formalism in \cite{blume2022taxonomy}. That is, the superoperator $\tilde{\mathcal{G}}(G)$ for each gate is modelled as
\begin{equation}
    \tilde{\mathcal{G}}(G) =  \exp\left(\sum_{i=1}^{12} v_i(G) \mathcal{V}_i\right) \mathcal{G}(G)
\end{equation}
where the $\mathcal{V}_i$ form a basis for the space of Markovian errors. For this example, the $12$ error generators consist of three each of $C, A, S$ and $H$ type generators defined in \cite{blume2022taxonomy}---where $S$ and $H$ are Pauli stochastic and Hamiltonian error generators respectively, while the remaining ``Pauli-correlation'' and ``active'' error generators augment the stochastic errors. For Hamiltonian errors, the leading order contribution to a gate's infidelity is at second order, whereas for stochastic errors it is at first order. Therefore, the Hamiltonian error rates (the $v_i$) are chosen uniformly at random from  choose $[0, \sqrt{\varepsilon}]$, while the rates (the $v_i$) for all other errors are chosen uniformly from $[0,\varepsilon]$. We do this for $785$ different error models, varying $\varepsilon$ from $\varepsilon=.001$ to $\varepsilon=.01$. The results of the numerical experiments are shown in Figs.~\ref{fig:markovianscatter} and~\ref{fig:markovian}.

For each error model, Fig.~\ref{fig:markovianscatter} shows the prediction for $r_{\Omega}$ from the $\mathscr{L}$ matrix theory (which is $r_{\gamma}$) versus an estimate $\hat{r}_{\Omega}$ for $r_{\Omega}$ obtained by simulating direct RB circuits, and fitting the data to $S_d = A + Bp^d$. Error bars are $1\sigma$, and they are computed using a standard bootstrap. We observe good agreement between $r_{\gamma}$ (the theory's prediction) and $r_{\Omega}$. However, quantifying any small underlying discrepancies between $r_{\Omega}$ and $r_{\gamma}$ is difficult due to the uncertainties in the $\hat{r}_{\Omega}$---which arise from simulating a finite set of direct RB circuits. To remove this uncertainty, we computed the exact $S_d$ values, using our exact $\mathscr{R}$ matrix theory for direct RB, and we then fit these exact values for $S_d$ to $S_d=A+Bp^d$ to obtain $r_{\Omega}$. Figure~\ref{fig:markovian} shows $r_{\Omega}$ versus $r_{\gamma}$, and a histogram (inset) of the relative error $(r_{\Omega} - r_{\gamma})/r_{\Omega}$. We find extremely close agreement between $r_{\Omega}$ and $r_{\gamma}$, with no more than $\pm 1\%$ relative error. This therefore demonstrates the accuracy of our $\mathscr{L}$ matrix theory for direct RB.

\section{Conclusion}\label{sec:conclusions}
Direct RB \cite{proctor2018direct} is a method for benchmarking a set of quantum gates that does not require the inflexible circuit sampling and the complex gate-compilation of standard Clifford group RB \cite{magesan2011scalable}. This means that direct RB can benchmark more qubits than Clifford group RB, and direct RB can reliably extract information about the performance of the native gates of a device. In this paper we have presented a general definition for the direct RB protocol, and theories that explain how and why direct RB is broadly reliable. Our theory shows that direct RB is reliable as long as, whenever an error occurs, it almost certainly spreads on to many qubits before another error occurs. This proof that direct RB is reliable does not rely on convergence of a sequence of gates to a uniformly element from a unitary 2-design. This is crucial for proving that direct RB can be reliable on $n\gg1$ qubits with realistic error rates, as the number of layers required for such convergence necessarily increases with $n$. Furthermore, our theories show that the direct RB error rate can be interpreted as the average infidelity of the benchmarked gates, building on similar results for Clifford group RB~\cite{proctor2017randomized, wallman2017randomized}.

Clifford group RB has been extended to an entire suite of methods, including methods for estimating the error rates of individual gates \cite{magesan2012efficient, harper2017estimating, chasseur2017hybrid}, for quantifying the coherent component of gate errors \cite{wallman2015estimating, feng2016estimating, sheldon2016characterizing}, for estimating leakage and loss \cite{wood2017quantification,chasseur2015complete,wallman2015robust}, for calibrating gates \cite{rol2017restless, kelly2014optimal}, and for quantify crosstalk \cite{gambetta2012characterization, McKay2020-no}. We anticipate that all of these methods can be adapted to the more flexible direct RB framework, and fully developing these techniques is an interesting direction for future research. One limitation of direct RB is that it is not indefinitely scalable---as shown in Fig.~\ref{Fig:compiler}---because initialization and measurement of a random stabilizer state requires $O(n^2/\log n)$ two-qubit gates. Recent work on mirror RB \cite{proctor2021scalable, hines2022demonstrating, mayer2021theory} has shown that it is possible to adapt direct RB to remove these expensive steps, and much of the theory and many of the analysis techniques used in this paper are also applicable to that more scalable protocol.

\section*{Acknowledgements}
The authors thank Jahan Claes for providing a simplification to the proof of Proposition~\ref{prop:preservedsubspace}. This material is based upon work supported by the U.S. Department of Energy, Office of Science, Office of Advanced Scientific Computing Research, Quantum Testbed Pathfinder. This research was funded, in part, by the Office of the Director of National Intelligence (ODNI), Intelligence Advanced Research Projects Activity (IARPA). Sandia National Laboratories is a multi-program laboratory managed and operated by National Technology and Engineering Solutions of Sandia, LLC., a wholly owned subsidiary of Honeywell International, Inc., for the U.S. Department of Energy’s National Nuclear Security Administration under contract DE-NA-0003525. All statements of fact, opinion or conclusions contained herein are those of the authors and should not be construed as representing the official views or policies of the U.S. Department of Energy, IARPA, the ODNI, or the U.S. Government. AMP acknowledges funding from NSF award DMR-1747426 and a NASA Space Technology Graduate Research Opportunity award. 

\bibliographystyle{quantum}
\bibliography{Bibliography}

\newpage
\appendix
\clearpage

\section{Proof of Proposition~\ref{prop:lmatdep}}\label{app:proof}

\begin{proof} By linearity $\mathscr{L}_{\mathbb{G}, \tilde{\mathbb{G}}, \Omega}\left(\mathcal{E}_{1}+\mathcal{E}_{\gamma}\right)=\mathcal{E}_{1}+\gamma \mathcal{E}_{\gamma}$. If $\gamma=1$ then the proposition holds trivially, so consider the case of $\gamma<1$. If $\gamma<1$ then, because $\tilde{\mathbb{G}}$ is a low-error implementation of $\mathbb{G}$, both 1 and $\gamma$ are non-degenerate eigenvalues of $\mathscr{L}_{\mathbb{G}, \tilde{\mathbb{G}}, \Omega}$. Therefore, up to constant scalings, $\mathcal{E}_{1}$ and $\mathcal{E}_{\gamma}$ are unique. Define
\begin{equation}
\tilde{\mathcal{G}}_{\Omega-\mathrm{avg}} \equiv \sum_{\mathcal{G} \in \mathbb{G}} \Omega(\mathcal{G}) \tilde{\mathcal{G}}, \quad \mathcal{G}_{\Omega-\mathrm{avg}} \equiv \sum_{\mathcal{G} \in \mathbb{G}} \Omega(\mathcal{G}) \mathcal{G},
\end{equation}
where we abuse notation and consider all gates as these as superoperators. Let $\sbra{V}$ denote some left eigenvector of the Hilbert-Schmidt space matrix $\tilde{\mathcal{G}}_{\Omega-\text { avg }}$ with eigenvalue $1$, i.e., $\sbra{V} \tilde{\mathcal{G}}_{\Omega-\text { avg }}=\sbra{V}$. This eigenvector exists because the elements of $\tilde{\mathbb{G}}$ are trace-preserving maps, which implies that $\tilde{\mathcal{G}}_{\Omega-\text {avg}}$ is also a trace-preserving map, and all trace-preserving maps have $1$ as an eigenvalue. Because the the elements of $\mathbb{G}$ are the superoperators for unitary gates, $\mathcal{G}^{-1}\sket{B_{0}}=\sket{B_{0}}$ for all $\mathcal{G} \in \mathbb{G}$, where $B_{0}=\mathbb{1}_{d} / \sqrt{d}$ (here we are using the basis for Hilbert-Schmidt space introduced in Section \ref{sec:notation}). This is the statement that unitary maps are unital. Hence we have that
\begin{equation}
\begin{aligned}
\mathscr{L}_{\mathbb{G}, \tilde{\mathbb{G}}, \Omega}(\sket{B_{0}}\sbra{V}) &=\sum_{\mathcal{G} \in \mathbb{G}} \Omega(\mathcal{G})[\mathcal{G}^{-1}\sket{B_{0}}\langle V| \tilde{\mathcal{G}}] \\
&=\sket{B_{0}}\sbra{V} \tilde{\mathcal{G}}_{\Omega-\mathrm{avg}},\\
&=\sket{B_{0}}\sbra{V}
\end{aligned}
\end{equation}
and therefore
\begin{equation}\label{eq:e1}
\mathcal{E}_{1}=\sket{B_{0}}\sbra{V}.
\end{equation}
Using the ``stacked'' representation discussed in Section~\ref{sec:notation}, we may turn $\mathscr{L}_{\mathbb{G}, \tilde{\mathbb{G}}, \Omega}(\mathcal{E}_{\gamma})=\gamma \mathcal{E}_{\gamma}$ into a more standard eigenvector equation.

Let $\mathcal{G}_{\mathrm{u}}^{-1}=\mathcal{G}^{-1}-\sket{B_{0}}\sbra{B_{0}}$ for all $\mathcal{G} \in \mathbb{G}$, which is known as the unital component of $\mathcal{G}^{-1}$. Because the elements in $\mathbb{G}$ are unitary gates
\begin{equation} \sbra{B_{0}}\mathcal{G}_{\mathrm{u}}^{-1}=\mathcal{G}_{\mathrm{u}}^{-1}\sket{B_{0}}=0,
\end{equation}
for all $\mathcal{G} \in \mathbb{G}$. Therefore, for any superoperator $\mathcal{X}$,
\begin{equation}
\begin{aligned}
\operatorname{vec}(\mathscr{L}_{\mathbb{G}, \tilde{\mathbb{G}}, \Omega}(\mathcal{X})) &=\sum_{\mathcal{G} \in \mathbb{G}} \Omega(\mathcal{G})[\tilde{\mathcal{G}}^{T} \otimes \mathcal{G}^{-1}] \operatorname{vec}(\mathcal{X}) \\
&=\sum_{\mathcal{G} \in \mathbb{G}} \Omega(\mathcal{G})[\tilde{\mathcal{G}}^{T} \otimes(\sket{B_{0}}\langle B_{0}|+\mathcal{G}_{\mathrm{u}}^{-1})] \operatorname{vec}(\mathcal{X}) \\
&=[\mathcal{L}_{0}+\mathcal{L}_{\perp}] \operatorname{vec}(\mathcal{X})
\end{aligned}
\end{equation}
with $\mathcal{L}_{0}=\tilde{\mathcal{G}}^{T}_{\Omega-\mathrm{ avg }} \otimes\sket{B_{0}}\langle B_{0}|$ and
\begin{equation}
\mathcal{L}_{\perp}=\sum_{\mathcal{G} \in \mathbb{G}} \Omega(\mathcal{G})(\tilde{\mathcal{G}}^{T} \otimes \mathcal{G}_{\mathrm{u}}^{-1}).
\end{equation}
Because $\mathcal{G}_{\mathrm{u}}^{-1}\sket{B_{0}}\langle B_{0}| = \sket{B_{0}}\langle B_{0}|\mathcal{G}_{\mathrm{u}}^{-1} = 0$, we have that the $\mathcal{L}_{0}$ and $\mathcal{L}_{\perp}$ matrices satisfy
\begin{equation}
\mathcal{L}_{0} \mathcal{L}_{\perp}=\mathcal{L}_{\perp} \mathcal{L}_{0}=0 .
\end{equation}
As such, the set of eigenvalues of $\mathscr{L}$ is the union of the sets of eigenvalues of $\mathcal{L}_{0}$ and $\mathcal{L}_{\perp}$. Moreover, if $\mathcal{V}$ is an eigen-operator of $\mathscr{L}$ and if the associated eigenvalue is not an eigenvalue of $\mathcal{L}_{0}$, then
\begin{equation}
\operatorname{vec}(\mathcal{V})=\sum_{j} \sum_{k>0} v_{j k}\sket{B_{j}} \otimes\sket{B_{k}},
\end{equation}
for some $v_{j k}$, and so $\sbra{B_{0}} \mathcal{V}=0$.

We now show that $\gamma$ is not an eigenvalue of $\mathcal{L}_{0}$, which then implies that $\sbra{B_{0}} \mathcal{E}_{\gamma}=0$. The eigenvalues of $\mathcal{L}_{0}$ are the eigenvalues of $\tilde{\mathcal{G}}_{\Omega-\mathrm{avg}}$, together with $0$. $\mathcal{G}_{\Omega-\mathrm{avg}}$ is a sequence-asymptotic unitary 2-design, and thus has $1$ as a non-degenerate eigenvalue. We will show this is also true for $\tilde{\mathcal{G}}_{\Omega-\mathrm{avg}}$.

In \cite{wallman2017randomized}, the author shows that, if the generating set is a 2-design, $\tilde{\mathcal{G}}_{\Omega-\mathrm{avg}}$ will generally have a unique largest eigenvalue. We extend this argument to sequence-asymptotic unitary 2-designs. The limiting map of the generator twirl is a 2-design twirl, and so in the limit the result from \cite{wallman2017randomized} holds. Now, suppose that it did not hold that the generator twirl had a non-degenerate 1-eigenspace. At no point in the limiting sequence could it acquire one, thus it has to have one to begin with.

Now, suppose by way of contradiction that $\gamma$ is an eigenvalue of $\tilde{\mathcal{G}}_{\Omega-\mathrm{avg}}$. Since $\gamma$ is a perturbation of $1$, in the limit that the perturbation goes to $0$, $\tilde{\mathcal{G}}_{\Omega-\mathrm{avg}}$ would acquire $1$ as a degenerate eigenvalue, thus $\gamma$ cannot be an eigenvalue of $\tilde{\mathcal{G}}_{\Omega-\mathrm{avg}}$.

Thus, $\mathcal{E}_{\gamma}$ is not an eigenvector of $\mathcal{L}_0$ and hence
\begin{equation}
\sbra{B_{0}} \mathcal{E}_{\gamma}=0.
\end{equation}
This equality implies that
\begin{equation}\label{eq:gamma_decomp}
\mathcal{E}_{\gamma}=\sum_{i} \sum_{j>0} \alpha_{i j}\sket{B_{j}}\sbra{B_{i}},
\end{equation}
for some $\alpha_{i j}$. A depolarization channel $\mathcal{D}_{\lambda} \operatorname{maps} \rho \rightarrow \rho$ with probability $\lambda$ and $\rho \rightarrow \mathbb{1} / d$ with probability $1-\lambda$. As such, in Hilbert-Schmidt space with the basis considered here
\begin{equation}
\mathcal{D}_{\lambda}=\sket{B_{0}}\sbra{B_{0}}+\lambda \sum_{i \neq 0}\sket{B_{i}}\sbra{B_{i}}.
\end{equation}
Hence, from Eqs.~\eqref{eq:gamma_decomp} and \eqref{eq:e1} we have that $\mathcal{E}_{1}+\gamma \mathcal{E}_{\gamma}=\mathcal{D}_{\gamma}(\mathcal{E}_{1}+\mathcal{E}_{\gamma})$. Finally, using this relation and considering the action of $\mathscr{L}_{\mathbb{G}, \tilde{\mathbb{G}}, \Omega}$ on $\mathcal{E}=\mathcal{E}_{1}+\mathcal{E}_{\gamma}$, we have that
\begin{equation}
\begin{aligned}
\mathscr{L}_{\mathbb{G}, \tilde{\mathbb{G}}, \Omega}(\mathcal{E}) &=\mathcal{E}_{1}+\gamma \mathcal{E}_{\gamma}, \\
&=\mathcal{D}_{\gamma}(\mathcal{E}_{1}+\mathcal{E}_{\gamma}), \\
&=\mathcal{D}_{\gamma} \mathcal{E},
\end{aligned}
\end{equation}
which concludes the proof.
\end{proof}

Note that the $\mathcal{E}$ in this proposition is not necessarily a completely positive map, and generically it is not. 

\section{Sampling distributions}\label{app:samplers}
In this appendix we present one possible family of sampling distributions ($\Omega$), called the \emph{edge grab} sampler and first introduced in~\cite{proctor2020measuring}. This sampling distribution is designed for sampling from an $n$-qubit gate set $\mathbb{G}$ containing gates consisting of a single layer of parallel one- and two-qubit gates, with one-qubit gates from some set $\mathbb{G}_1$ and two-qubit gates only between a connected qubits as specified by an edge list $E$ (that can contain directed edges). The sampling distribution is parameterized by the mean two-qubit gate of the sampled gates $\bar{\xi}$. The two-qubit gate density of an $n$-qubit layer is simply $\xi = \nicefrac{2\alpha}{n}$ where $\alpha$ is the number of two-qubit gates in the layer. The edge grab sampling distribution $\Omega_{\bar{\xi}}$ is most easily described by an algorithm for drawing a sample from  $\Omega_{\bar{\xi}}$. Drawing a sampling consists of the following procedure:
\begin{enumerate}
\item \emph{Select a candidate set of edges} $E_{c}$. Initialize $E_{c}$ to the empty set, and initialize $E_{r}$ to the set of all edges $E$. Then, until $E_{r}$ is the empty set:
\begin{enumerate} 
\item[1.1] Select an edge $v$ uniformly at random from $E_{r}$.
\item[1.2] Add $v$ to $E_{c}$ and remove all edges that have a qubit in common with $v$ from $E_{r}$.
\end{enumerate}
\item \emph{Select a subset of the candidate edges}. For each edge in $E_{c}$, include it in a final edge set $E_{f}$ with a probability of $n\bar{\xi}/(2|E_{c}|)$ where $|E_{c}|$ is the total number of edges in $E_{c}$.  If $n\bar{\xi}/(2|E_{c}|) > 1$ construct a new edge set sample $E_c$.
\item \emph{Construct the sampled gate}. The sampled gate $G \in \mathbb{G}$ is then constructed by adding a two-qubit gate on each edge in $E_{f}$ and for all remaining qubits independently and uniformly sample a single-qubit gate from $\mathbb{G}_1$ to include in $G$.
\end{enumerate}

The expected number of two-qubit gates in $G$ is $n \bar{\xi}/2$, so this sampler generates an $n$-qubit layer with an expected two-qubit gate density of $\bar{\xi}$. One useful property of this sampling is that the probability of sampling any particular $n$-qubit gate $G$ is non-zero for every $G \in \mathbb{G}$ (except if $\bar{\xi}=0$ or $\bar{\xi} = 1$). Finally, note that the edge grab algorithm is invalid if $n\bar{\xi}/(2|E_{c}|) > 1$ for any possible candidate edge set $E_c$. For an even number of fully-connected qubits, $\bar{\xi}$ can take any value between $0$ and $1$. But for any other connectivity the maximum achievable value of $\bar{\xi}$ is smaller. In our experiments, we set $\bar{\xi} = \nicefrac{1}{8}$. This is an achievable value of $\bar{\xi}$ in the edge grab algorithm for all the processors that we benchmarked.
\end{document}